\documentclass[fleqn,usenatbib]{mnras}

\usepackage{newtxtext,newtxmath}
\usepackage[T1]{fontenc}

\DeclareRobustCommand{\VAN}[3]{#2}
\let\VANthebibliography\thebibliography
\def\thebibliography{\DeclareRobustCommand{\VAN}[3]{##3}\VANthebibliography}

\usepackage{graphicx}	% Including figure files
\usepackage{amsmath}	% Advanced maths commands
\usepackage{pdflscape}  % Landscape pages
\usepackage{caption}
\usepackage{subcaption}
\usepackage{multirow}
\usepackage{multicol}
\usepackage[normalem]{ulem}
\usepackage{threeparttable}

\defcitealias{2024MNRAS.532.1991B}{Paper I}
\newcommand{\paperone}{\citetalias{2024MNRAS.532.1991B}}

\title[SPOTS: First results]{Spectro-Polarimetric Observations of TeV Sources (SPOTS): First results}

\author[J. Barnard et al.]{
Joleen Barnard,$^{1}$\thanks{E-mail: elsj@ufs.ac.za}
B. van Soelen,$^{1}$\thanks{E-mail: vansoelenb@ufs.ac.za}
I.P. van der Westhuizen,$^{1}$
M. B\"{o}ttcher,$^{2}$
A. Martin-Carrillo,$^{3}$ \newauthor
H.M. Schutte,$^{4}$
S.\ van der Merwe,$^{5}$
and M. Zacharias$^{6,2}$
\\
$^{1}$Department of Physics, University of the Free State, Bloemfontein, 9301, South Africa\\
$^{2}$Centre for Space Research, North West University, Potchefstroom, 2520, South Africa\\
$^{3}$School of Physics and Centre for Space Research, University College Dublin, Belfield, Dublin 4, Ireland\\
$^{4}$Michigan Technological University, Houghton, Michigan, 49931, United States of America\\
$^{5}$Department of Mathematical Statistics and Actuarial Science, University of the Free State, Bloemfontein 9301, South Africa \\
$^{6}$Landessternwarte, Universität Heidelberg, Königstuhl 12, 69117 Heidelberg, Germany
}

\date{Accepted XXX. Received YYY; in original form ZZZ}

\pubyear{\the\year{}}

\begin{document}
\label{firstpage}
\pagerange{\pageref{firstpage}--\pageref{lastpage}}
\maketitle

\begin{abstract}
Blazars are jetted active galactic nuclei, with the jet aligned along the observer's line of sight. Their spectral energy distributions are dominated by non-thermal emission, with an underlying thermal component at optical/ultraviolet wavelengths. However, the underlying jet magnetic field structure and particle acceleration mechanisms requirements for the non-thermal emission are still under debate. Polarization measurements can provide critical insights, and we investigate the optical polarization properties of TeV-emitting blazars using long-term optical monitoring. We present results from the first 21-months of the Spectro-Polarimetric Observations of TeV Sources (SPOTS) campaign, using the Southern African Large Telescope, of 14 blazars. Overall, observations of the sample during this campaign showed a low average optical polarization ($\Pi\lesssim10\%$). While some sources exhibited smooth polarization angle rotations on timescales of days to weeks, others showed stochastic variations consistent with turbulent magnetic fields. The average ordering of the magnetic field was low ($F_B\lesssim0.10$), consistent with emission arising in extended, turbulent regions of the jet. For individual sources, correlations between polarization and its frequency dependence were found, but were not found across the entire sample. The nature of the frequency dependence varied between observations, indicating that simple one-zone models are insufficient and that $F_B$ must be wavelength dependent. This study highlights the complex nature of blazar jets and underscores the importance of long-term, multi-wavelength polarimetric monitoring. This comprehensive dataset enables detailed modelling of individual sources, and provides valuable context for future X-ray polarimetry observations.
\end{abstract}

\begin{keywords}
galaxies: active -- galaxies: jets -- polarization -- techniques: polarimetric
\end{keywords}

%%%%%%%%%%%%%%%%%%%%%%%%%%%%%%%%%%%%%%%%%%%%%%%%%%

%%%%%%%%%%%%%%%%% BODY OF PAPER %%%%%%%%%%%%%%%%%%

\section{Introduction}

Active galactic nuclei (AGN), found at the centre of some galaxies, are powered by the accretion of material onto a central, supermassive black hole \citep[SMBH;][]{2017A&ARv..25....2P, 1995PASP..107..803U}. Blazars are part of the jetted (radio-loud) class of AGN in which the direction of motion of the highly collimated jet is closely aligned with the observer's line of sight. These jets are composed of highly relativistic particles and originate either from accretion driven processes \citep[see e.g.][]{1982MNRAS.199..883B} or from the extraction of rotational energy from a SMBH through Poynting flux of the rotating magnetosphere \citep[e.g.][]{1977MNRAS.179..433B}. Due to the viewing geometry and the relativistic motion of these jets, the observed emission is highly Doppler-boosted. Beaming effects and/or rapid magnetic reconnection events close to the black hole may result in extreme variability, with timescales ranging from minutes \citep[e.g.][]{2009MNRAS.395L..29G, 2009A&A...502..749A, 2005ApJ...630..130B, 2007ApJ...669..862A} to years \citep{2019ARep...63..316K, 2007A&A...467..465C, 2021MNRAS.505.6103P}. Additionally, beaming effects boost the emission, resulting in blazars being the most numerous $\gamma$-ray sources in the extragalactic sky \citep{2020ApJS..247...33A}.

Blazars can be classified into two different types -- flat-spectrum radio quasars (FSRQs) and BL Lacertae type objects (BLLs). This classification is based on the presence or absence of spectral features, where FSRQs display strong, broad (equivalent widths, $\vert W_\lambda \vert > 5$\,\rm{\AA}) emission lines, and BLLs have weak (or absent), narrow ($\vert W_\lambda \vert < 5$\,\rm{\AA}) emission lines or featureless spectra \citep[see e.g.][and references therein]{2004MNRAS.351...83L}. BLLs are also subdivided based on their synchrotron peak frequency ($\nu_{\rm sy}$) into low-synchrotron peaked ($\nu_{\rm sy} < 10^{14}$; LBL), intermediate-synchrotron peaked ($10^{14} \leq \nu_{\rm sy} \leq 10^{15}$; IBL), and high-synchrotron peaked sources \citep[$\nu_{\rm sy} > 10^{15}$; HBL;][]{1995ApJ...444..567P, 1998MNRAS.299..433F}.

The spectral energy distributions (SEDs) of blazars are dominated by two broad, non-thermal radiation components originating in the jet. The low-energy component ranges from radio to optical/UV, sometimes extending to soft X-rays, and is produced by leptonic synchrotron radiation from the relativistic electrons, as evident from the high degrees of polarization observed in the radio and optical regimes \citep[see e.g.][and references therein]{2023MNRAS.523.4504O, 2014A&A...566A..59A}. Within the infrared to ultraviolet wavelengths there is also an underlying thermal contribution from a combination of the accretion disc, dust torus, broad-line region, and the host galaxy itself. The low-energy component is, therefore, a superposition of thermal and non-thermal emission contributions. Due to the variable nature of the jet emission, the non-thermal flux and polarization levels fluctuate over time \citep[e.g.][]{2017Galax...5...52B}, while the thermal component remains comparatively steady, and contributes to a dilution of the overall observed polarization. 

Various models have been proposed for the mechanisms producing the high-energy component (ranging from X-ray to $\gamma$-rays; extending into GeV and sometimes $>$TeV energies). These fall into two main scenarios, namely leptonic and hadronic. In the leptonic scenario the high-energy component is inverse-Compton (IC) emission produced by the same population of electrons that produces the synchrotron radiation. This occurs either through synchrotron self-Compton (SSC) where the electrons up-scatter the synchrotron photons produced by the same electron population, or through external Compton (EC) where the target photon field originates externally to the jet \citep[see e.g.][and references therein]{2013ApJ...768...54B, 2016ApJ...830...94F}. Hadronic models, however, argue that the emission is produced by proton-initiated synchrotron radiation and/or photo-pion processes, in which pions and muons decay and cause cascade emission \citep{1993A&A...269...67M, 2003APh....18..593M, 2019Galax...7...85Z}. 

Single zone models are commonly used to explain blazar SEDs, which assume that the bulk of the observed emission arises from a single, compact region or ``blob'' of material. Such models have been successfully used to explain the correlated multi-wavelength (MWL) variability observed in some blazars \citep[see e.g.][and references therein]{2023MNRAS.526.5054L}. However, the rapid variability that some TeV blazars exhibit cannot be sufficiently explained by single-zone models \citep{2013ApJ...768...54B}. These models also fail to reproduce the ``orphan flares'' that have been observed in some blazars \citep[see e.g.][]{2017ATel10791....1T}. Although hadronic models can explain this phenomenon, as well as the detection of high-energy neutrinos, they require much more energy in the jet \citep[see e.g.][]{2013ApJ...768...54B}. This indicates that the emission is more likely leptonic, originating in multiple emission regions along the jet, which often provides a better explanation for the observed behaviour \citep[see e.g.][]{2023MNRAS.519.6349D, 2025A&A...698L..19L}. An example of multi-zone modelling has recently been shown for the FSRQ PKS~1510-089, which exhibited a significant flux drop in the high energy $\gamma$-ray and optical bands, while the X-ray and very high-energy (VHE) bands remained steady \citep{2023ApJ...952L..38A}. It was argued that the best explanation was that the emission came from at least two emission regions, and that the primary emission zone, producing the GeV and optical emission, vanished. Such events demonstrate that the jet structure and emission processes are more complex than single-zone interpretations suggest. Consequently, differentiating between the various scenarios requires having good constraints on the particle population producing the low-energy emission. However, a full picture of the non-thermal jet emission cannot be constructed by spectral modelling alone.

Polarization measurements can provide additional information that can remove the ambiguity in the spectral modelling. For example, optical polarization can be used to disentangle the thermal (unpolarized) emission for the host galaxy, accretion disc and torus, from the non-thermal (polarized) emission originating from the jet \citep{2022ApJ...925..139S, 2017Galax...5...52B}. This, in turn, provides insight into the jet's magnetic field geometry, structure and orientation, as well as the particle acceleration mechanisms producing the high-energy component. The measurement of polarization at higher energies could also distinguish between leptonic and hadronic $\gamma$-ray emission \citep{2013ApJ...774...18Z, 2014ApJ...789...66Z, 2022Natur.611..677L}.

Polarization measurements, therefore, are crucial for investigating how very high energy TeV $\gamma$-ray emission is produced in blazars, as different polarization levels are expected based on how and where the emission is produced in the jet, as well as what the structure and strength of the jet magnetic field is \citep[see e.g.][and references therein]{2019Galax...7...85Z, 2014ApJ...788..104Z}. As a result, many polarimetric monitoring campaigns have been undertaken at radio and optical wavelengths, which include (but are not limited to) MOJAVE \citep{2018ApJS..234...12L}, POLAMI \citep{2018MNRAS.474.1427A}, Boston University's radio BEAM-ME and optical MOBPol projects \citep[][]{2016Galax...4...47J, 2021Galax...9...27M, 2017ApJ...846...98J, 2022ApJS..260...12W, 2021Galax...9...27M}, the Kanata Telescope polarimetry monitoring campaign \citep{2018Galax...6...16I}, the Liverpool Telescope's polarimetry with the RINGO3 polarimeter \citep{2004SPIE.5489..679S}, the Steward Observatory's Ground-based Observational Support of the \textit{Fermi} Gamma-ray Space Telescope \citep{2009arXiv0912.3621S}, and RoboPol \citep[e.g.][]{2014MNRAS.442.1693P}.

More recently the Imaging X-ray Polarimetry Explorer \citep[IXPE;][]{2022JATIS...8b6002W} has, since its launch in 2021, detected statistically significant X-ray polarization from six HBLs and obtained upper limits on the X-ray polarization of various IBLs and LBLs \citep[see e.g.][and references therein]{2024Galax..12...50M}. In HBLs, the X-ray emission is produced by the highest-energy electrons, which cool rapidly, and their emission shifts to lower energies as they move away from the acceleration site. The observed X-ray polarization is, therefore, expected to originate close to regions where the particle acceleration is most efficient and the magnetic field is most ordered. Consequently, IXPE measurements have provided insight into the acceleration mechanism at work within the jet, as different mechanisms would provide different levels of polarization. Two main acceleration mechanisms have been proposed: shocks in the jet or magnetic reconnection \citep{2024Galax..12...50M, 2025ApJ...988L..50B, 2022ApJ...938L...7D, 2022Natur.611..677L}.

In the shock-in-jet scenario \citep{1985ApJ...298..114M}, particles are accelerated in a small region and move downstream along the jet while they rapidly lose energy. Close to the shock front, the magnetic field is highly ordered and leads to higher degrees of polarization observed at higher energies. The magnetic field rapidly becomes more turbulent downstream, and the particles diffuse into more/larger regions leading to lower degrees of polarization observed at lower energies, with the polarization angle aligning closely to the jet position angle \citep{2018MNRAS.480.2872T}. Alternatively, in the magnetic reconnection scenario \citep{2015MNRAS.450..183S}, the magnetic energy is dissipated along the jet with a highly disordered magnetic field which causes lower degrees of optical and X-ray polarization and randomly distributed polarization angles \citep{2025arXiv250814168C}.

Early IXPE observations of several blazars produced X-ray polarization measurements and MWL constraints that are consistent with a leptonic, energy-stratified, shock-acceleration mediated SSC origin of their X-ray emission. In these cases, the degree of polarization is strongly chromatic, with the X-ray polarization exceeding the optical polarization ($\langle \Pi_{\rm X} / \Pi_{\rm O} \rangle \sim 1.7$ - $7.5$) and the polarization angle remaining stable and aligned with the jet direction \citep[see e.g.][]{2023ApJ...953L..28M, 2023ApJ...942L..10M, 2023ApJ...948L..25P, 2025arXiv250814168C, 2024Galax..12...50M, 2022Natur.611..677L}.

However, more recent IXPE results across a larger blazar sample have revealed some inconsistencies with the shock scenario. These inconsistencies include weak chromaticity in the degree of polarization, polarization angle rotations, and mismatches between the X-ray and optical polarization angles \citep{2025ApJ...988L..50B, 2025ApJ...986..230M, 2024ApJ...963....5E}. These signatures point toward scenarios involving magnetic reconnection, turbulence, or multiple emission zones.  Relativistic magnetic reconnection modelling and simulations have reproduced the observed X-ray flux and polarization properties and, therefore, remain a viable competing mechanism \citep{2025arXiv251013776D}. Consequently, while the initial IXPE results hinted towards a leptonic SSC interpretation for some sources, the underlying particle acceleration may be driven by magnetic reconnection and/or turbulence rather than shocks alone.

In light of the discussion above, optical spectropolarimetric observations of blazars that provide the degree (percentage) and angle of polarization resolved across optical wavelengths are crucial to provide supporting data for the IXPE findings and for MWL modelling. It is an extremely useful diagnostic tool for disentangling ordered and chaotic magnetic field components in the optical regime \citep{2015MNRAS.453.1669B}, shedding light on the frequency dependence of the polarization. This, in turn, allows for investigating shock propagation and turbulence in the jet, polarization angle rotations, as well as the change in the thermal spectral features arising from the line regions and host galaxy. This information is used to model the contribution of the thermal accretion disc component along with the presence of shocks in the jet, as the flux-polarization correlation gives an indication of the ordered/chaotic structure of the jet's magnetic field \citep[e.g.][]{2020MNRAS.498..599T}. This was shown with simultaneous SED and spectropolarimetry modelling of the FSRQ 4C +01.02 \citep{2022ApJ...925..139S}, which indicated an accretion-disc dominated spectrum and polarization decreasing towards higher frequencies. Alternatively, an increase in the degree of polarization towards higher frequencies, characteristic of ``shock-in-jet'' signatures \citep[i.e. polarization increasing towards higher frequencies][]{2016MNRAS.463.3365A}, has been found in other blazars in \citet[][hereafter Paper I]{2024MNRAS.532.1991B}.

\paperone\ presented results for a sample of 18 blazars observed during flaring and quiescent states, taken with the Southern African Large Telescope \citep[SALT;][]{2006_SALT_Buckley}, located at the South African Astronomical Observatory (SAAO). While target-of-opportunity (ToO) observations have given valuable insight into the flaring state of blazars \citep[e.g.][]{2004ApJ...601..151K, 2017ICRC...35..652S, 2017AIPC.1792e0029C}, they fail to provide a comprehensive, unbiased view of quiescent states and do not reveal the full range of blazar behaviour. Therefore, a more robust, long-term monitoring campaign is necessary. Many studies have been undertaken on individual TeV-emitting blazars \citep[including, but not limited to,][]{2023A&A...679A..28P, 2020ApJ...891..170V, 2022MNRAS.515.2633P}, as well as the broader TeV blazar population \citep[e.g.][]{2016RAA....16..103L, 2023PASP..135h4103L, 2024RAA....24i5005S}. However, large-scale, dedicated surveys that combine optical spectra and polarization with MWL observations of these sources remain limited. 

To this end, a long-term optical spectropolarimetric monitoring campaign of a selection of TeV emitting blazars is being undertaken with SALT. This campaign -- Spectro-Polarimetric Observations of TeV Sources, or SPOTS -- is a crucial step towards a more holistic understanding of TeV emitting blazars. While $>$1400 BLLs, $>$700 FSRQs, and $>$1400 blazar candidates are identified at GeV energies by \textit{Fermi}-LAT \citep{2022ApJS..260...53A}, the TeV emitting population is significantly smaller, with only 93 known AGN, as listed in the TeVCat\protect\footnote{\url{https://tevcat.org/}} catalogue \citep{2024RAA....24i5005S, 2008ICRC....3.1341W}. Of these, only eleven are FSRQs, and the vast majority of the rest are HBLs. The SPOTS campaign is uniquely capable of probing TeV blazar features and place better constraints on the high-energy emission, especially during quiescent states.

This paper provides an overview of the optical data taken with the SPOTS campaign, including the source selection criteria (Section~\ref{sec:sourceselection}) and observations and data reduction processes (Section~\ref{sec:obs}). Furthermore, it will provide a summary of the observations taken since the start of the campaign in 2023 November to 2025 August, along with the observed properties of the sources in Section~\ref{sec:results}. Section~\ref{sec:concl} will provide a discussion on these results.

\section{Source Selection}
\label{sec:sourceselection}

The sources included in the SPOTS campaign were selected from known TeV emitting blazars. This list was filtered by selecting all sources that were visible with SALT (${\rm Dec} < 11^\circ$), and had visual magnitudes V $< 17$\,mag to ensure feasible SALT observation times. Sources were also filtered to include those that were previously part of RoboPol, MOBPol, and the Steward observatory polarimetry monitoring campaigns. Finally, the sources were ranked based on their \textit{Fermi}-LAT variability, where the more variable sources took precedence. Table~\ref{tab:blazar_properties} summarizes the properties of the fourteen (14) targets observed between 2023 November and 2025 August. This includes 9 HBLs, 1 IBL, 2 LBLs, and 2 FSRQs.

%%%%%%%%%%%%%%%%%%%%%%%%%%%%%%%%%%%%%%%%%%%%%%%%%%%%%%%%%%%%%%%%%%%%%%%%%%%%%%%%%%%%%%%%%%%%%%%%%%%%%%%%%%%%%%
\begin{table*}
\begin{center}
\small
\begin{threeparttable}
\caption{A summary of the blazars that form part of the SPOTS monitoring campaign, including coordinates, type, redshift, the average measured apparent visual magnitude during this campaign, and synchrotron peak frequency. Most values were obtained from the TeVCat database \citep{2008ICRC....3.1341W}, Simbad database \citep{2000A&AS..143....9W}, and the NASA/IPAC Extragalactic Database\tnote{a}. The synchrotron peak frequencies ($\nu_{\rm sy}$) were taken from the Fourth Catalog of Active Galactic Nuclei detected by the LAT \citep[4LAC;][]{2020arXiv201008406L, 2020ApJ...892..105A}.}
\label{tab:blazar_properties}
\begin{tabular}{lcccccc}
\hline
Target & RA (J2000) & Dec. (J2000) & Type & $V$-mag & Redshift ($z$) & $\nu_{\rm sy}$ \cr
& [hh:mm:ss] & [$^\circ:\arcmin:\arcsec$] & & & & [Hz] \cr
\hline
RBS 0248 & 01:52:33.5 & +01:46:40.3 & HBL & 15.76 & 0.08 & 3.80$\times 10^{16}$ \cr
1RXS J023832.6-311658 & 02:38:32.5 & -31:16:58 & HBL & 16.71 & 0.23 & 2.02$\times 10^{16}$ \cr
PKS 0301-243 & 03:03:23.49 & -24:07:35.86 & HBL & 16.10 & 0.27 & 2.51$\times 10^{15}$ \cr
1ES 0414+009 & 04:16:52.96 & +01:05:20.4 & HBL & 17.57 & 0.29 & 4.90$\times 10^{16}$ \cr
PKS 0447-439 & 04:49:28.2 & -43:50:12 & HBL & 14.02 & 0.11 & 4.47$\times 10^{15}$ \cr
TXS 0506+056 & 05:09:25 & +05:42:09 & IBL & 15.59 & 0.34 & 3.55$\times 10^{14}$ \cr
PKS 0736+017 & 07:39:17.0 & +01:36:12 & FSRQ & 16.34 & 0.19 & 1.51$\times 10^{13}$ \cr
PKS 1440-389 & 14:44:00.2 & -39:08:21 & HBL & 14.96 & 0.14 & 4.47$\times 10^{15}$ \cr
PKS 1510-089 & 15:12:52.2 & -09:06:21.6 & FSRQ & 16.64 & 0.36 & 1.10$\times 10^{13}$ \cr
AP Lib & 15:17:41.8 & -24:22:19.5 & LBL & 15.06 & 0.05 & 9.06$\times 10^{13}$ \cr
PKS 1749+096 & 17:51:32.82 & +09:39:00.73 & LBL & 17.02 & 0.32 & 7.94$\times 10^{12}$ \cr
PKS 2005-489 & 20:09:27.0 & -48:49:52 & HBL & 13.95 & 0.07 & 2.00$\times 10^{15}$ \cr
PKS 2155-304 & 21:58:52.7 & -30:13:18 & HBL & 13.83 & 0.12 & 5.69$\times 10^{15}$ \cr
1ES 2322-409 & 23:24:48.00 & -40:39:36.0 & HBL & 16.00 & 0.17 & 5.69$\times10^{15}$ \cr
\hline
\end{tabular}

\begin{tablenotes}
\item[a] The NASA/IPAC Extragalactic Database (NED) is funded by the National Aeronautics and Space Administration and operated by the California Institute of Technology.
\end{tablenotes}

\end{threeparttable}
\end{center}
\end{table*}
%%%%%%%%%%%%%%%%%%%%%%%%%%%%%%%%%%%%%%%%%%%%%%%%%%%%%%%%%%%%%%%%%%%%%%%%%%%%%%%%%%%%%%%%%%%%%%%%%%%%%%%%%%%%%%

\section{Observations and Data Reduction}
\label{sec:obs}

\subsection{Optical spectropolarimetry}
\label{sec:specpol}

\subsubsection{Observational setup}

All optical spectropolarimetric observations were taken using the Robert Stobie Spectrograph \citep[RSS;][]{2003SPIE.4841.1463B, 2003SPIE.4841.1634K} in \textsc{linear} mode, using the PG0900 grating at a grating angle of 14.375$^{\circ}$, and with a slit width of $1.5"$. This configuration provides a wavelength coverage of $\lambda \sim 3920$ to $6990$\,\rm{\AA}, with a resolving power of $R \sim 770$ to $1140$. An Argon arc frame was taken immediately after every science observation cycle. For each target, the slit orientation was chosen so that a comparison star also lay within the slit. This was done in order to compare the polarization measured for the target to the general degree (and slope) of polarization within that region of the sky.

Observations were taken once every two weeks, and 325 SALT observations have been taken up to 2025 August. A more detailed summary of all the successfully executed observations is provided in Table~\ref{tab:SALT_obs_summary} in Appendix~\ref{app:SALT_summary}.

\subsubsection{Data reduction}

Data reduction was performed using an automated, modified version of the \textsc{pysalt}/\textsc{polsalt} data reduction pipeline \citep{2010SPIE.7737E..25C, 2022heas.confE..56C}, in which the wavelength calibration has been performed using \textsc{python}, and additional cosmic ray cleaning has been performed using \textsc{lacosmic}.\protect\footnote{\url{https://github.com/larrybradley/lacosmic}} The spectral extraction window was kept fixed for all of the observations. For each observation, the comparison star spectrum was also extracted. All of the uncertainties in the degree of linear polarization and equatorial polarization angle reported in this paper were derived using the \textsc{polsalt} pipeline and are, therefore, statistical in nature. The RSS is, however, capable of achieving polarimetric precision at a $\lesssim0.1\%$ level under favourable observing conditions \citep[see e.g.][]{2003SPIE.4841.1463B, 2004SPIE.5489..679S, 2006SPIE.6267E..0ZB}.

An important caveat is that the measured degree of polarization is biased towards higher values for low signal-to-noise observations. For the blazars investigated in this work, a wide range of S/N ratios has been obtained, and such biases may be present for the lower S/N observations. For completeness, the S/N ratios of the observations for each source are given in Table~\ref{tab:S/N_properties} in Appendix~\ref{app:SALT_summary}. A correction for the bias has not been applied to this work \citep[see e.g.][]{1985A&A...142..100S, 2006PASP..118.1340V, 2014MNRAS.439.4048P}.

Due to the changing aperture size of the SALT pupil, absolute flux calibration of the spectropolarimetric data is not possible. However, relative flux calibration has been performed with the use of spectrophotometric standard star observations and is sufficient to obtain the shape of the optical spectra. For this purpose, the standard star LTT 7987 was observed using the same RSS setup as the targets.

\subsubsection{Effects of interstellar polarization}

Following a similar argument as in \paperone, the observations have not been corrected for the interstellar polarization (ISP), since its effect on the degree of polarization and its frequency dependence should be negligible (typically, $\Pi_{\rm ISP} \sim 0.1-0.5\%$, whereas the observed polarization of the targets is, on average, $\Pi \gtrsim 1\%$). The Galactic extinction, as well as the typical ISP polarization levels along the line of sight of each source are summarized in Table~\ref{tab:ISP_properties} in Appendix~\ref{app:ISP}.

\subsubsection{Host galaxy correction}
\label{sec:host_galaxy_correction}

As already mentioned, the observed emission at optical wavelengths is a superposition of non-thermal jet emission and thermal emission from other components of the blazar. This thermal contribution dilutes the observed degree of polarization and could potentially introduce an artificial chromatic dependency in the polarization. Therefore, it is necessary to correct the spectropolarimetry data for the host galaxy contribution. In order to implement a correction to the observations of the HBL sources, we assume that the optical emission is dominated by the jet component with a smaller contribution from the host galaxy and a negligible contribution from other components such as the accretion disc. This should be a good approximation for HBLs, where the synchrotron emission peaks above the frequency range of the optical observations reported here. This allows us to estimate the relative contribution of the jet (synchrotron) and thermal (host galaxy) contribution to the observed flux. However, for intermediate and low synchrotron-peaked sources (where the synchrotron emission peaks at wavelengths below or in the optical range) the optical emission can be more dominated by the thermal contributions from the accretion disc and line regions \citep[see e.g.][]{2017A&A...607A..49S, 2020MNRAS.496.1295W, 1999ASPC..159..399K}, which requires more detailed modelling of the SED of each source, for every observation. For this work we have only applied a correction to the HBLs and leave the correction of the IBLs, LBLs, and FSRQs as a future work on individual sources. 

The observed degree of polarization as a function of wavelength, $\lambda$, is determined by 
\begin{equation}
    \Pi_{\rm obs}(\lambda) = \frac{\Pi_{\rm jet}(\lambda) F_{\rm jet}(\lambda) }{F_{\rm jet} (\lambda) + F_{\rm galaxy}(\lambda)}
\end{equation}
where $\Pi_{\rm jet}(\lambda)$ is the intrinsic degree of polarization from the jet, and $F_{\rm jet}(\lambda)$ and $F_{\rm galaxy}(\lambda)$ are the fluxes from the jet and host galaxy, respectively. The total observed optical flux is then given by
\begin{equation}
    F_{\rm total}(\lambda) = F_{\rm jet}(\lambda) + F_{\rm galaxy}(\lambda).
\end{equation}
We approximated the jet contribution by a log-parabola of the form, 
\begin{equation}
    F_{\rm jet}(\lambda) = f_0 \left( \frac{\lambda}{\lambda_0} \right)^{\alpha + \beta \ln \left( \lambda/\lambda_0 \right)},
\end{equation}
to reproduce the curvature in the flux-calibrated spectra. Here, $f_0$ is a normalisation constant at the reference wavelength $\lambda_0$, and $\alpha$ and $\beta$ are the slope and curvature of the spectrum, respectively. The host galaxy contribution was fitted as
\begin{equation}
    F_{\rm galaxy}(\lambda) = {\rm ratio} \times f_0 \times T_{\rm galaxy} \left( \lambda,z \right),
\end{equation}
where the ratio is the fractional contribution of the host galaxy at $\lambda_0$, and $T_{\rm galaxy}$ is an elliptical galaxy template, which is dependent on both the redshift ($z$) and wavelength. We used the elliptical galaxy template by \citet{1996ApJ...467...38K}. The corrected degree of polarization was then obtained as
\begin{equation}
    \Pi_{\rm corr}(\lambda) = \frac{F_{\rm total}(\lambda)}{F_{\rm jet}(\lambda)} \Pi_{\rm obs}(\lambda).
\end{equation}

\begin{figure}
    \centering
    \includegraphics[width=\columnwidth]{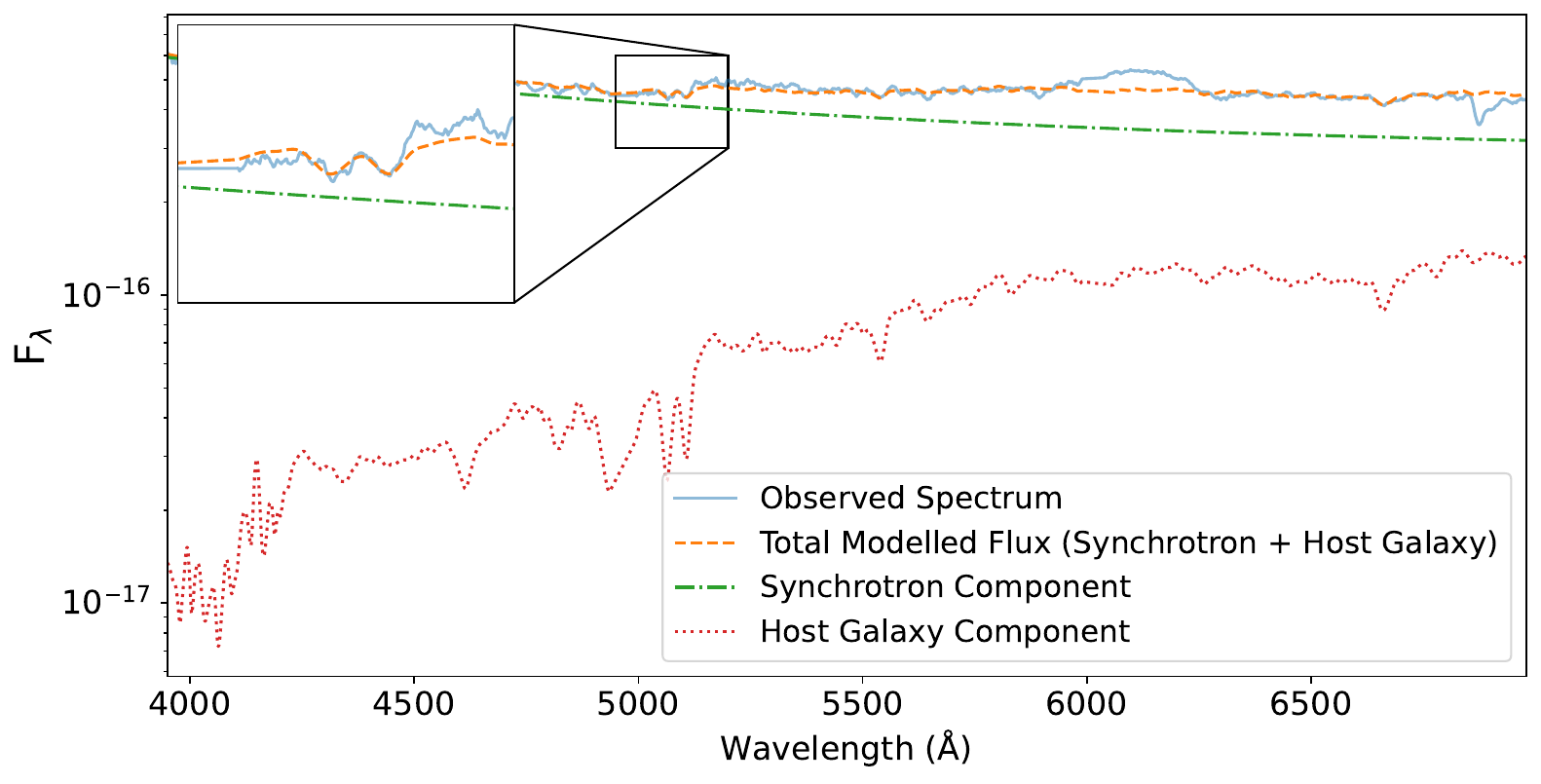}
    \caption{Example of host galaxy + log-parabola fit to an observed spectrum of the HBL 1ES 0414+009. The blue solid line shows the observed spectrum. The green dash-dotted and red dotted lines shows the fitted synchrotron and host galaxy spectra, respectively, whereas the dashed orange line indicates the total model fit to the spectrum. The inset highlights the Ca II H\&K break region, showing how the host galaxy template reproduces the absorption feature and constrains the host contribution to the total flux.}
    \label{fig:HG_fit}
\end{figure}

An example of the fit to an optical spectrum is shown in Fig.~\ref{fig:HG_fit}. The figure highlights the Ca II H\&K break region, where the absorption of the host galaxy is most prominent.

\subsubsection{A note on the calculation of the degree and angle of polarization}
\label{sec:pol_calculations}

Throughout this paper, the averages of the degree of polarization as well as the polarization angle are quoted. These were calculated in wavelength bins chosen to avoid (a) noisy regions of the spectra where the S/N ratio was low, and (b) any spectral features (emission/absorption lines) or telluric lines that may have a depolarizing effect on the data. Since both the degree and angle of polarization are non-linear quantities derived from the Stokes $Q$ and $U$ parameters -- they do not follow Gaussian distributions. The weighted averages of the Stokes $Q$ and $U$ parameters were calculated within each wavelength bin and, from them, the average degree and angle of polarization were calculated \citep[see e.g.][]{2015MNRAS.453.1669B, 2016MNRAS.463.3365A, 2016MNRAS.457.2252B, 2016A&A...590A..10K}.

\subsection{Optical photometry}

The optical photometric observations were taken with the Sinistro CCD cameras on the various 1.0-m telescopes in the Las Cumbres Observatory Global Telescope network \cite[LCOGT,][]{2013PASP..125.1031B}. Since both brightness and spectral (i.e., colour) variations are important to diagnose in blazar emission, observations were undertaken in three different filters: Bessell B, V, and R. The photometric observations serve as a complement to the optical spectropolarimetry data, as it is important to investigate the flux and polarization behaviour simultaneously. 

The LCOGT observations were taken with a $\sim$ 3-day cadence in order to observe any short-term variability as well as the long-term evolution of the optical flux. Since the start of the LCOGT monitoring (2023 August), more than 1300 observations have been taken.

The initial data reduction, including bad pixel masking, CCD reduction, and astrometric calibration was performed using the standard \textsc{banzai} pipeline.\footnote{\url{https://github.com/LCOGT/banzai}} Standard aperture photometry was performed with an automated \textsc{python}-based pipeline, which was discussed in \paperone. The resulting flux-values were then corrected for Galactic extinction \citep{2011ApJ...737..103S}.\footnote{\url{https://irsa.ipac.caltech.edu/applications/DUST/}} The errors quoted for the optical flux values are statistical in nature.

\section{Results (2023 -- 2025)}
\label{sec:results}

In this section, the data analysis and results from the observations taken with the SPOTS campaign will be summarized, including the methods used to place constraints on the ordering of the jet magnetic field. Additionally, some properties of the sources in the sample will be discussed. It is important to note that none of the sources in the sample exhibited any flaring activity during the period of monitoring. Hence, this is an overview of the quiescent state behaviour of blazars. 

Table~\ref{tab:full_results} provides a summary of the spectropolarimetric and photometric results. On average, the observed degree of polarization for all of the sources remained low ($\lesssim10\%$), but significant variation for some sources was found, varying by as much as $\sim 18\%$. Since the spectral features (emission/absorption lines) are thermal in nature, it depolarized the emission at those wavelengths. To account for this, all spectral features and atmospheric telluric lines have been excluded when calculating the averages in the degree of polarization and polarization angle. Additionally, the averages in the degree of polarization were calculated in wavelength ranges where the S/N ratios were high. For six sources, smooth, long-term polarization angle rotations were observed, whereas stochastic variability in the polarization angle was found for the rest. Some notes on individual sources are provided in Appendix~\ref{app:individualsources}.

%%%%%%%%%%%%%%%%%%%%%%%%%%%%%%%%%%%%%%%%%%%%%%%%%%%%%%%%%%%%%%%%%%%%%%%%%%%%%%%%%%%%%%%%%%%%%%%%%%%%%%%%%%%%%%
 \begin{table*}
 \begin{center}
  \caption{A summary of the observational results of the first two years of the SPOTS campaign. The SALT spectropolarimetry data is summarized as the average, minimum, and maximum degree of polarization ($\langle \Pi \rangle$, $\Pi_{\rm min}$, and $\Pi_{\rm max}$, respectively), the average, minimum, and maximum of the equatorial polarization angle ($\langle \rm PA \rangle$, $\rm PA_{\rm min}$, and $\rm PA_{\rm max}$, respectively) wrapped to be between [-90;90], as well the average frequency dependence ($\langle m_\Pi \rangle$) of the polarization. The average ordering of the magnetic field ($\langle F_B \rangle$) is given as estimated from the optical spectropolarimetry, unless otherwise indicated by an asterisk (*). In these cases, $\langle F_B \rangle$ was determined from the optical photometry data. In this table, $\langle F_B \rangle$ is indicated as an estimated percentage. Lastly, the average V-band optical flux ($\langle F_V \rangle$) from the LCOGT observations are given. For the frequency dependence of the polarization and V-band optical flux values, the uncertainties are given as the standard deviation of each dataset.}
  \label{tab:full_results}
  \begin{tabular}{lccccccccc}
    \hline
    Target & $\langle\Pi\rangle$ & $\Pi_{\rm min}$ & $\Pi_{\rm max}$ & $\langle\rm{PA}\rangle$ & $\rm{PA}_{\rm min}$ & $\rm{PA}_{\rm max}$ & $\langle m_{\Pi} \rangle$ & $\langle F_B \rangle$ & $\langle F_V \rangle$ \\
     & [$\%$] & [$\%$] & [$\%$] & [$^{\circ}$] & [$^{\circ}$] & [$^{\circ}$] & [$10^{-14}$\,\%\,/\,Hz] & [$\%$] & [$10^{-4}$ Jy] \\
    \hline
    RBS 0248 & 6.97 & 2.12\,$\pm$\,0.48 & 11.36\,$\pm$\,0.33 & 38.86 & 7.34\,$\pm$\,1.25 & 47.93\,$\pm$\,0.79 & 1.28\,$\pm$\,0.09 & 8.74 & 19.28\,$\pm$\,0.04 \\
    1RXS J023832.6--311658 & 0.14 & 0.06\,$\pm$\,0.13 & 3.43\,$\pm$\,0.15 & -77.03 & -88.24\,$\pm$\,8.46 & 52.28\,$\pm$\,3.00 & 1.11\,$\pm$\,0.05 & 1.42 & 7.70\,$\pm$\,0.02 \\
    PKS 0301--243 & 5.33 & 2.45\,$\pm$\,0.07 & 9.09\,$\pm$\,0.06 & 22.15 & 7.92\,$\pm$\,1.17 & 38.62\,$\pm$\,0.54 & 0.25\,$\pm$\,0.02 & 7.927 & 13.53\,$\pm$\,0.01 \\
    1ES 0414+009 & 3.66 & 0.81\,$\pm$\,0.24 & 8.08\,$\pm$\,0.37 & -37.95 & -74.59\,$\pm$\,4.24 & 74.02\,$\pm$\,1.33 & 0.87\,$\pm$\,0.08 & 6.92 & 4.52\,$\pm$\,0.02 \\
    PKS 0447--439 & 5.35 & 1.34\,$\pm$\,0.04 & 16.53\,$\pm$\,0.04 & -43.47 & -78.80\,$\pm$\,0.18 & 53.59\,$\pm$\,0.47 & -0.05\,$\pm$\,0.01 & 10.27 & 91.00\,$\pm$\,0.55 \\
    TXS 0506+056 & 5.16 & 4.15\,$\pm$\,1.07 & 21.65\,$\pm$\,0.19 & -40.03 & -76.46\,$\pm$\,0.22 & 78.00\,$\pm$\,0.33 & 0.77\,$\pm$\,0.06 & 11.46* & 24.92\,$\pm$\,0.06 \\
    PKS 0736+017 & 0.79 & 0.39\,$\pm$\,0.14 & 7.77\,$\pm$\,0.16 & 15.93 & -51.92\,$\pm$\,7.30 & 80.83\,$\pm$\,0.65 & 0.11\,$\pm$\,0.04 & 5.53* & 10.56\,$\pm$\,0.02 \\
    PKS 1440--389 & 2.75 & 0.62\,$\pm$\,0.11 & 9.32\,$\pm$\,0.06 & -71.08 & -84.70\,$\pm$\,0.19 & 85.04\,$\pm$\,0.81 & -0.10\,$\pm$\,0.02 & 7.22 & 48.15\,$\pm$\,0.04 \\
    PKS 1510--089 & 1.91 & 0.06\,$\pm$\,0.14 & 6.78\,$\pm$\,0.11 & 52.62 & -89.32\,$\pm$\,8.84 & 80.87\,$\pm$\,0.45 & 0.04\,$\pm$\,0.02 & 3.60* & 9.94\,$\pm$\,0.02 \\
    AP Lib & 4.88 & 1.70\,$\pm$\,0.12 & 11.12\,$\pm$\,0.13 & -10.40 & -20.63\,$\pm$\,1.96 & -0.30\,$\pm$\,0.89 & 0.82\,$\pm$\,0.05 & 6.24* & 45.96\,$\pm$\,0.09 \\
    PKS 1749+096 & 3.23 & 1.63\,$\pm$\,0.21 & 18.48\,$\pm$\,0.19 & -7.88 & -76.48\,$\pm$\,0.57 & 83.58\,$\pm$\,1.04 & 0.40\,$\pm$\,0.04 & 10.31* & 7.13\,$\pm$\,0.03 \\
    PKS 2005--489 & 1.93 & 0.93\,$\pm$\,0.06 & 10.40\,$\pm$\,0.11 & 46.68 & -20.70\,$\pm$\,0.58 & 88.78\,$\pm$\,0.30 & 0.43\,$\pm$\,0.04 & 6.36 & 98.78\,$\pm$\,0.14 \\
    PKS 2155--304 & 5.52 & 2.09\,$\pm$\,0.10 & 10.54\,$\pm$\,0.20 & -84.40 & -88.52\,$\pm$\,0.37 & 89.93\,$\pm$\,0.41 & -0.06\,$\pm$\,0.03 & 8.97 & 106.98\,$\pm$\,0.14 \\
    1ES 2322--409 & 12.64 & 6.64\,$\pm$\,0.13 & 15.77\,$\pm$\,0.17 & 5.88 & -1.27\,$\pm$\,0.23 & 15.11\,$\pm$\,0.47 & 0.36\,$\pm$\,0.03 & 17.43 & 13.45\,$\pm$\,0.02 \\
    \hline
  \end{tabular}
  \end{center}
 \end{table*}
%%%%%%%%%%%%%%%%%%%%%%%%%%%%%%%%%%%%%%%%%%%%%%%%%%%%%%%%%%%%%%%%%%%%%%%%%%%%%%%%%%%%%%%%%%%%%%%%%%%%%%%%%%%%%%

\subsection{Confirming significance of variability}

As part of the SALT data reduction process, a comparison star (which lay on the slit and was observed at the same time as the target) was also extracted. It was noted that the variability in the measured polarization of the comparison star, which should remain stable, was larger than the reported statistical errors in polarization measurements, suggesting additional systematic uncertainties. In order to test that the observed changes in the degree of polarization of the targets were statistically significant, we undertook a Bayesian analysis to confirm that the variability in the targets was larger than that of the comparison stars.

It was noted that the reported measurement errors do not fully capture the true uncertainty in the polarization measurements. This motivated the inclusion of an additional variance component in the statistical model described below, which was favoured by Bayesian model comparison.

The data consists of measurements with gaps in time. The measurements \((y_{ij})\) were assumed to be conditionally independent given the target \((j)\) and reported error value \((s_{ij})\). This assumption is reasonable for observations taken multiple days apart. Each target may have a different number of observed values \((I_j)\) and the times do not need to be evenly spaced or related to the times of measurement of other targets. The control stars are to be considered a random sample from the population of stable stars in general. In this scenario a Gaussian random effects model with conditional heteroscedasticity can be applied.

Various models of the measurements of the control stars were fitted using the {\sc brms} package \citep{JSSv080i01}. The models were compared using the leave-out-one cross validation information criterion. Of the considered models, it was found that the most parsimonious model considered is given by
\[
y_{ij} \sim N(\mu_j, \sigma_{ij}), 
\]
 where $\sigma_{ij}$ is the residual standard deviation, and
 \[
 \mu_j \sim N(\mu, \tau) 
 \]
where $\tau$ is the between-star standard deviation, and the residual standard deviation (on the log scale) is given by
\[
\log\sigma_{ij} = \beta_0 + \beta_1\log s_{ij} + \gamma_j
\]
where $\beta_0$ and $\beta_1$ are constants for the data and $\gamma_j$ are star specific random effects.

In order to test whether a specific target varied more than the corresponding control star, we considered measurements from the control star $(x)$ and from the target $(y)$ with a number of independent observations from each. These were fitted with the model 
\begin{eqnarray}
y_{i} &\sim N(\mu^{\rm target}, \sigma_{i}^{\rm target})\\
\log\sigma_{i}^{\rm target} &= \beta_{\rm target} + \log s_{i}^{\rm target}
\end{eqnarray}
and
\begin{eqnarray}
x_{i} &\sim N(\mu^{\rm control}, \sigma_{i}^{\rm control})\\
\log\sigma_{i}^{\rm control} &= \beta_{\rm control} + \log s_{i}^{\rm control}
\end{eqnarray}
 and the probability $P[\beta_{\rm target} > \beta_{\rm control}]$ was calculated. This was done for the observed pairs of target--star to produce a standard Bayesian probability result. The probabilities are summarized in Table~\ref{tab:bayesian}. The comparison star of RBS 0248 was too faint for meaningful comparison, while the probability of variability for 1RXS\,J023832.6-311658 ($P=0.665$) and 1ES\,0414+009 ($P=0.941$) were lower. The remaining targets are variable. 1RXS\,J023832.6-311658 showed a low level of polarization over the full observing campaign, which accounts for this non-variability compared to the comparison star.

\begin{table}
\centering
\caption{Results of the Bayesian analysis calculating the probability that $P[\beta_{\rm target} > \beta_{\rm control}]$. }
\label{tab:bayesian}
\begin{tabular}{lc}
\hline
Target & Value \\
\hline
RBS\,0248            & --    \\
1RXS\,J023832.6-311658 & 0.665 \\
PKS\,0301-243        & 1     \\
1ES\,0414+009        & 0.941 \\
PKS\,0447-439        & 1     \\
TXS\,0506+056        & 1     \\
PKS\,0736+017        & 1     \\
PKS\,1440-389        & 1     \\
PKS\,1510-089        & 1     \\
AP\,Lib              & 1     \\
PKS\,1749+096        & 1     \\
PKS\,2005-489        & 1     \\
PKS\,2155-304        & 1     \\
1ES\,2322-409        & 1     \\
\hline
\end{tabular}
\end{table}

\subsection{Ordering  of the magnetic field}

The observed degree of polarization found for each source is due to the non-thermal electron spectrum and the ordering of the magnetic field in the emission region of the jet, as well as the dilution due to any unpolarized thermal component. Below we discuss how the spectropolarimetric observations, which provide details of the frequency dependence of the polarization, can be used to place constraints on the ordering of the magnetic field in the optical emission region.

\subsubsection{IBLs, LBLs, and FSRQs}

For the lower synchrotron-peaked sources in the sample (LBLs, IBLs, FSRQs), the optical spectra show stronger emission features (originating in the accretion disc, dust torus, and line regions), which complicates the modelling of these spectra. However, assuming that the general shape of the spectra are still dominated by the non-thermal emission, a power-law fit to the optical photometric data will still be sufficient for a first-order approximation. For these sources, the ordering of the magnetic field ($F_B$) was estimated by combining LCOGT photometry with SALT polarization measurements.

\begin{figure}
    \centering
    \includegraphics[width=\linewidth]{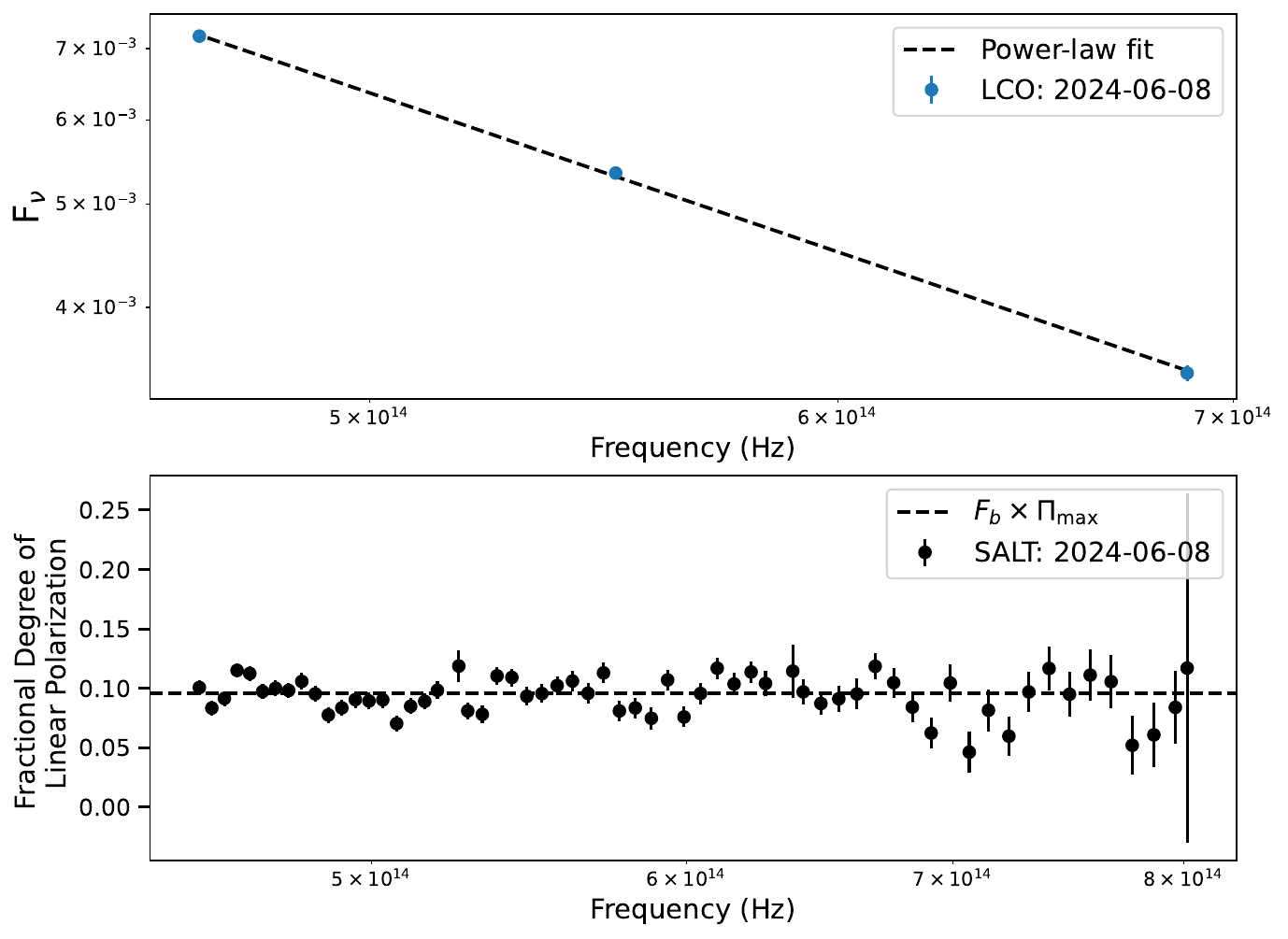}
    \caption{Example of the $F_B$ model fitting procedure (for IBLs, LBLs, and FSRQs) applied to an observation of the LBL AP Librae. \textit{Top:} A power-law fit (black dashed line) to the LCOGT flux points (blue points). \textit{Bottom:} The observed degree of polarization (black points) and the predicted degree of polarization based on the derived $F_B$ value (black dashed line).}
    \label{fig:Pol_fit_phot}
\end{figure}

For a power-law distribution of electrons the maximum degree of polarization for synchrotron radiation is given by 
\begin{equation}
    \Pi_{\rm max} = \frac{p + 1}{p + \frac{7}{3}},
    \label{eq:max_pi_power_law}
\end{equation}
where the index of the electron distribution, $p$, is related to the spectral index by $p = 2\alpha + 1$. Assuming the overall shape of the spectrum is determined by the non-thermal emission, we fit a power-law ($F^\prime_\nu \propto \nu^{\prime \alpha}$) to the B, V, and R flux (after correcting for redshift), to determine $\alpha$. The flux values were found by interpolating between the closest LCOGT observations dates, to the date  of the SALT observation.
%A power law was fitted to the B, V, and R flux points, from which $\alpha$ and the subsequent $\Pi_{\rm max}$ values were estimated for each observation. 
The ordering of the magnetic field was then obtained for each observations by 
\begin{equation}
    F_B = \frac{\langle \Pi_{\rm obs} \rangle}{\Pi_{\rm max}},
\end{equation}
where $\langle \Pi_{\rm obs} \rangle$ is the mean degree of polarization found from for the spectropolarimetry observations. An example of this fitting procedure is shown in Fig.~\ref{fig:Pol_fit_phot}. Note that $F_B = 1$ indicates a perfectly ordered magnetic field, and $F_B = 0$ a completely chaotic magnetic field.

\subsubsection{HBLs}

For the HBLs, we assume the optical emission is dominated by the thermal emission from the host galaxy and the non-thermal emission from the jet, and other contributions can be considered negligible. We, therefore, use the log-parabola fit that was obtained as part of the host-galaxy correction (section~\ref{sec:host_galaxy_correction}), to constrain the ordering of the magnetic field. We calculate the spectral gradient of the log-parabola as a function of wavelength, $\alpha(\lambda)$, and then estimate the maximum degree of polarization at each wavelength assuming synchrotron emission from a power-law electron distribution that would produce the observed spectral slope (Eq.~\ref{eq:max_pi_power_law}). This gives an estimate of the maximum polarization as a function of wavelength. We have tested this approximation, using a full calculation of the synchrotron flux and polarization, and the approximation is accurate to within a few per cent, as long as the spectrum does not deviate dramatically from a power-law, and the wavelength range considered is not close to a cut-off induced due to the minimum/maximum electron energy in the electron spectrum. (see Appendix~\ref{app:polarization_approximation}).

The observed degree of polarization is then related to the maximum as
\begin{equation}
    \Pi_{\rm obs} = F_B \times \Pi_{\rm max},
    \label{eqn:FB_pol}
\end{equation}
This approximate $F_B$ is assumed to be constant over the optical wavelength range. 

\begin{figure}
    \centering
    \includegraphics[width=\linewidth]{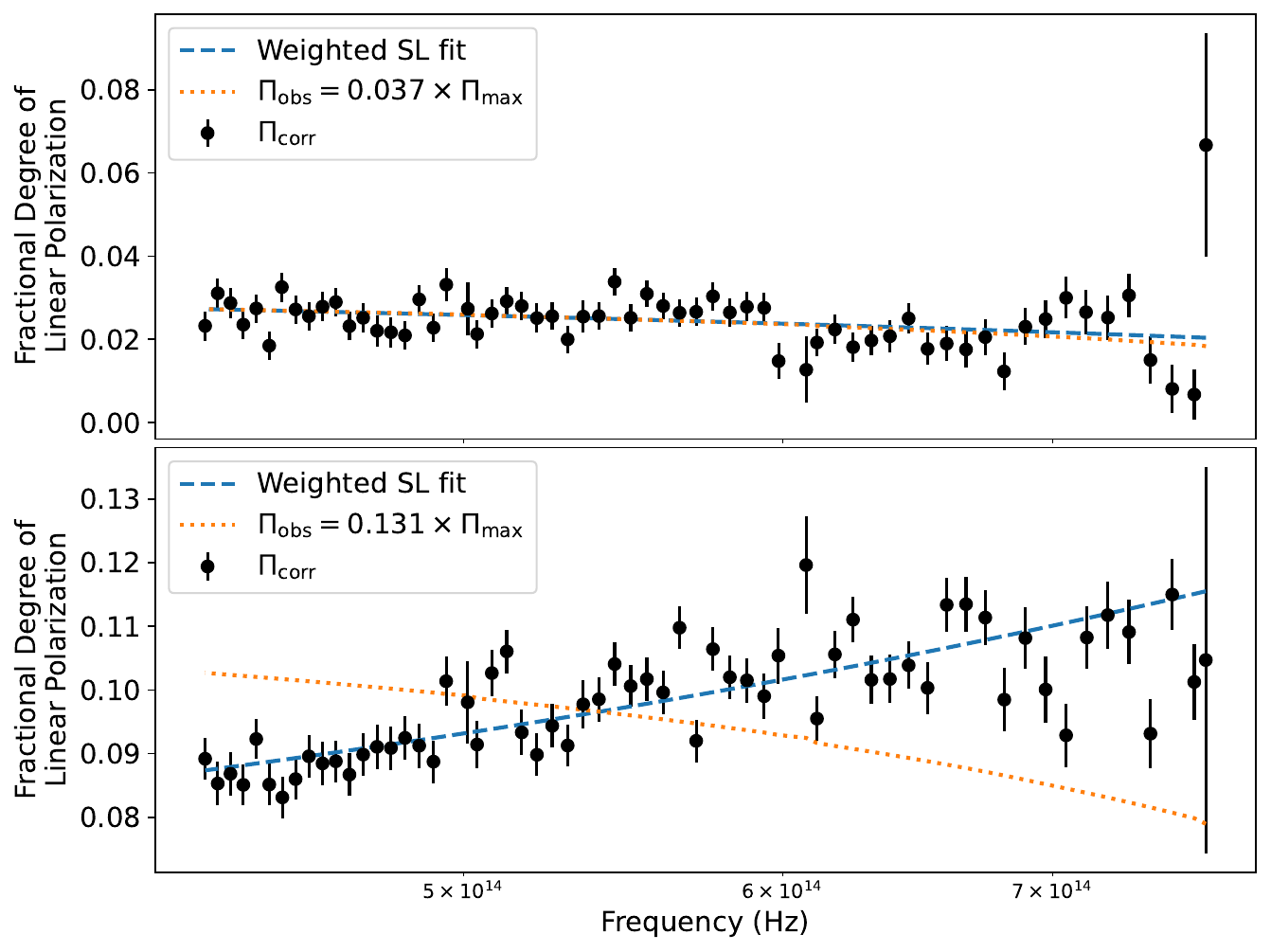}
    \caption{Two cases of the $F_B$ model fit (for HBLs) applied to observations of the HBL PKS 0301--243. The black points show the observed (host galaxy corrected) degree of polarization. The blue dashed line indicates a straight-line fit to the observed polarization, and the orange dotted line indicates the degree of polarization predicted by the $F_B$ model. \textit{Top:} An observation in which the $F_B$ fit reproduces the frequency dependence of the observed polarization, and the degree of polarization is low. \textit{Bottom:} An observation in which the $F_B$ fit deviates and the degree polarization is higher, indicating that the observed frequency dependence arises from additional processes.}
    \label{fig:Pol_fit}
\end{figure}

From the resulting $F_B$ fit and $\Pi_{\rm max}$ estimates for the HBLs, a model for the observed degree of polarization can be reconstructed to assess its frequency dependence. If the frequency dependence of $\Pi_{\rm obs}$ is well reproduced by the $F_B$-derived model (as shown in the top plot of Fig.~\ref{fig:Pol_fit}), the slope can be attributed to the underlying particle distribution, within a single zone. However, if the observed degree of polarization deviates significantly from the $F_B$ model (as in the bottom plot of Fig.~\ref{fig:Pol_fit}), the frequency dependence must arise from additional physical processes, such as a shock-in-jet scenario, or the presence of multiple emission zones, which would imply that $F_B$ is wavelength dependent.

To investigate the degree to which the $F_B$ model matches the observed degree and frequency dependence of the polarization, a comparison was drawn between a weighted straight-line (SL) fit (with greater weight assigned to data points with smaller uncertainties) and the $F_B$ model fit. The SL fit is, therefore, just a representative of the observational data. For each observation, the reduced chi-squared values of both models were calculated, and the difference between them was quantified as
\begin{equation}
    \Delta \chi^2 = \chi^2_{\rm SL} - \chi^2_{\rm F_B}.
\end{equation}
By construction, $\Delta \chi^2 > 0$ indicates that the $F_B$ model provides a better representation of the data, while $\Delta \chi^2 < 0$ indicates that the SL fit performs better. In the ideal case of $\Delta \chi^2 \approx 0$, both models perform equally well, which would suggest that the frequency dependence and degree of polarization can be explained solely by the underlying particle spectrum.

\begin{figure}
    \centering
    \includegraphics[width=\linewidth]{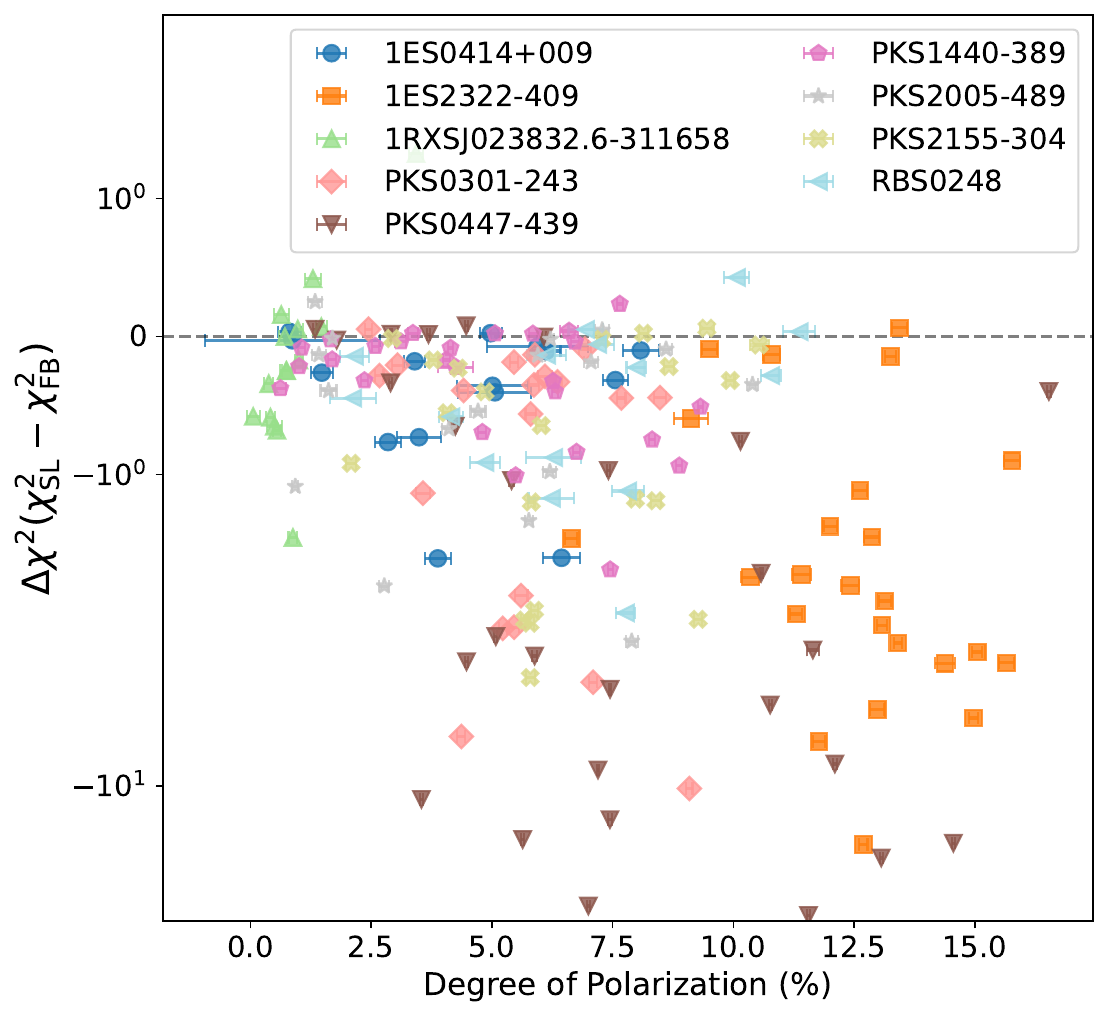}
    \caption{Comparison of the $F_B$ and SL model fits to the observed degree of polarization across the full sample of HBLs. Here, postive values indicate cases where the $F_B$ model provides a better fit, and negative values indicate where the SL model provides a better fit. This trend shows that the $F_B$ model only represents the data well at low polarization levels, suggesting the influence of additional physical processes beyond a simple one-zone synchrotron model.}
    \label{fig:model_mismatch}
\end{figure}

Fig.~\ref{fig:model_mismatch} shows the distribution of $\Delta \chi^2$ versus the degree of polarization for all observations. The median value of $\Delta \chi^2$ is $-0.44$, indicating that most observations are compatible with a single zone emission model. However, a significant portion of the observations show a deviation from this. A Spearman's rank correlation between the model mismatch parameter ($\Delta \chi^2$) and the observed degree of polarization yields $\rho = -0.38$ (p-value $= 1.44 \times 10^{-7}$), indicating a statistically significant trend.\footnote{Throughout this paper, correlations are considered statistically significant when p-value $\leq10^{-3}$.} Thus, $\Delta \chi^2$ decreases (growing increasingly negative) with increasing polarization.

This result suggests that the $F_B$ model better captures the observed frequency dependence of the polarization only when the degree of polarization is low. The $F_B$ model is, therefore, consistent with the expectations from simplistic, one-zone synchrotron emission. It fails, however, to match the observed frequency dependence at higher degrees of polarization, which implies that additional physical processes (e.g. shock-in-jet scenarios or multiple emission zones) contribute to the observed polarization behaviour. This is consistent with the case shown in Fig.~\ref{fig:Pol_fit}. 

\subsubsection{Distribution of $F_B$}

\begin{figure}
    \centering
    \includegraphics[width=\linewidth]{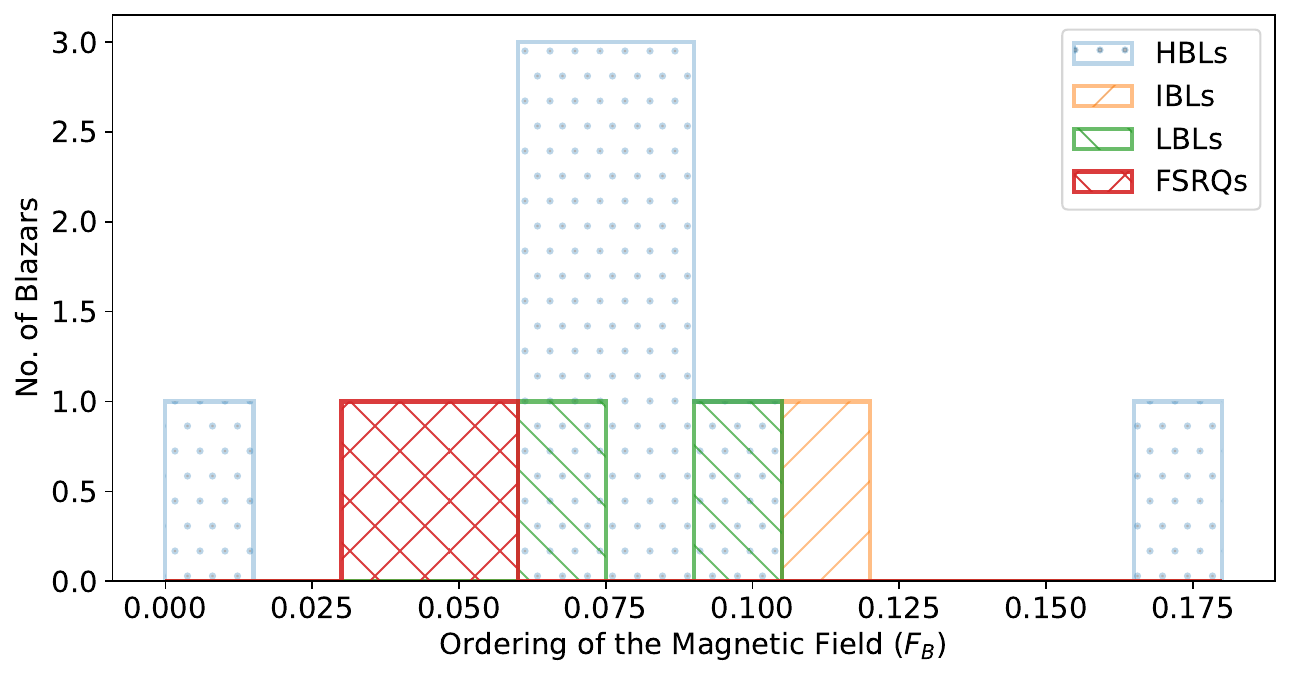}
    \caption{Histogram showing the distribution of the ordering of the magnetic field across the different blazar subtypes (where HBLs are represented in blue, IBLs in orange, LBLs in green, and FSRQs in red).}
    \label{fig:FB_stats}
\end{figure}

Fig.~\ref{fig:FB_stats} shows the distribution of $F_B$ for all the blazars observed in the SPOTS campaign, and the inferred $F_B$ values are consistent with weakly ordered magnetic fields for all of the sources. The distribution of $F_B$ indicates systematic differences between FSRQs and BLLs: the magnetic field of FSRQs tend to be less ordered than in BLLs (though a wide range of $F_B$ values are found for the HBLs).

\subsection{Polarization angle behaviour and rotations}

There is an inherent $180^\circ$ ambiguity in the polarization angle which can -- especially in cases where the data is sparse -- obscure rotations and prevent an accurate depiction of its behaviour. In this work we have followed polarization angle shifting method 2 outlined in \citet{2013EPJWC..6106003K}, in which the difference in polarization angle between two adjacent measurements is estimated as, 
\[
\Delta \theta_{\rm est} = \lvert \Delta \theta \rvert - \sqrt{\sigma^{2}(\theta_{i+1}) + \sigma^{2}(\theta_{i})},
\]
where $\lvert \Delta \theta \rvert$ is the absolute difference between the points, and $\sigma(\theta_{n})$ is the error in the measurement. The deviation is then evaluated under three scenarios: no shift, or shifts by $\pm180^\circ$. The solution that minimises $\lvert \Delta \theta \rvert$ is selected, such that a $\pm180^\circ$ adjustment is only applied when the deviation is reduced. This procedure is applied cumulatively to ensure a continuous evolution of the polarization angle.

In this paper, a polarization angle rotation is defined as a continuous, monotonic change in the polarization angle measurements over a minimum of three data points ($N$), with an overall $|\Delta \theta| \geq 30^\circ$ across the full rotation segment, similar to the procedure followed in \citet{2016MNRAS.462.1775B}. However, due to the typically noisy nature of spectropolarimetric data (especially at low S/N ratios), small fluctuations in monotonicity within measurement uncertainties were allowed. For each polarization angle dataset, candidate rotation segments were identified. For these, we assessed whether these rotations could be caused by stochastic processes by performing Monte-Carlo simulations under a random-walk null hypothesis, following the method outlined in \citet{2016A&A...590A..10K}. These simulations provided a measure of the significance of the rotation ($p$-value; $p_{|\Delta \theta|}$). Thus, a polarization angle rotation was identified when the following three conditions were met:

\begin{itemize}
    \item $N \geq 3$,
    \item $|\Delta \theta| \geq 30^\circ$, and
    \item $p_{|\Delta \theta|} < 0.2$.
\end{itemize}
Rotations were then further classified as weak ($30^\circ < |\Delta \theta| \leq 60^\circ$), moderate ($60^\circ < |\Delta \theta| \leq 90^\circ$), or strong ($|\Delta \theta| \geq 90^\circ$).

Across all of the sources, the polarization angle was found to be highly variable even in quiescent states. For most of the sources in this campaign, stochastic variability in the polarization angle was found. However, for four sources, smooth, long-term rotations spanning timescales of months were detected.

These smooth rotations could reflect ordered or curved magnetic field structures within the jet and/or gradual changes in the dominant emission regions \citep[see e.g.][]{2010Natur.463..919A, 2014ApJ...780...87M, 2015ApJ...804...58Z}. \citet{2018MNRAS.474.1296B} found that there is a correlation between polarization angle rotations and $\gamma$-ray flares. In this project, observations were taken during quiescent states for all sources. The detection of these long-term polarization angle swings/rotations is not exclusively linked to enhanced flux states, but may also trace slow magnetic field evolution in jets.

Table \ref{tab:PA_swings} characterises the polarization angle swings exhibited in the aforementioned sources. The timescale, amplitude, and rotation strength of the rotation is provided from the definition and tests outlined above. Additionally, as a check to see whether these smooth rotations are associated with any increases in optical flux or polarization, it was characterised as follows: a weak increase in flux/polarization is indicated if it remains within one standard deviation of the mean; a moderate increase is indicated if it falls within three standard deviations of the mean; and a high increase is indicated if it lies three standard deviations or more above the mean. We note that this method is simple and only intended as an initial diagnostic. A more in-depth analysis of flux/polarization correlations with the polarization angle rotations is left to future work.

\begin{table*}
\centering
\caption{A summary of the smooth polarization angle rotations detected in four of the sources in the SPOTS monitoring campaign. For these rotations, the timescale over which they occur, their amplitude and strength, as well as whether or not they correlate with an increase in flux or polarization is provided.}
\label{tab:PA_swings}
\begin{tabular}{lccccc}
\hline
Target & Timescale & Amplitude  & Rotation & Increase   & Increase  \\
       & [d]       & [$^\circ$] & strength & in Flux? & in $\Pi$?   \\
\hline
1ES\,0414+009        & 146 & 161.10 & Strong   & Weak     & Moderate \\
1RXSJ023832.6-311658 & 209 & 256.30 & Strong   & Moderate & Weak     \\
PKS\,1440-389        & 131 & 177.82 & Moderate & Weak     & Moderate \\
PKS\,1749+096        & 51  & 96.71  & Strong   & Weak     & Moderate \\
\hline
\end{tabular}
\end{table*}

\subsection{Polarization properties by source}

\begin{figure}
    \centering
    \includegraphics[width=\linewidth]{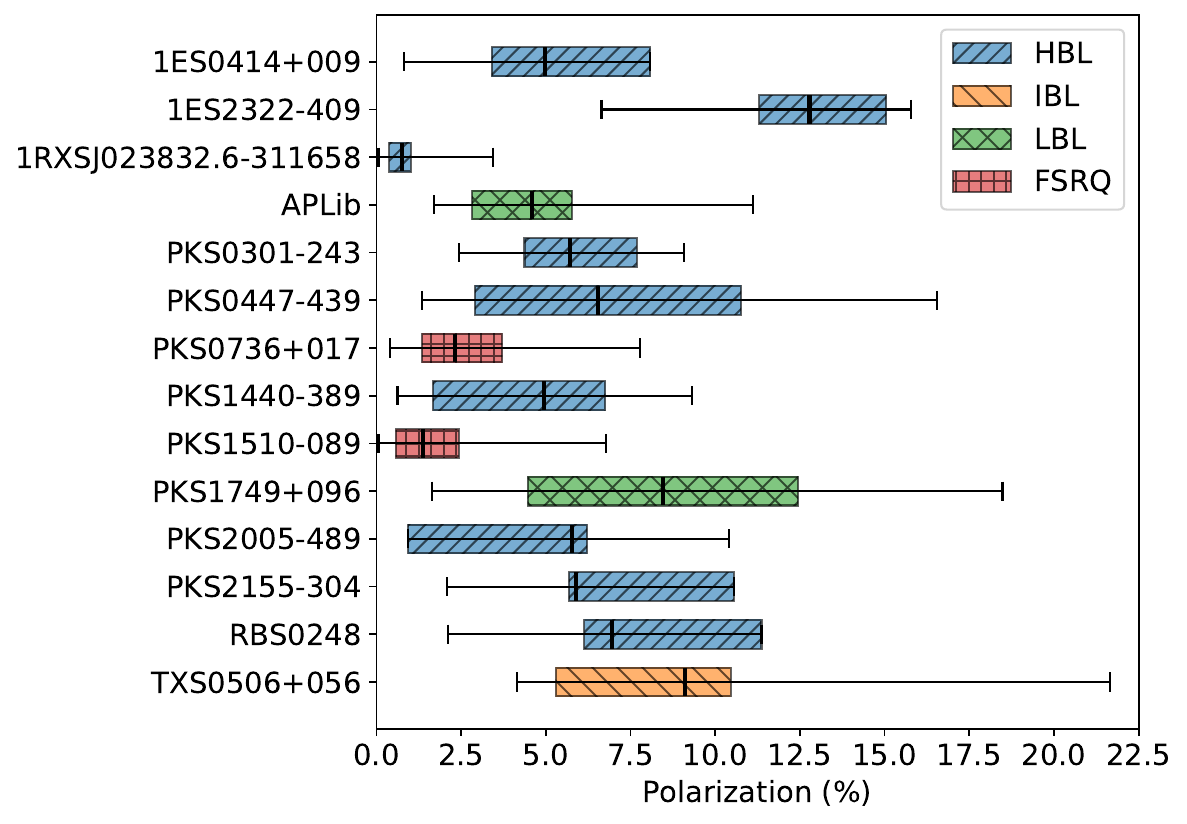}
    \caption{Boxplot representing the host galaxy corrected degree of linear polarization observed in the sample (where HBLs are represented in blue, IBLs in orange, LBLs in green, and FSRQs in red). The black vertical lines in the boxes show the median value over all observations for each source. The boxes give the shortest interval in which 65 per cent of the observations lie, and the horizontal lines represent the full range of observed values.}
    \label{fig:pol_boxplot}
\end{figure}

Fig.~\ref{fig:pol_boxplot} provides a summary of the behaviour of the degree of polarization (host galaxy corrected) for each blazar in our sample. From this, it is clear that, even in quiescence or low states of activity, the BLLs in the sample display a wide range of behaviour, but are overall substantially polarized and highly variable. The two FSRQs, however, display low degrees of polarization with a narrow range of behaviour, and are much less variable than the BLLs. This is consistent with what was reported by \citet{2025A&A...703A..19C}, who found that high synchrotron-peaked sources generally show higher degrees of polarization than their lower-peaked counterparts.

\begin{figure}
    \centering
    \includegraphics[width=\linewidth]{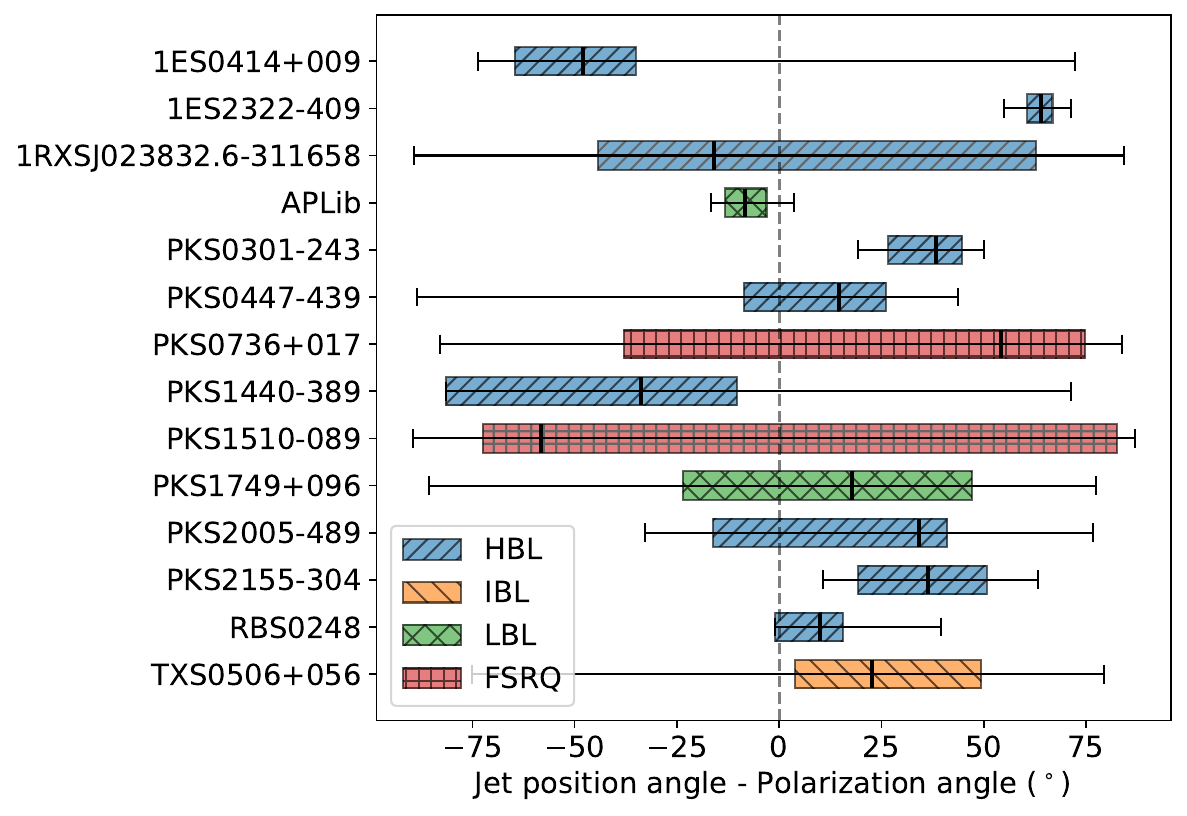}
    \caption{Boxplot representing the angular difference between the jet direction and the polarization angle of the sources in the sample (where HBLs are represented in blue, IBLs in orange, LBLs in green, and FSRQs in red). The black vertical lines in the boxes show the median value over all observations for each source. The boxes give the shortest interval in which 65 per cent of the observations lie, and the horizontal lines represent the full range of observed values. The dashed gray line indicates a difference of $0^\circ$.}
    \label{fig:pa_boxplot}
\end{figure}

The alignment of the observed polarization angles to the jet position angles was investigated for the sample, with the jet position angles obtained from the MOJAVE \citep{2018ApJS..234...12L} and TANAMI \citep{2010A&A...519A..45O} surveys. The average of the jet position angle measured over all available epochs was adopted. Fig.~\ref{fig:pa_boxplot} shows the offset between the observed polarization angle and jet position angles. The median offset measured across all sources is $\sim 13^\circ$, with a standard deviation of $\sim 30^\circ$. For both the BLL and FSRQ samples, a large distribution in this offset is observed. During the monitoring period, the polarization angle was (overall) highly variable and not necessarily aligned with the jet position angle. \citet{2025A&A...703A..19C} found that the offset between the polarization angle and jet position is consistent with zero for both high and low synchrotron-peaked sources, which (combined with the chromaticity found in the degree of polarization) aligns with a shock-in-jet scenario. The results from the SPOTS campaign are in contrast to what is predicted from the shock-in-jet model, as the poorly aligned polarization angle might suggest that the optical emission is produced by multiple weakly polarized zones, where the magnetic field is dominated by turbulence \citep{2014ApJ...780...87M}. This scenario naturally produces both the misalignment between the jet position angle and polarization angle as well as the large scatter in polarization angle between observations. However, since the observations were made during quiescence, the low degrees of polarization likely increase the uncertainty in the measured polarization angles, potentially amplifying its variations. 

\subsection{Colour trends}

\begin{figure*}
    \centering
    \includegraphics[width=0.8\linewidth]{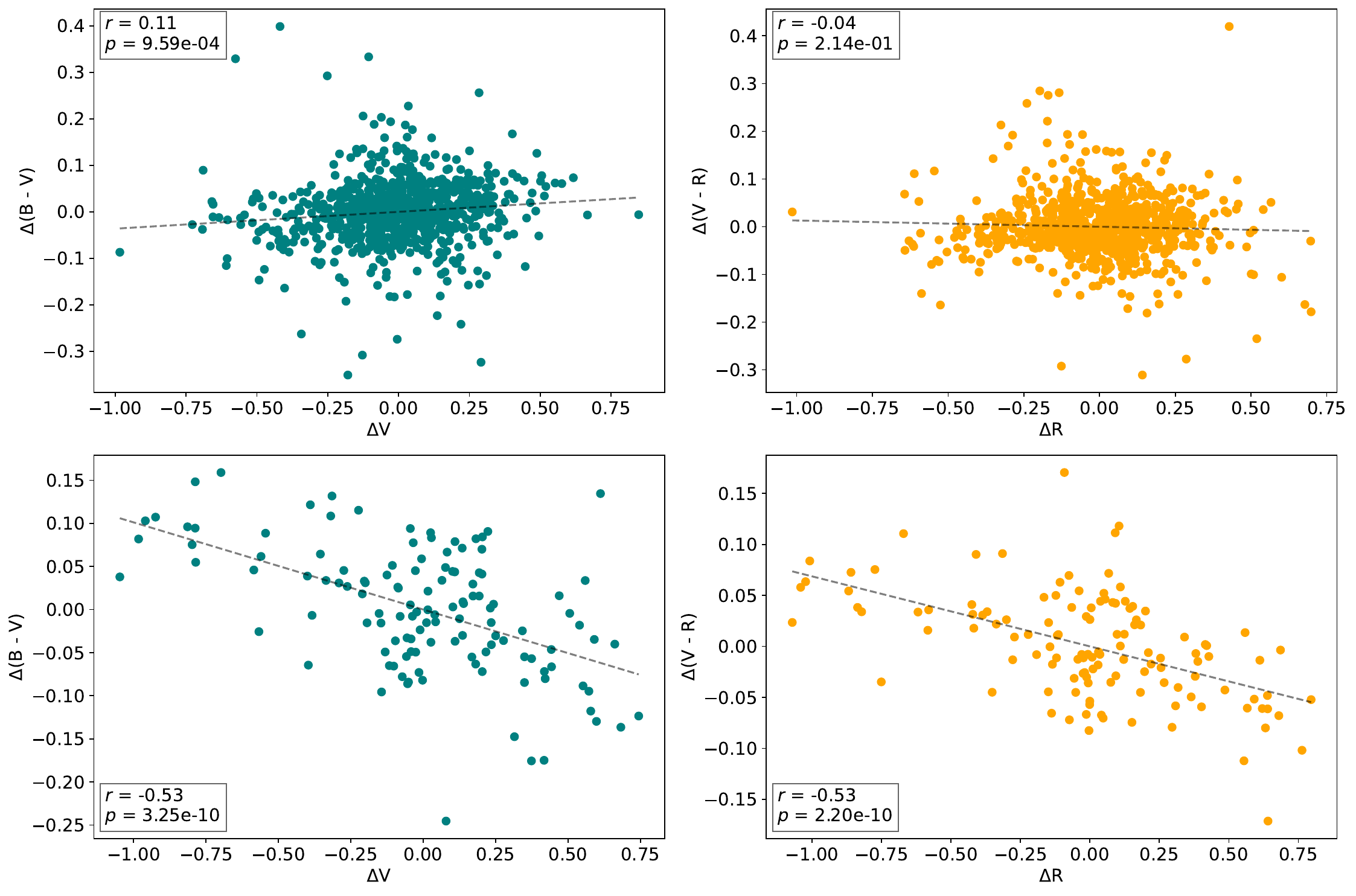}
    \caption{The colour-magnitude diagrams for the all BLLs (\textbf{top row}) and FSRQs (\textbf{bottom row}) in the sample. The colour-magnitudes diagrams are shown for $\Delta(B - V)$ versus $\Delta V$ (\textbf{left column}), and $\Delta(V - R)$ versus $\Delta R$ (\textbf{right column}). The dashed gray lines give the best-fitting straight lines to the data, with the Pearson correlation coefficients (r) and p-values given in the text boxes.}
    \label{fig:combined_cmds}
\end{figure*}

From the optical photometry data, the colour-magnitude diagrams could be extracted for both BLLs and FSRQs (top and bottom rows of Fig.~\ref{fig:combined_cmds}, respectively). The BLLs show a moderately strong brighter-when-bluer (BWB) trend in $B - V$, with no significant correlation (or a very weak redder-when-brighter trend) in $V - R$. The FSRQs display a strong redder-when-brighter (RWB) trend. For BLLs, a BWB trend is generally expected due to the weak or no contribution from the thermal accretion disc. However, for HBLs in quiescent state, a weak or no correlation in colour and magnitude is also consistent with what has been found in literature \citep[see e.g.][]{2012MNRAS.425.3002G, 2018ApJS..237...30M, 2023MNRAS.519.5263Z}. FSRQs, however, have very luminous, stable accretion disc components that are dominant when the optical flux levels are low. As the optical flux increases, the synchrotron emission becomes more dominant, leading to the RWB trend observed here \citep{2023MNRAS.519.5263Z, 2016Ap&SS.361..345M}.

\subsection{Flux-polarization correlation}

%%%%%%%%%%%%%%%%%%%%%%%%%%%%%%%%%%%%%%%%%%%%%%%%%%%%%%%%%%%%%%%%%%%%%%%%%%%%%%%%%%%%%%%%%%%%%%%%%%%%%%%%%%%%%%
\begin{table}
\begin{center}
\small
\caption{A summary of the correlation between the optical flux and degree of polarization for each source. The time lags (days), correlation coefficients, and confidence levels of the observed correlations are given.}
\label{tab:correlations}
\begin{tabular}{lccc}
\hline
Target & Time         & Correlation       & Confidence \\
       & lag ($\tau$) & coefficient ($r$) & level \\
\hline
RBS 0248 & -0.87 & 0.40 & $>68\%$ \\
1RXS J023832.6--311658 &  -26.72 & 0.19 & not significant \\
PKS 0301--243 & 69.67 & 0.40 & $>68\%$ \\
1ES 0414+009 & -34.56 & 0.27 & $>68\%$ \\
PKS 0447--439 & -49.96 & 0.73 & $>95\%$ \\
TXS 0506+056 & -6.92 & 0.25 & not significant \\
PKS 0736+017 & -37.58 & 0.78 & $>95\%$ \\
PKS 1440--389 & 15.44 & 0.70 & $>95\%$ \\
PKS 1510--089 & 7.80 & 0.42 & $>68\%$ \\
AP Lib & 40.96 & 0.67 & $>95\%$ \\
PKS 1749+096 & -24.49 & 0.39 & $>68\%$ \\
PKS 2005--489 & 24.19 & 0.53 & $>68\%$ \\
PKS 2155--304 & 30.66 & 0.79 & $>95\%$ \\
1ES 2322--409 & -38.92 & 0.62 & $>95\%$ \\
\hline
\end{tabular}
\end{center}
\end{table}
%%%%%%%%%%%%%%%%%%%%%%%%%%%%%%%%%%%%%%%%%%%%%%%%%%%%%%%%%%%%%%%%%%%%%%%%%%%%%%%%%%%%%%%%%%%%%%%%%%%%%%%%%%%%%%

To investigate correlations (or lack thereof) between the optical polarization and photometric variability of each target, a z-transformed discrete correlation function \citep[ZDCF;][]{1997ASSL..218..163A} was employed using the \textsc{python}-based package \textsc{pyZDCF} \citep{jankov_2022_7253034}. The significance of the observed correlations was assessed through a Monte-Carlo bootstrap procedure, whereby artificial light curves were generated by random resampling of the original data. For each randomized pair, the ZDCF was recomputed, and the distribution of the correlation coefficients at each lag was used to estimate the $1\sigma$, $2\sigma$, and $3\sigma$ significance intervals. 

The ZDCF correlation results are summarised in Table~\ref{tab:correlations}. Here, a strong correlation coefficient is classified as $|r| \geq 0.6$, a moderate correlation as $0.3 \lesssim |r| < 0.6$, and a weak/no correlation as $|r| < 0.3$. Additionally, a positive correlation coefficient indicates that the optical flux and degree of polarization increases/decreases in the same sense. A negative correlation coefficient signifies an anti-correlation between the flux and polarization, whereby the polarization increases as the flux decreases, and vice versa. In the ZDCF framework adopted here, the polarization data is treated as the first time series and the optical flux as the second. Therefore, a positive time lag ($\tau$) implies that flux variations lag behind those in the polarization, while a negative time lag indicates that flux variations lead the polarization.

For six sources, a strong and significant correlation (at a $3\sigma$ confidence level) was found, while for another five, a moderate but statistically significant correlation (at a $2\sigma$ confidence level) was detected. All correlations were found with a non-zero time lag, except for RBS 0248, which showed a near-zero time lag. A caveat of the ZDCF is that large temporal gaps in the data can lead to poorly populated lag bins, which in turn affects both the correlation and uncertainty estimates. All targets become unobservable for several months in a year due to seasonal visibility constraints, resulting in naturally fragmented light curves.

In previous long-term blazar monitoring campaigns, anti-correlations and/or weak correlations between the optical flux and degree of polarization have been reported. The KANATA telescope's polarimetry monitoring has shown (similar to what was found in this study) that the relationship between flux and polarization is source- and epoch-dependent, with individual sources exhibiting reduced polarization during high-flux states \citep{2011PASJ...63..639I}. Similarly, the Liverpool Telescope's monitoring campaign found both correlated and anti-correlated trends in the flux/polarization behaviour during different states of activity \citep{2016MNRAS.462.4267J}.

Within the framework of a very simplistic shock-in-jet scenario, correlated changes in flux and polarization may arise if a propagating shock compresses the magnetic field and increases its ordering, as this simultaneously enhances synchrotron emissivity and increases the degree of polarization. However, it is important to note that shock-in-jet models do not generally predict a correlation between flux and polarization. For example, if there is turbulence downstream of the shock, multiple emission regions, or complex magnetic field geometries, shock-driven flux variations may still be observed, but stochastic magnetic field fluctuations might produce polarization levels that do not correlate with the total intensity.

The presence of non-zero lags in the correlations further suggests that, even when shocks are involved, the flux-polarization variations do not necessarily arise from the same region or probe the same physical aspects of the jet. Such time delays may, in principle, reflect the evolution of the shock as it propagates through the jet. We note, however, that any physical interpretation is limited by the sparse and uneven sampling of the light curves. Both optical flux and polarization variability can occur on shorter timescales than the typical cadence of these observations. Hence, the measured time lags may reflect non-simultaneous variability patterns or sampling effects.

\subsection{Frequency dependence of the polarization}

\begin{figure}
    \centering
    \includegraphics[width=\linewidth]{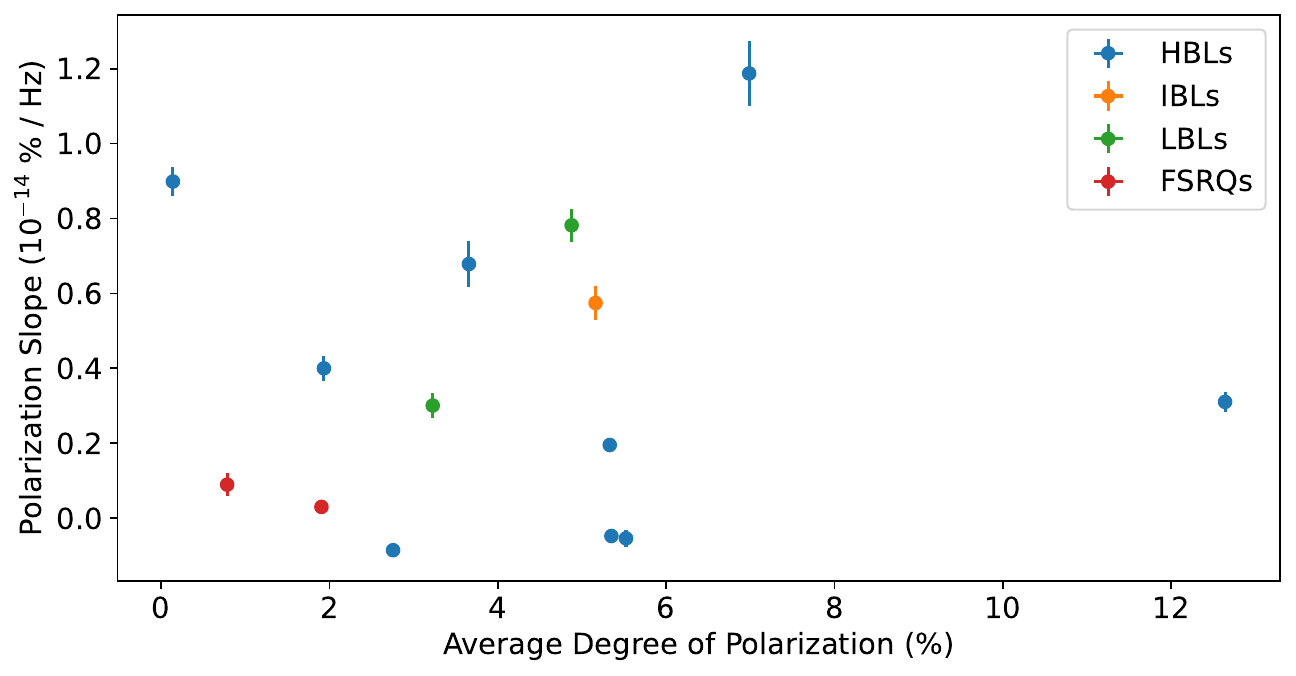}
    \caption{The averaged frequency dependence of the degree of polarization versus the averaged degree of polarization for each of source, indicated by blazar type. No significant correlation is found.}
    \label{fig:pol_slope_type}
\end{figure}

In \paperone\ a slight trend was found that the polarized emission became redder (i.e.\ polarization increased towards longer wavelengths) when the degree of polarization increased for blazars in quiescent states. However, in our expanded sample of low state blazars, no correlation was found when considering the average slope and polarization per source, when considering all sources ($\rho = 0.02$, p-value $= 0.95$), or grouping by blazar type (Fig.~\ref{fig:pol_slope_type}).

\begin{figure}
    \centering
    \includegraphics[width=\linewidth]{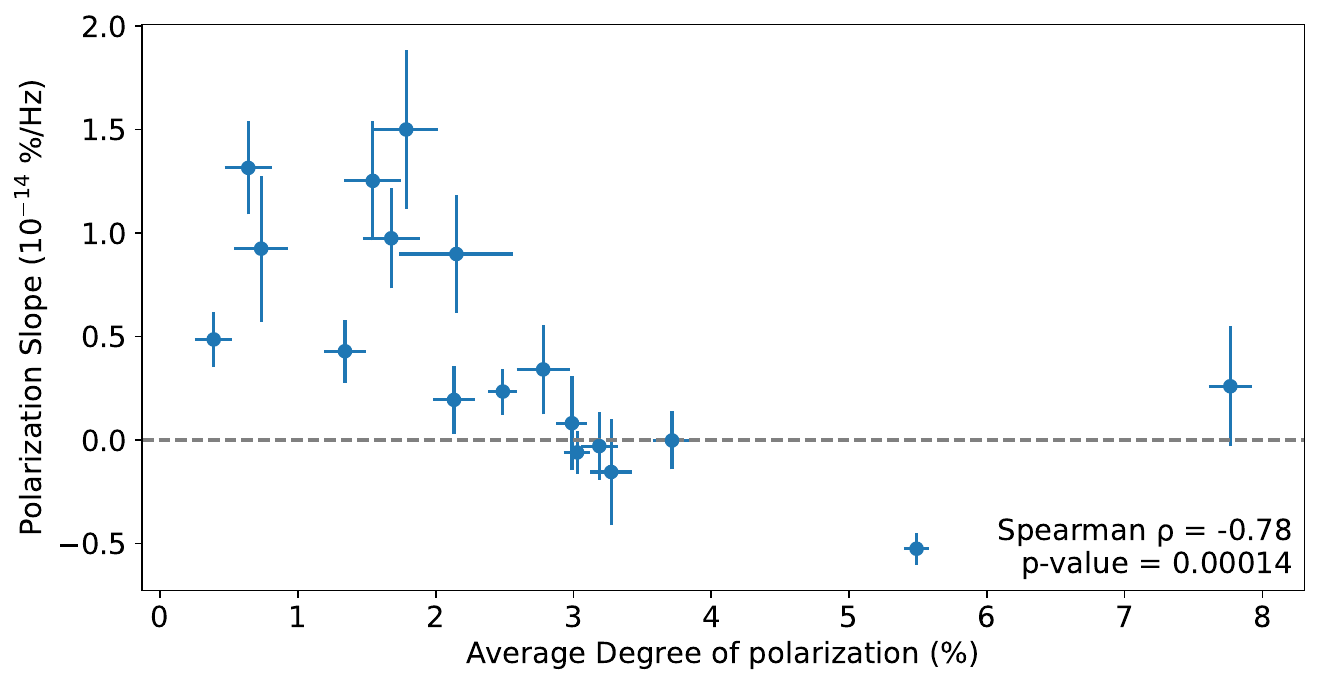}
    \includegraphics[width=\linewidth]{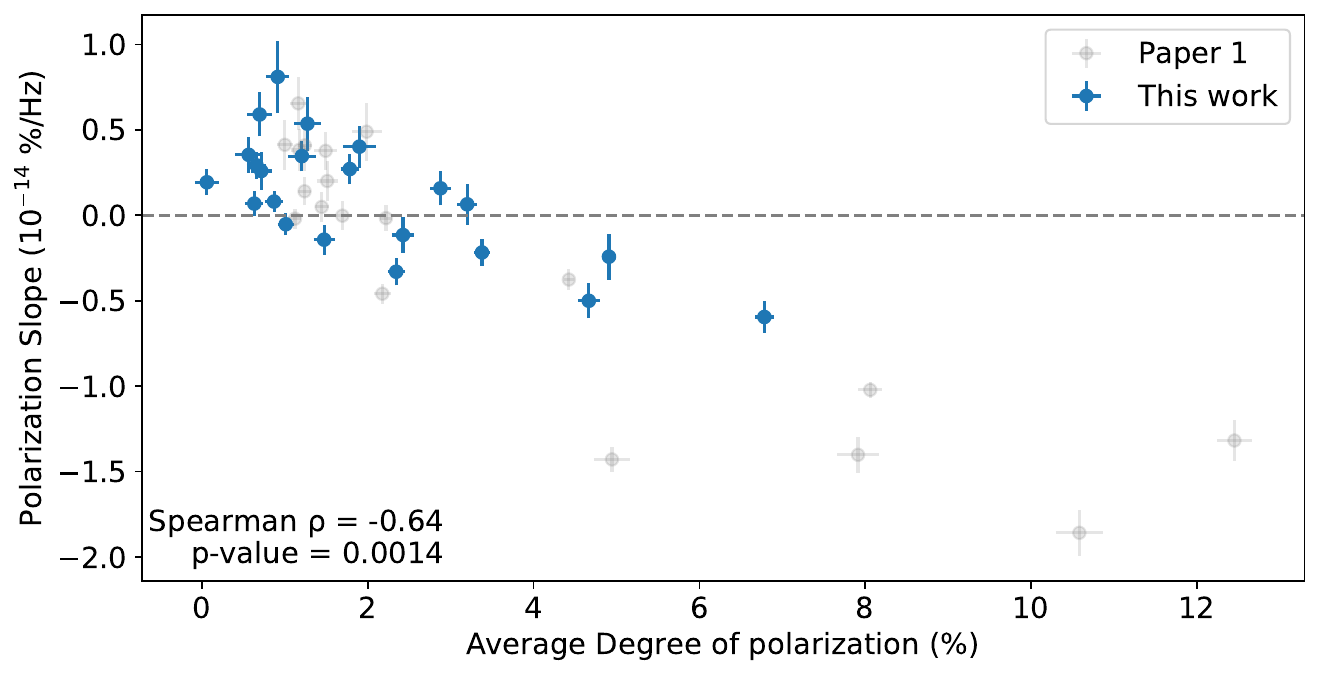}
    \includegraphics[width=\linewidth]{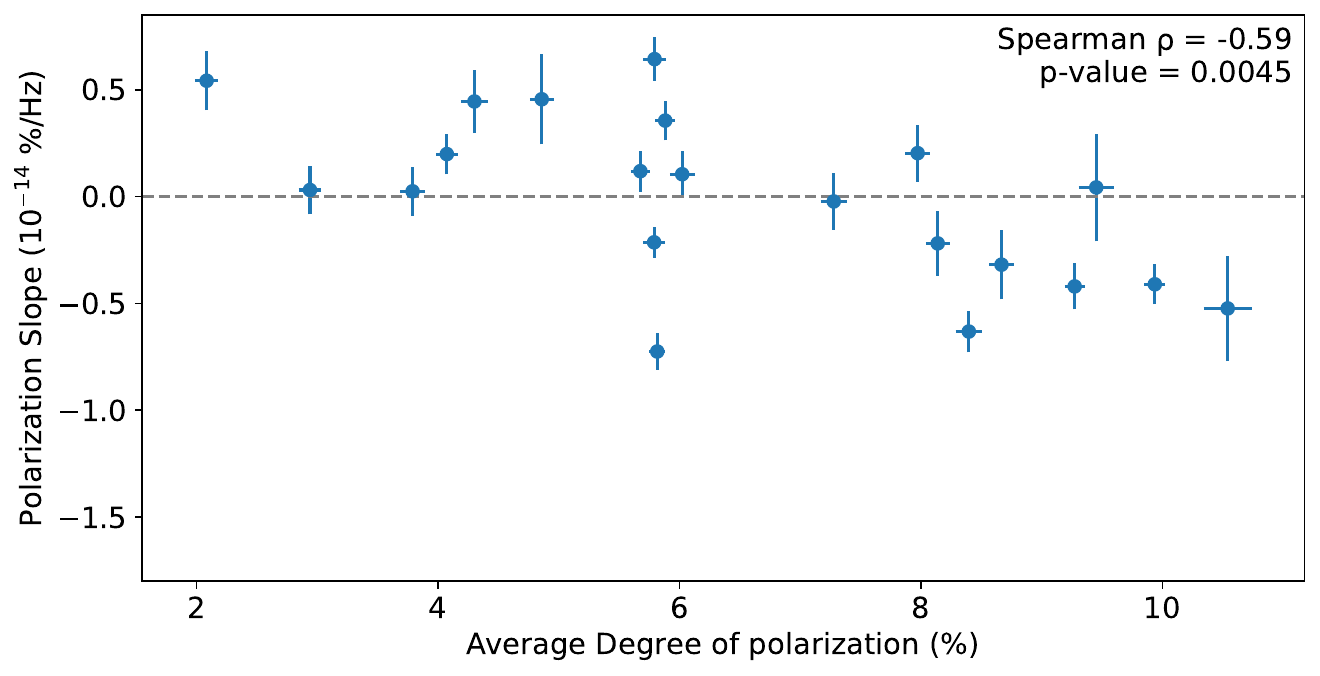}
    \includegraphics[width=\linewidth]{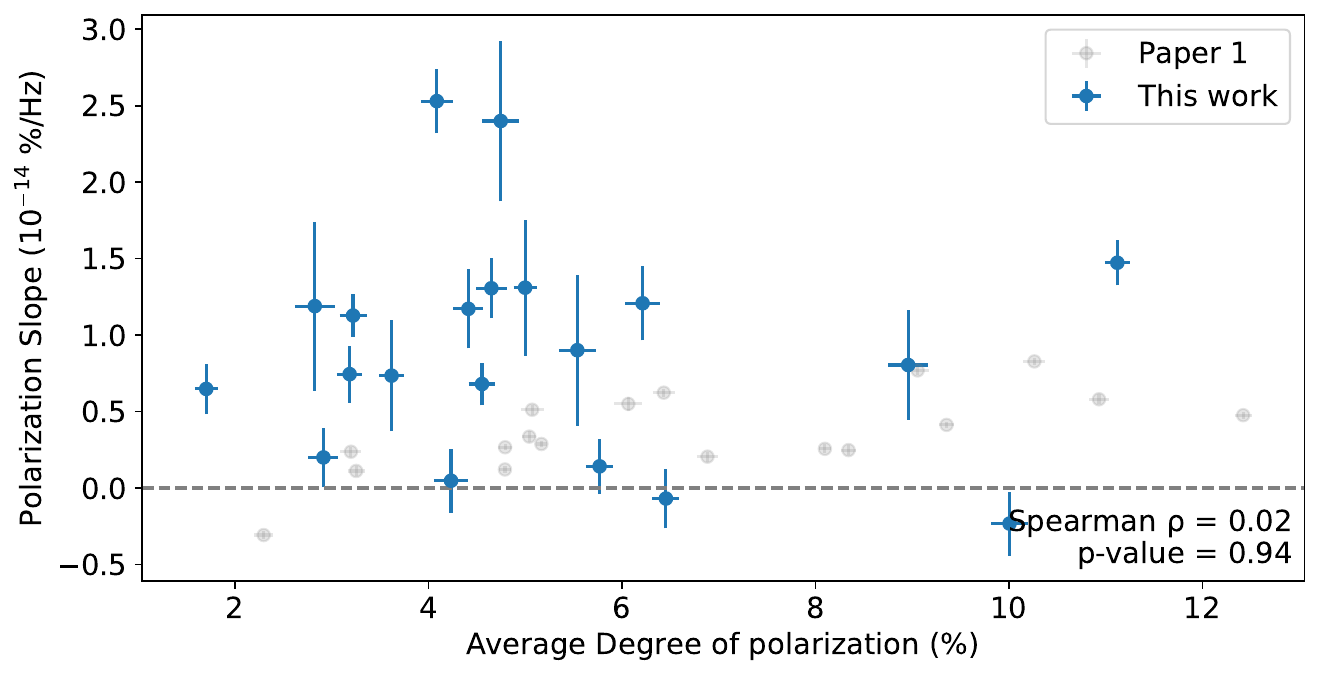}
    \caption{The frequency dependence of the degree of polarization versus the average degree of polarization for individual observations of four sources (\textit{Top:} The FSRQ PKS 0736+017, \textit{upper middle:} the FSRQ PKS 1510-089, \textit{lower middle:} the HBL PKS 2155-304, and \textit{bottom:} the LBL AP Lib). Blue datapoints indicate data taken in this work, while grey datapoints show earlier measurements from \paperone.}
    \label{fig:pol_v_slope_per_source}
\end{figure}

However, a statistically significant anti-correlation is found when considering the frequency dependence and degree of polarization of individual observations for some sources, namely PKS 0736+017, PKS 1510-089, and PKS 2155-304 (shown in the top, upper and lower middle panels of Fig.~\ref{fig:pol_v_slope_per_source}, respectively). The Spearman's rank correlation test results for all of these sources are given in their respective plots in Fig.~\ref{fig:pol_v_slope_per_source}. Thus, for these sources, the frequency dependence becomes more negative (i.e. decreasing towards higher frequencies) as the degree of polarization increases. The frequency dependence of the polarization for all sources is discussed in more detail in Appendix~\ref{app:individualsources}.

In \paperone, the nature of the frequency dependence was shown for three individual sources, which included PKS 1510-089 and AP Lib. In \paperone, an anti-correlation between the frequency dependence and degree of polarization of $\rho = -0.782$ (p-value = $7.519 \times 10^{-5}$) was found for PKS 1510-089. For the PKS 1510-089 observations taken in this work, an anti-correlation of $\rho = -0.64$ (p-value = $0.001$) was found. However, when combining these two datasets (as shown in the upper middle panel of Fig.~\ref{fig:pol_v_slope_per_source}), the strength and significance of the anti-correlation increased significantly to $\rho = -0.71$ (p-value = $1.81 \times 10^{-7}$). This indicates that the correlation is consistently present across epochs. It is possible that the correlation may arise due to the relative contribution between the thermal and non-thermal emission components, or due to changes in the observed synchrotron emission spectrum within the measured wavelength range. The polarization observed within the optical wavelength ranges depends on the particle spectrum, magnetic the magnetic field strength (as this changes the critical frequency; equations~\ref{equ:synchrotron_polarization} \& \ref{eqn:critical_freq}). Additionally, changes in the ordering of the magnetic field with wavelength would change the frequency dependence of the polarization, as this will change the gradient found for a best-fitted straight line fit to the data (equation~\ref{eqn:FB_pol}). No further correlations were found between, for example, the optical flux and degree of polarization, or the flux and frequency dependence of the polarization. It is, however, unlikely that the correlation is driven by changes in the emitting particle population, as PKS 1510-089 has been in a low/quiescent state since 2021, with very limited flux variability and a stable spectrum. Multi-epoch polarimetric monitoring with higher cadence would be required to distinguish between different magnetic-field evolution scenarios.

For AP Lib, a strong, significant, positive correlation between the frequency dependence and degree of polarization of $\rho = 0.618$ (p-value $= 6.251 \times 10^{-3}$) was found in \paperone. For the observations taken in this work, however, no correlation was found ($\rho = 0.02$, p-value $= 0.94$) (bottom panel of Fig.~\ref{fig:pol_v_slope_per_source}). When combining these two datasets, the correlation found in \paperone disappears, with $\rho = -0.01$ (p-value = $0.94$). This indicates that the correlation found in \paperone was likely due to flux variability in a more active state.

\subsection{Relation between the degree of polarization and the synchrotron peak frequency}

\begin{figure}
    \centering
    \includegraphics[width=\columnwidth]{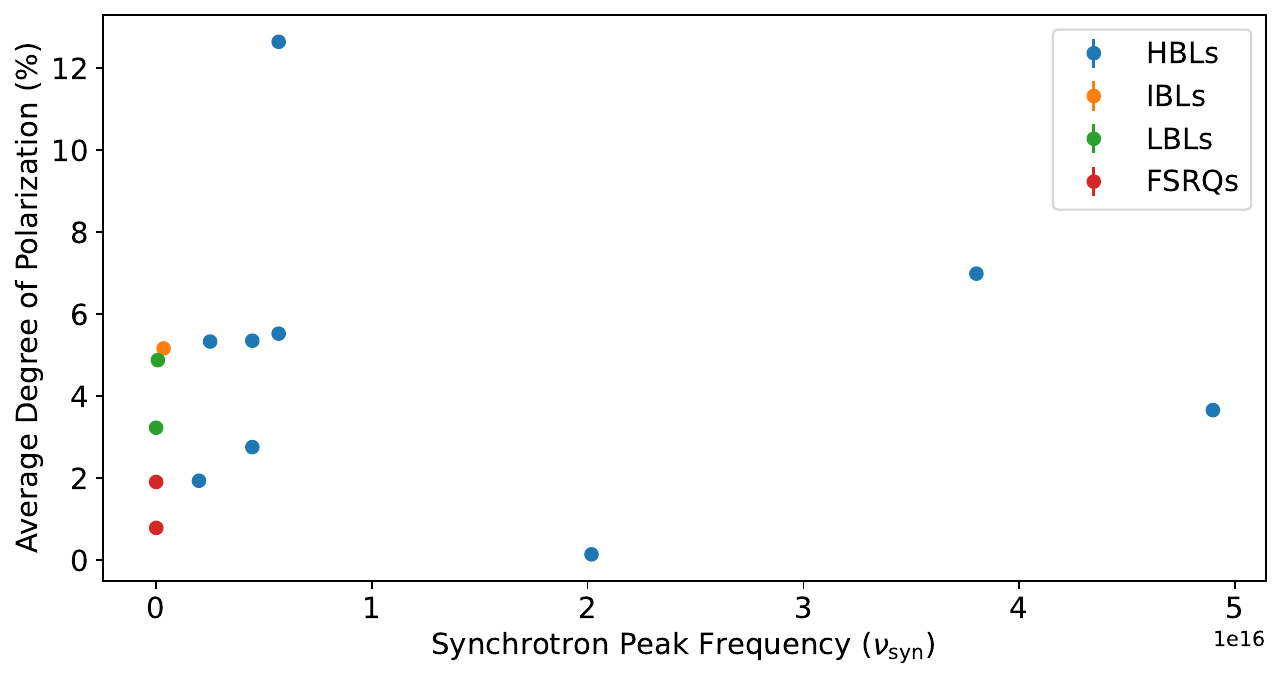}
    \caption{The averaged degree of polarization versus the synchrotron peak frequencies ($\nu_{\rm syn}$) for all sources, grouped by type.}
    \label{fig:all_pol_vs_nusyn}
\end{figure}

Unlike reported by \citet{2016MNRAS.463.3365A}, or marginally found in in \paperone, no correlation was found between the average degree of polarization and synchrotron peak frequency (Fig.~\ref{fig:all_pol_vs_nusyn}; $\rho = 0.39$, p-value $=0.16$ for the entire sample, and $\rho = 0.23$, p-value $=0.56$ for the HBLs only) for our new sample. However, this lack of correlation is likely due to the far smaller number of data points used here, and not necessarily because no correlation exists.

\section{Discussion and Conclusions}
\label{sec:concl}

The observations presented here provide a comprehensive overview of the optical photometric and spectropolarimetric properties of TeV blazars over a 21 month timescale. Across all sources, a low level of polarization ($\lesssim10\%$), and low ordering of the magnetic field ($F_B\lesssim0.3$) was found. This is consistent with the sources being in a quiescent state; showing no strong flares, low flux levels, and possible contributions of an underlying thermal emission component.

The relatively low degree of polarization suggests that the optical emission is likely coming from a larger region within the jet. Within the context of an energy-stratified emission model, this would correspond to lower magnetic field ordering ($F_B$). On average, the observed level of polarization is around 10 per cent, consistent with expectations from a stratified scenario, although substantial variability is observed across all sources as well as for individual sources. This is similar to what was found with the IXPE, which reported relatively low X-ray polarization (but higher than at radio/optical wavelengths) in several blazars. This supports the idea that the emission arises from extended regions in the jet with turbulent magnetic fields \citep[e.g.][]{2024ApJ...963....5E,2024Galax..12...50M}.

The polarization angle exhibited significant variations over time, with the average orientation broadly misaligned with the jet position angle. We also detected several polarization angle swings in our dataset. In an IXPE/MWL campaign on the blazar PG 1553+113, such a polarization angle swing was observed in the optical regime, without a corresponding rotation in the X‑ray or radio bands \citep{2023ApJ...953L..28M}. This ``orphan'' optical polarization angle swing suggests that the optical-emitting region can be separate from higher-energy emission regions in the jet. This provides further support to energy-stratified models explaining the MWL polarization behaviour.

It is clear from the frequency dependence of the degree of polarization that a simple one-zone model fails to capture the observed behaviour of these sources. This suggests that $F_B$ has to be wavelength dependent due to shocks and/or energy stratified emission regions within the jet. Alternatively, the frequency dependence of the polarization may be due to some underlying thermal contribution diluting the observed degree of polarization at specific wavelengths. However, no correlation between the observed flux and frequency dependence of the polarization was found in this study. This indicates that the frequency dependence is rather due to a contribution from the jet structure and magnetic field.

The behaviour of the FSRQs in the sample differed from that of the BLLs, showing much less variation in the observed polarization. For these sources, the thermal contribution is significant and the polarized jet emission is, therefore, diluted. Throughout this monitoring period, the FSRQ PKS 1510-089 remained in a low state of activity, exhibiting stable flux levels and low degrees of polarization.

The observed degree of polarization can be expressed as $\Pi = F_B \cdot \Pi_{\rm synch}$, where $\Pi_{\rm synch}$ depends on the electron spectrum and magnetic field strength, (equation~\ref{equ:synchrotron_polarization}). Variations in the frequency dependence of the degree of polarization can then be due to, for example, changes in the relative level of synchrotron flux to the flux of the thermal contribution, changes in the ordering of the magnetic field, or changes in the underlying particle spectrum. \paperone\ exhibited how the level of synchrotron flux and $F_B$ can affect the observed frequency dependence of the polarization. While there is no universal behaviour or overall correlation found between the flux and polarization, or between the polarization and its frequency dependence, some individual sources exhibited a clear correlation. For example, both of the FSRQs in the sample showed a strong correlation between the polarization and its frequency dependence. Since FSRQs are more thermally dominated, the correlation may reflect the influence of the thermal emission on the polarization. The lack of correlation in the BLLs suggests that the underlying particle distribution and/or jet magnetic field geometry causes the observed polarization behaviour.

Since the sources are in quiescent states, it is unlikely that the emission is dominated by a single emission zone. Rather, the observed fluctuations in the flux and polarization levels may be a superposition of multiple emission regions, each with its own distinct magnetic field structure and ordering. In cases that show a clear correlation between the flux and polarization, a single zone may be dominating the optical emission, which was the model described for PKS 1510-089 in \citet{2023ApJ...952L..38A}.

Overall, this observational campaign highlights the complex nature of blazar jets, particularly in quiescent states. The low polarization levels, significant polarization angle variations, and frequency-dependent behaviour all point to a structured jet with varying magnetic field configurations in different emission regions, rather than a single-zone, homogeneous emission region. These results underscore the importance of long-term, MWL polarimetric monitoring. This project provides a comprehensive dataset, enabling detailed modelling of individual sources as a future work.

\section*{Acknowledgements}

The authors would like to thank the anonymous referee for their valuable and insightful feedback, which greatly enhanced this paper.
JB acknowledges support by the National Research Foundation of South Africa (NRF, grant number PMDS22060619200). BvS acknowledges this work is based on the research supported in part by the National Research Foundation of South Africa (Ref Numbers 119430 and CSRP23041894484). Some of the observations reported in this paper were obtained with the Southern African Large Telescope (SALT) under program 2023-2-MLT-003 (PI: J. Barnard). This work makes use of observations from the Las Cumbres Observatory global telescope network. This research has made use of the SIMBAD database, operated at CDS, Strasbourg, France, as well as the NASA/IPAC Extragalactic Database (NED), which is funded by the National Aeronautics and Space Administration and operated by the California Institute of Technology. This research has made use of the VizieR catalogue access tool, CDS, Strasbourg, France. This work made use of ASTROPY:\footnote{http://www.astropy.org.} a community-developed core PYTHON package and an ecosystem of tools and resources for astronomy.

%%%%%%%%%%%%%%%%%%%%%%%%%%%%%%%%%%%%%%%%%%%%%%%%%%
\section*{Data Availability}

The optical data are available upon reasonable request from the authors. 



\bibliographystyle{mnras}
\bibliography{SPOTS_bib} % if your bibtex file is called example.bib

%%%%%%%%%%%%%%%%%%%%%%%%%%%%%%%%%%%%%%%%%%%%%%%%%%

%%%%%%%%%%%%%%%%% APPENDICES %%%%%%%%%%%%%%%%%%%%%

\appendix

\section{Summary of executed SALT observations}
\label{app:SALT_summary}

Table~\ref{tab:SALT_obs_summary} summarises all successfully completed SALT observations included in this study. For each target, the table lists the Modified Julian Date (MJD) range over which the observations were obtained, along with the total number of successful observations during that period, and the resulting average monitoring cadence, separated by SALT observing semester (2023-2 to 2025-1).

%%%%%%%%%%%%%%%%%%%%%%%%%%%%%%%%%%%%%%%%%%%%%%%%%%%%%%%%%%%%%%%%%%%%%%%%%%%%%%%%%%%%%%%%%%%%%%%%%%%%%%%%%%%%%%
\begin{landscape}
\begin{table}
\begin{center}
\small
\caption{A summary of the successfully completed SALT observations reported in this paper. It provides the date range (in MJD) in which the observations were taken, the number of successful observations, and average cadence obtained for each source, during each SALT observing semester (2023-2 to 2025-1).}
\label{tab:SALT_obs_summary}
\begin{tabular}{lcccccccccccc}
\hline
Target & \multicolumn{3}{c|}{2023-2} & \multicolumn{3}{c|}{2024-2} & \multicolumn{3}{c|}{2024-1} & \multicolumn{3}{c|}{2025-1} \cr
\hline
% \cline{2-10}
& Range of & No. of & Ave. & Range of & No. of & Ave. & Range of & No. of & Ave. & Range of & No. of & Ave. \cr
& obs. [MJD] & obs. & cadence [d] & obs. [MJD] & obs. & cadence [d] & obs. [MJD] & obs. & cadence [d] & obs. [MJD] & obs. & cadence [d] \cr
\hline
RBS 0248	&	60260 - 60281	&	3	&	18	&	60389 - 60442	&	5	&	22	&	60408 - 60458	&	3	&	20	&	60321 - 60335	&	8	&	3	\cr
1RXS J023832.6-311658	&	60254 - 60296	&	3	&	30	&	60383 - 60444	&	4	&	41	&	60407 - 60470	&	5	&	17	&	60312 - 60336	&	4	&	20	\cr
PKS 0301-243	&	60254 - 60328	&	5	&	24	&	60380 - 60441	&	7	&	20	&	60409 - 60474	&	6	&	17	&	60313 - 60340	&	6	&	11	\cr
1ES 0414+009	&	60253 - 60338	&	5	&	25	&	60403 - 60441	&	4	&	15	&	60409 - 60475	&	6	&	17	&	60317 - 60325	&	2	&	8	\cr
PKS 0447-439	&	60252 - 60373	&	9	&	18	&	60386 - 60444	&	6	&	19	&	60408 - 60507	&	9	&	14	&	60302 - 60316	&	7	&	4	\cr
TXS 0506+056	&	60256 - 60334	&	6	&	19	&	60400 - 60441	&	3	&	24	&	60394 - 60475	&	9	&	11	&	--	&	--	&	--	\cr
PKS 0736+017	&	60286 - 60356	&	10	&	15	&	--	&	--	&	--	&	60400 - 60526	&	9	&	16	&	--	&	--	&	--	\cr
PKS 1440-389	&	60311 - 60357	&	8	&	13	&	60369 - 60427	&	6	&	23	&	60450 - 60510	&	4	&	17	&	60288 - 60335	&	9	&	10	\cr
PKS 1510-089	&	60338 - 60363	&	8	&	11	&	60369 - 60414	&	6	&	21	&	60480	&	1	&	--	&	60272 - 60319	&	8	&	9	\cr
AP Lib	&	60309 - 60357	&	7	&	15	&	60369 - 60422	&	7	&	19	&	60442 - 60504	&	3	&	17	&	60272 - 60274	&	9	&	0.25	\cr
PKS 1749+096	&	60344 - 60363	&	4	&	9	&	60368 - 60421	&	7	&	19	&	--	&	--	&	--	&	60272 - 60318	&	10	&	4	\cr
PKS 2005-489	&	60253 - 60347	&	3	&	83	&	60370 - 60441	&	9	&	22	&	--	&	--	&	--	&	60272 - 60273	&	12	&	0.5	\cr
PKS 2155-304	&	60256 - 60347	&	2	&	162	&	60373 - 60441	&	9	&	22	&	60399 - 60430	&	2	&	29	&	60272 - 60300	&	12	&	7	\cr
1ES 2322-409	&	60257 - 60292	&	3	&	18	&	60374 - 60444	&	9	&	22	&	60400 - 60454	&	3	&	15	&	60278 - 60336	&	11	&	6	\cr
\hline
\end{tabular}
\end{center}
\end{table}
\end{landscape}
%%%%%%%%%%%%%%%%%%%%%%%%%%%%%%%%%%%%%%%%%%%%%%%%%%%%%%%%%%%%%%%%%%%%%%%%%%%%%%%%%%%%%%%%%%%%%%%%%%%%%%%%%%%%%%

Table~\ref{tab:S/N_properties} provides a summary of the achieved S/N ratios for the observations of each target in this work. Here, the average, minimum, and maximum S/N ratios are provided. The S/N ratios were calculated using the DER\_SNR method outlined in \citep{2008ASPC..394..505S}. Since it estimates the noise from pixel-to-pixel variations, the unbinned spectra was used to estimate the S/N ratios.

%%%%%%%%%%%%%%%%%%%%%%%%%%%%%%%%%%%%%%%%%%%%%%%%%%%%%%%%%%%%%%%%%%%%%%%%%%%%%%%%%%%%%%%%%%%%%%%%%%%%%%%%%%%%%%
\begin{table}
\begin{center}
\small
\caption{The average, minimum, and maximum signal-to-noise ratios obtained in the SALT observations of each target in the SPOTS monitoring campaign.}
\label{tab:S/N_properties}
\begin{tabular}{lccc}
\hline
Target & $\langle \rm{S/N} \rangle$ & $\rm{S/N}_{\rm{min}}$ & $\rm{S/N}_{\rm{max}}$ \\
\hline
RBS 0248 & 27 & 16 & 42 \cr
1RXS J023832.6--311658 & 50 & 31 & 67 \cr
PKS 0301--243 & 48 & 30 & 73 \cr
1ES 0414+009 & 14 & 6 & 21 \cr
PKS 0447--439 & 137 & 33 & 188 \cr
TXS 0506+056 & 21 & 6 & 34 \cr
PKS 0736+017 & 32 & 12 & 51 \cr
PKS 1440--389 & 59 & 12 & 88 \cr
PKS 1510--089 & 37 & 21 & 52 \cr
AP Lib & 31 & 21 & 42 \cr
PKS 1749+096 & 22 & 10 & 40 \cr
PKS 2005--489 & 34 & 23 & 66 \cr
PKS 2155--304 & 44 & 17 & 72 \cr
1ES 2322--409 & 33 & 14 & 59 \cr
\hline
\end{tabular}
\end{center}
\end{table}
%%%%%%%%%%%%%%%%%%%%%%%%%%%%%%%%%%%%%%%%%%%%%%%%%%%%%%%%%%%%%%%%%%%%%%%%%%%%%%%%%%%%%%%%%%%%%%%%%%%%%%%%%%%%%%

\section{A note on the effects of interstellar polarization}
\label{app:ISP}

As discussed in Sec.~\ref{sec:specpol}, the effect of Galactic extinction on the measured polarization can be considered negligible for the targets presented in this study (see also \paperone). Table \ref{tab:ISP_properties} provides the estimated interstellar extinction along each target's line of sight, along with the typical and maximum expected ISP values, as well as the median value of the maximum expected polarization angle change due to the ISP for each source.

%%%%%%%%%%%%%%%%%%%%%%%%%%%%%%%%%%%%%%%%%%%%%%%%%%%%%%%%%%%%%%%%%%%%%%%%%%%%%%%%%%%%%%%%%%%%%%%%%%%%%%%%%%%%%%
\begin{table*}
\begin{center}
\small
\caption{A summary of the Galactic extinction \citep{2011ApJ...737..103S} along each target's line of sight, along with the corresponding typical and maximum ISP levels \citep{1975ApJ...196..261S, Smith04}, and the maximum polarization angle change that can be expected due to the effects of the ISP \citep{2005MNRAS.363.1241H}.}
\label{tab:ISP_properties}
\begin{tabular}{lcccc}
\hline
Target & Galactic Extinction & ISP$_{\rm{typical}}$ & ISP$_{\rm{max}}$ & $\Delta \rm{PA}_{\rm{max}}$ \\
& $E(B-V)$ & $(\%)$ & $(\%)$ & ($^\circ$) \\
\hline
RBS 0248 & 0.028 & 0.084 & 0.251 & 0.004 \cr
1RXS J023832.6--311658 & 0.028 & 0.083 & 0.248 & 0.005 \cr
PKS 0301--243 & 0.019 & 0.056 & 0.169 & 0.002 \cr
1ES 0414+009 & 0.111 & 0.336 & 1.007 & 0.013 \cr
PKS 0447--439 & 0.012 & 0.035 & 0.10 & 0.001 \cr
TXS 0506+056 & 0.092 & 0.277 & 0.832 & 0.005 \cr
PKS 0736+017 & 0.116 & 0.359 & 1.048 & 0.017 \cr
PKS 1440--389 & 0.102 & 0.307 & 0.921 & 0.013 \cr
PKS 1510--089 & 0.085 & 0.256 & 0.768 & 0.019 \cr
AP Lib & 0.118 & 0.353 & 1.059 & 0.011 \cr
PKS 1749+096 & 0.151 & 0.454 & 1.363 & 0.011 \cr
PKS 2005--489 & 0.048 & 0.144 & 0.431 & 0.006 \cr
PKS 2155--304 & 0.019 & 0.056 & 0.167 & 0.001 \cr
1ES 2322--409 & 0.016 & 0.049 & 0.147 & 0.001 \cr
\hline
\end{tabular}
\end{center}
\end{table*}
%%%%%%%%%%%%%%%%%%%%%%%%%%%%%%%%%%%%%%%%%%%%%%%%%%%%%%%%%%%%%%%%%%%%%%%%%%%%%%%%%%%%%%%%%%%%%%%%%%%%%%%%%%%%%%

\section{Notes on Individual Sources}
\label{app:individualsources}

The observed optical photometric and spectropolarimetric observations of the individual sources are summarized below. Fig.~\ref{fig:all_lcs} shows, from top to bottom, the optical photometric light curve (LCOGT observations), the degree of polarization, the frequency dependence of the polarization, and the polarization angle (SALT data) for each source. The frequency dependence of the polarization refers to whether the polarization increases (+) or decreases (-) towards higher frequencies. A Spearman's rank correlation was applied to investigate the link between the degree and frequency dependence of the polarization. Because of the $180^\circ$ ambiguity in the polarization angle, the polarization angles are plotted by assuming  the smallest change in angle, as was done in e.g. \citet{2015MNRAS.453.1669B,2016MNRAS.462.1775B}.

\subsection{RBS 0248}

RBS 0248 remained fairly stable during the period of observation. The mean V-band flux was $1.93$\,mJy, ranging from $1.54$ to $2.27$\,mJy (standard deviation $0.16$\,mJy). The degree of polarization fluctuated between $2.12 \pm 0.48$ and $11.36 \pm 0.33\%$. The ZDCF between the V-band flux and polarization gives $\tau = -0.87$\,d, with a moderate correlation of $r = 0.40$ ($>68\%$ confidence). For all but one observation, the polarization increases towards higher frequencies, and the frequency dependence remained fairly constant with an average of $(1.28 \pm 0.09)\times10^{-14}\,\%/\rm{Hz}$. A moderate, but statistically insignificant anti-correlation was found between the degree of polarization and its frequency dependence for this source ($\rho = -0.57$, p-value $= 0.03$). The polarization angle remained fairly constant, ranging from $7.34 \pm 1.25$ to $47.93 \pm 0.79^{\circ}$. 

\subsection{1RXS J023832.6-311658}

The V-band flux of 1RXS J023832.6-311658 remained fairly constant, lying with $0.16$\,mJy of the mean flux,  $0.77$\,mJy. The degree of polarization remained very low, and fluctuated between $0.06 \pm 0.13$ and $3.43 \pm 0.15\%$. A weak, but statistically insignificant correlation was found between the optical V-band flux and degree of polarization ($\tau = -26.72$\,d, $r = 0.19$) for this source, with $r$ consistent with the noise levels at all time lags. All observations exhibited polarization increasing towards higher frequencies, and varied between $(7.42 \pm 1.31)\times10^{-15}$ and $(2.74 \pm 0.45) \times10^{-14}\,\%/\rm{Hz}$, with an average of $(1.11 \pm 0.05) \times10^{-14}\,\%/\rm{Hz}$. No statistically significant correlation was found between the degree of polarization and its frequency dependence for this source ($\rho = 0.10$, p-value $= 0.71$). A smooth polarization angle rotation was observed between MJD 60492 and MJD 60701, ranging from $51.82 \pm 2.98$ to $-204.48 \pm 3.63^{\circ}$.

\subsection{PKS 0301-243}

For PKS 0301-243, some variability was observed. The V-band flux varied between $1.12$ to $1.80$\,mJy, with a mean value of $1.35$\,mJy (standard deviation $0.16$\,mJy). The degree of polarization varied between $2.45 \pm 0.07$ and $9.09 \pm 0.06\%$. The ZDCF between the polarization and V-band flux shows a peak at $\tau = 69.67$\,d, with a moderate correlation of $r = 0.40$ ($>68\%$ confidence). The frequency dependence of the polarization fluctuated between increasing and decreasing towards higher frequencies between observations. It varied from $(-2.86 \pm 0.70) \times 10^{-15}$ to $(1.54 \pm 0.38) \times 10^{-14}\,\%/\rm{Hz}$, with an average of $(2.50 \pm 0.22) \times 10^{-15}\,\%/\rm{Hz}$. No correlation was found between the degree of polarization and its frequency dependence for this source ($\rho = -0.01$, p-value $= 0.97$). Stochastic variability in the polarization angle was detected, varying from $7.92 \pm 1.17$ to $38.62 \pm 0.54^{\circ}$.

\subsection{1ES 0414+009}

During the period of observation, the optical flux of 1ES 0414+009 remained within $\pm 0.05$\,mJy  of the mean flux of $0.45$\,mJy in the V-band. The degree of polarization remained low and fairly constant, fluctuating between $0.81 \pm 0.24$ and $8.08 \pm 0.37\%$. No significant correlation was found between the optical V-band flux and degree of polarization for this source ($\tau = -34.56$\,d, $r = 0.27$, $>68\%$ confidence). The frequency dependence of the polarization fluctuated between increasing and decreasing towards higher frequencies between observations. It varied from $(-1.45 \pm 0.28) \times 10^{-14}$ to $(5.23 \pm 0.77) \times 10^{-14}\,\%/\rm{Hz}$, with an average of $(8.73 \pm 0.80) \times 10^{-15}\,\%/\rm{Hz}$. No correlation was found between the degree of polarization and its frequency dependence for this source ($\rho = 0.01$, p-value $= 0.97$). A smooth polarization angle rotation was observed between MJD 60564 and MJD 60710, rotating from $127.71 \pm 4.58$ to $-33.40 \pm 1.31^{\circ}$.

\subsection{PKS 0447-439}

PKS 0447-439 was found to be variable, with distinct peaks (around MJD 60350, MJD 60600, and MJD 60900, respectively) in the optical light curves. The R-band fluxes ranged from $5.70$ to $19.69$\,mJy, with a mean flux of $10.41$\,mJy (standard deviation $3.07$\,mJy). The degree of polarization also showed larger fluctuations, changing between $1.34 \pm 0.04$ and $16.53 \pm 0.04\%$. The ZDCF between the polarization and R-band flux shows a peak at $\tau = -49.96$\,d, with a strong correlation of $r = 0.73$ ($>95\%$ confidence). This suggests that the polarization is strongly correlated, but lags behind the optical flux changes. The frequency dependence of the polarization fluctuated between increasing and decreasing towards higher frequencies between observations. It varied from $(-1.78 \pm 0.13) \times 10^{-14}$ to $(3.71 \pm 2.34) \times 10^{-15}\,\%/\rm{Hz}$, with an average of $(-5.32 \pm 0.79) \times 10^{-16}\,\%/\rm{Hz}$. No correlation was found between the degree of polarization and the frequency dependence ($\rho = -0.24$, p-value $= 0.22$). A polarization angle rotation was seen between MJD 60269 to 60398, rotating from $18.09 \pm 0.24$ to $125.41 \pm 0.12^{\circ}$.

\subsection{TXS 0506+056}

TXS 0506+056 exhibited some flux variability around the mean V-band flux of $2.49$\,mJy (standard deviation $0.64$\,mJy) ranging from $1.53$ to $4.66$\,mJy. The degree of polarization fluctuated between $4.15 \pm 1.07$ and $21.64 \pm 0.19\%$, which was the highest level of polarization detected across all sources during the period of observation. No significant correlation was found between the optical V-band flux and degree of polarization ($\tau = -6.92$\,d, $r = 0.25$), with correlation coefficients consistent with the noise levels at all time lags. The frequency dependence of the polarization varied between $(-7.22 \pm 3.87) \times 10^{-15}$ to $(2.75 \pm 0.26) \times 10^{-14}\,\%/\rm{Hz}$, with an average of $(7.68 \pm 0.60) \times 10^{-15}\,\%/\rm{Hz}$. No correlation was found between the degree of polarization and its frequency dependence ($\rho = -0.11$, p-value $= 0.67$). The polarization angle varied stochastically between $77.77 \pm 0.68$ and $216.72 \pm 0.63^{\circ}$.

\subsection{PKS 0736+017}

Two distinct peaks were observed around MJD 60250 and MJD 60720 in the optical light curves of PKS 0736+017, with the V-band flux ranging from $0.59$ to $3.09$\,mJy. The mean flux was found to be $1.06$\,mJy, with a standard deviation of $0.62$\,mJy. The degree of polarization fluctuated between $0.39 \pm 0.14$ and $7.77 \pm 0.16\%$. The ZDCF between the V-band flux and polarization shows a peak at $\tau = -37.58$\,d, with a strong correlation of $r = 0.78$ ($>95\%$ confidence). Thus, the polarization is strongly correlated to, and lags behind the observed flux changes. The frequency dependence of the polarization varied between $(-5.26 \pm 0.78) \times 10^{-15}$ to $(1.50 \pm 0.38) \times 10^{-14}\,\%/\rm{Hz}$, with an average of $(1.06 \pm 0.38) \times 10^{-15}\,\%/\rm{Hz}$. A strong, statistically significant anti-correlation was found between the degree of polarization and its frequency dependence for this source ($\rho = -0.78$, p-value $= 1.38 \times 10^{-4}$). From the data, it can be suggested that the polarization angle underwent a long-term, smooth rotation between MJD 60398 and MJD 60428, ranging from $30.18 \pm 3.74$ to $-28.72 \pm 3.86^{\circ}$.

\subsection{PKS 1440-389}

PKS 1440-389 exhibited flux variability during the monitoring period, fluctuating from $3.37$ to $7.04$\,mJy. The average V-band flux was $4.82$\,mJy (standard deviation $0.76$\,mJy). The degree of polarization remained low, and fluctuated between $0.62 \pm 0.11$ and $9.32 \pm 0.06\%$. The ZDCF between the V-band flux and polarization gives $\tau = 15.44$\,d, with a strong correlation of $r = 0.70$ ($>95\%$ confidence), suggesting that the optical fluxes are correlated to, but lag behind the changes observed in the polarization. The frequency dependence of the polarization fluctuated between increasing and decreasing towards higher frequencies between observations. It varied from $(-7.35 \pm 1.20) \times 10^{-15}$ to $(1.14 \pm 0.38) \times 10^{-14}\,\%/\rm{Hz}$, with an average of $(-9.79 \pm 1.61) \times 10^{-16}\,\%/\rm{Hz}$. No correlation was found between the degree of polarization and its frequency dependence for this source ($\rho = -0.34$, p-value $= 0.09$). The polarization angle rotated smoothly from $84.92 \pm 0.81$ to $262.74 \pm 0.83^{\circ}$ between MJD 60416 and MJD 60547.

\subsection{PKS 1510-089}

PKS 1510-089 exhibited some low-level variability, with a mean V-band flux of $0.99$\,mJy (standard deviation of $0.14$\,mJy), ranging from $0.79$ to $1.55$\,mJy. The degree of polarization remained low, and fluctuated between $0.06 \pm 0.14$ and $6.78 \pm 0.11\%$. The ZDCF between the V-band flux and polarization finds a maximum at $\tau = 7.8$\,d, with a moderate correlation of $r = 0.42$ ($>68\%$ confidence). The frequency dependence of the polarization fluctuated between increasing and decreasing towards higher frequencies between observations. It varied from $(-5.95 \pm 0.91) \times 10^{-15}$ to $(8.10 \pm 2.12) \times 10^{-15}\,\%/\rm{Hz}$, with an average of $(4.03 \pm 1.95) \times 10^{-16}\,\%/\rm{Hz}$. A strong, statistically significant correlation was found between the degree of polarization and its frequency dependence for this source ($\rho = -0.64$, p-value $= 0.001$). Stochastic variability in the polarization angle was observed, fluctuating between $25.69 \pm 0.81$ and $90.68 \pm 8.84^{\circ}$.

\subsection{AP Librae}

AP Lib exhibited some variability, with a mean V-band flux of $4.60$\,mJy, ranging from $3.45$ to $7.03$\,mJy, with a standard deviation of $0.80$\,mJy. The degree of polarization fluctuated between $1.70 \pm 0.12$ and $11.12 \pm 0.13\%$. The ZDCF between the polarization and V-band flux shows a peak at $\tau = 40.96$\, d, with a strong correlation of $r = 0.67$ ($>95\%$ confidence). This suggests that the polarization is strongly correlated to the optical flux, but increases before the optical flux does. The frequency dependence of the polarization fluctuated between increasing and decreasing towards higher frequencies between observations. It varied from $(-2.33 \pm 2.09) \times 10^{-15}$ to $(2.53 \pm 0.21) \times 10^{-14}\,\%/\rm{Hz}$, with an average of $(8.20 \pm 0.47) \times 10^{-15}\,\%/\rm{Hz}$. No correlation was found between the degree of polarization and the frequency dependence  ($\rho = 0.02$, p-value $= 0.94$). The polarization angle remained fairly constant, with stochastic variation between $159.37 \pm 1.96$ and $179.69 \pm 0.89^{\circ}$.

\subsection{PKS 1749+096}

Low-level variability was observed for PKS 1749+096, with a mean V-band flux of $0.71$\,mJy, ranging from $0.33$ to $1.81$\,mJy. The standard deviation of the fluxes was $0.29$\,mJy. The degree of polarization fluctuated between $1.63 \pm 0.21$ and $18.48 \pm 0.19\%$. No significant correlation was found between the optical V-band flux and degree of polarization ($\tau = -24.49$\,d, $r = 0.39$, $>68\%$ confidence). The frequency dependence of the polarization fluctuated between increasing and decreasing towards higher frequencies between observations. It varied from $(-4.28 \pm 0.90) \times 10^{-15}$ to $(3.31 \pm 0.29) \times 10^{-14}\,\%/\rm{Hz}$, with an average of $(3.97 \pm 0.44) \times 10^{-15}\,\%/\rm{Hz}$. No correlation was found between the degree of polarization and its frequency dependence for this source ($\rho = 0.22$, p-value $= 0.33$). The polarization angle rotated smoothly from $83.58 \pm 1.04$ to $143.73 \pm 0.64^{\circ}$ between MJD 60403 and 60488, and from $180.26 \pm 0.59$ to $83.55 \pm 0.30^{\circ}$ between MJD 60831 and MJD 60882.

\subsection{PKS 2005-489}

PKS 2005-489 was found to be highly variable in the optical regime. A distinct peak in the optical flux was observed around MJD 60620. During the period of observation, the V-band fluxes ranged from $7.05$\,mJy to $18.86$\,mJy. The mean flux was found to be $9.88$\,mJy (standard deviation of $2.86$\,mJy). The degree of polarization fluctuated between $0.93 \pm 0.06$ and $10.40 \pm 0.10\%$. The ZCDF between the V-band flux and polarization peaked at $\tau = 24.19$\,d with a moderate correlation of $r = 0.53$ ($>68\%$ confidence). The frequency dependence of the polarization fluctuated between increasing and decreasing towards higher frequencies between observations. It varied from $(-7.59 \pm 1.91) \times 10^{-15}$ to $(1.74 \pm 0.31) \times 10^{-14}\,\%/\rm{Hz}$, with an average of $(4.28 \pm 0.35) \times 10^{-15}\,\%/\rm{Hz}$. No correlation was found between the degree of polarization and its frequency dependence for this source ($\rho = -0.38$, p-value $= 0.13$). The polarization angle varied stochastically between $-20.70 \pm 0.58$ and $88.78 \pm 0.29^{\circ}$.

\subsection{PKS 2155-304}

PKS 2155-304 exhibited variability, with V-band fluxes ranging from $7.03$ to $18.30$\,mJy. The mean flux was $10.70$\,mJy, with a standard deviation of $2.02$\,mJy. The degree of polarization fluctuated between $2.09 \pm 0.10$ and $10.54 \pm 0.20\%$. The ZDCF between the V-band flux and polarization gives $\tau = 30.66$\,d, with $r = 0.79$ ($>95\%$), indicating that the optical flux lags behind the changes in polarization, but are strongly correlated. The frequency dependence of the polarization fluctuated between increasing and decreasing towards higher frequencies between observations. It varied from $(-7.25 \pm 0.85) \times 10^{-15}$ to $(6.43 \pm 1.02) \times 10^{-15}\,\%/\rm{Hz}$, with an average of $(-6.07 \pm 2.45) \times 10^{-16}\,\%/\rm{Hz}$. A strong, statistically significant anti-correlation was found between the degree of polarization and its frequency dependence for this source ($\rho = -0.59$, p-value $= 0.004$). The polarization angle remained fairly constant, ranging from $64.83 \pm 0.75$ and $117.22 \pm 0.77^{\circ}$.

\subsection{1ES 2322-409}

The optical flux of 1ES 2322-409 exhibited some variability during the monitoring period, with a mean V-band flux of $1.34$\,mJy, a standard deviation of $0.40$\,mJy, and fluxes ranging from $0.77$ to $2.28$\,mJy. The degree of polarization varied between $6.64 \pm 0.13$ and $15.77 \pm 0.17\%$. The ZDCF between the polarization and V-band light curve shows a peak at $\tau = -38.92$\,d, with a strong correlation coefficient of $r = 0.62$. Bootstrap simulations indicate that this correlation exceeds the $95\%$ confidence level, suggesting that the polarization is strongly correlated to, but lags behind the flux changes. The frequency dependence of the polarization fluctuated between increasing and decreasing towards higher frequencies between observations. It varied from $(-2.19 \pm 0.24) \times 10^{-14}$ to $(2.21 \pm 0.27) \times 10^{-14}\,\%/\rm{Hz}$, with an average of $(3.64 \pm 0.32) \times 10^{-15}\,\%/\rm{Hz}$. No correlation was found between the degree of polarization and its frequency dependence ($\rho = 0.01$, p-value $= 0.96$). Stochastic variability was observed in the polarization angle, ranging from $-1.27 \pm 0.23$ to $15.11 \pm 0.47^{\circ}$.

\begin{figure*}
    \centering
    \begin{subfigure}[t]{0.42\textwidth}
        \includegraphics[width=\linewidth]{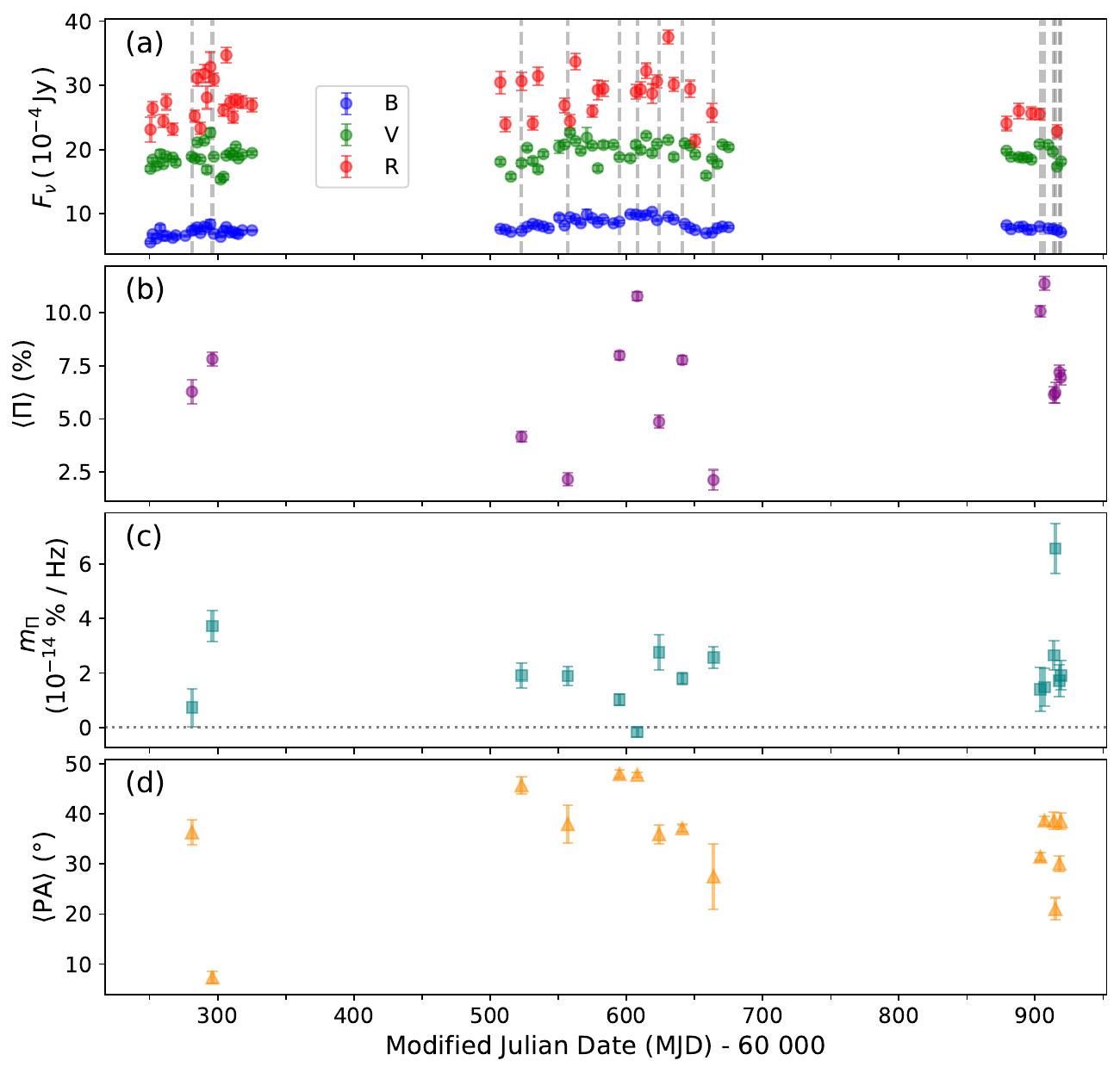}
        \caption{RBS 0248}
    \end{subfigure}
    \hfill
    \begin{subfigure}[t]{0.42\textwidth}
        \includegraphics[width=\linewidth]{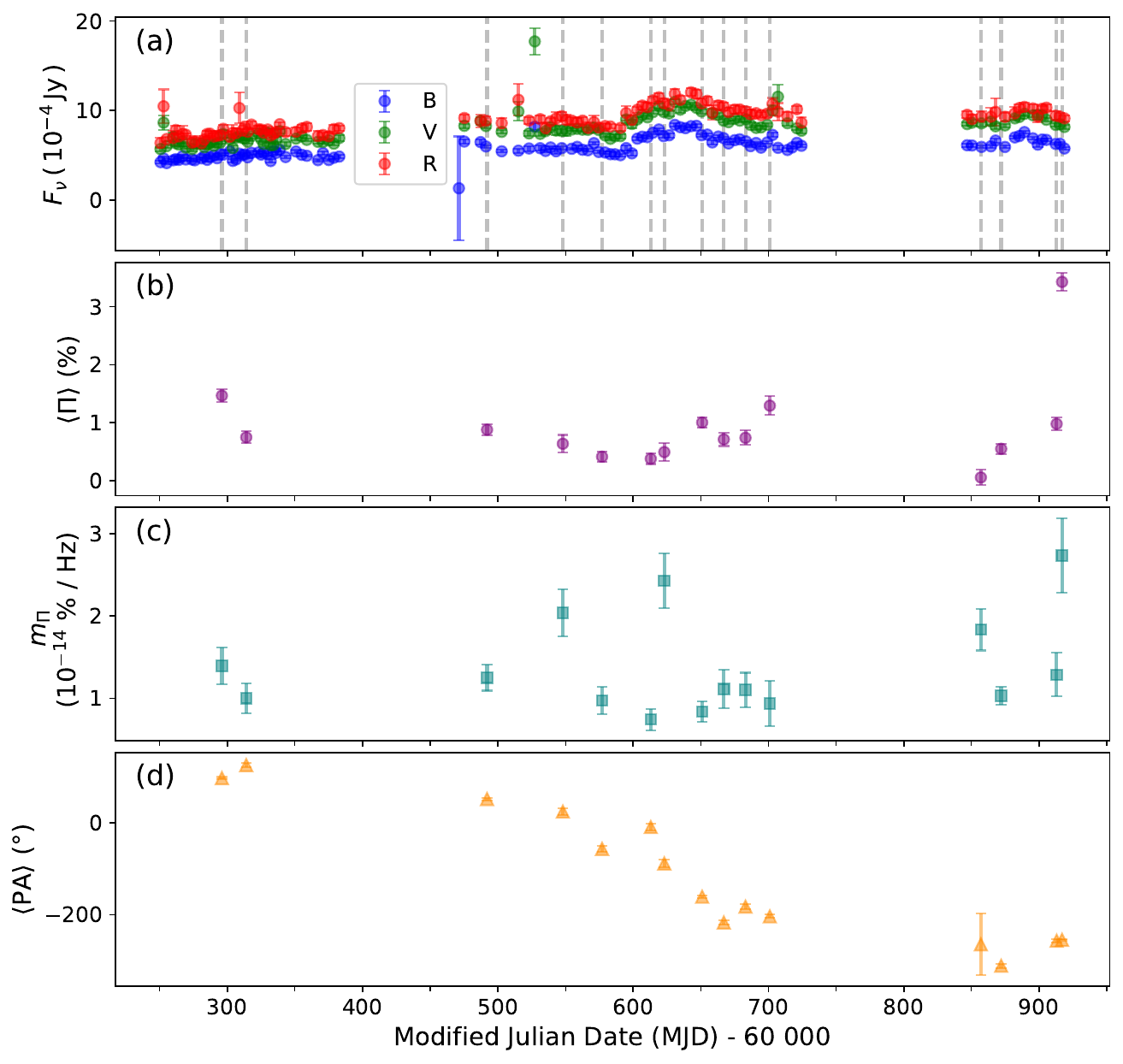}
        \caption{1RXS J023832.6-311658}
    \end{subfigure}

    \vspace{0.3cm}

    \begin{subfigure}[t]{0.42\textwidth}
        \includegraphics[width=\linewidth]{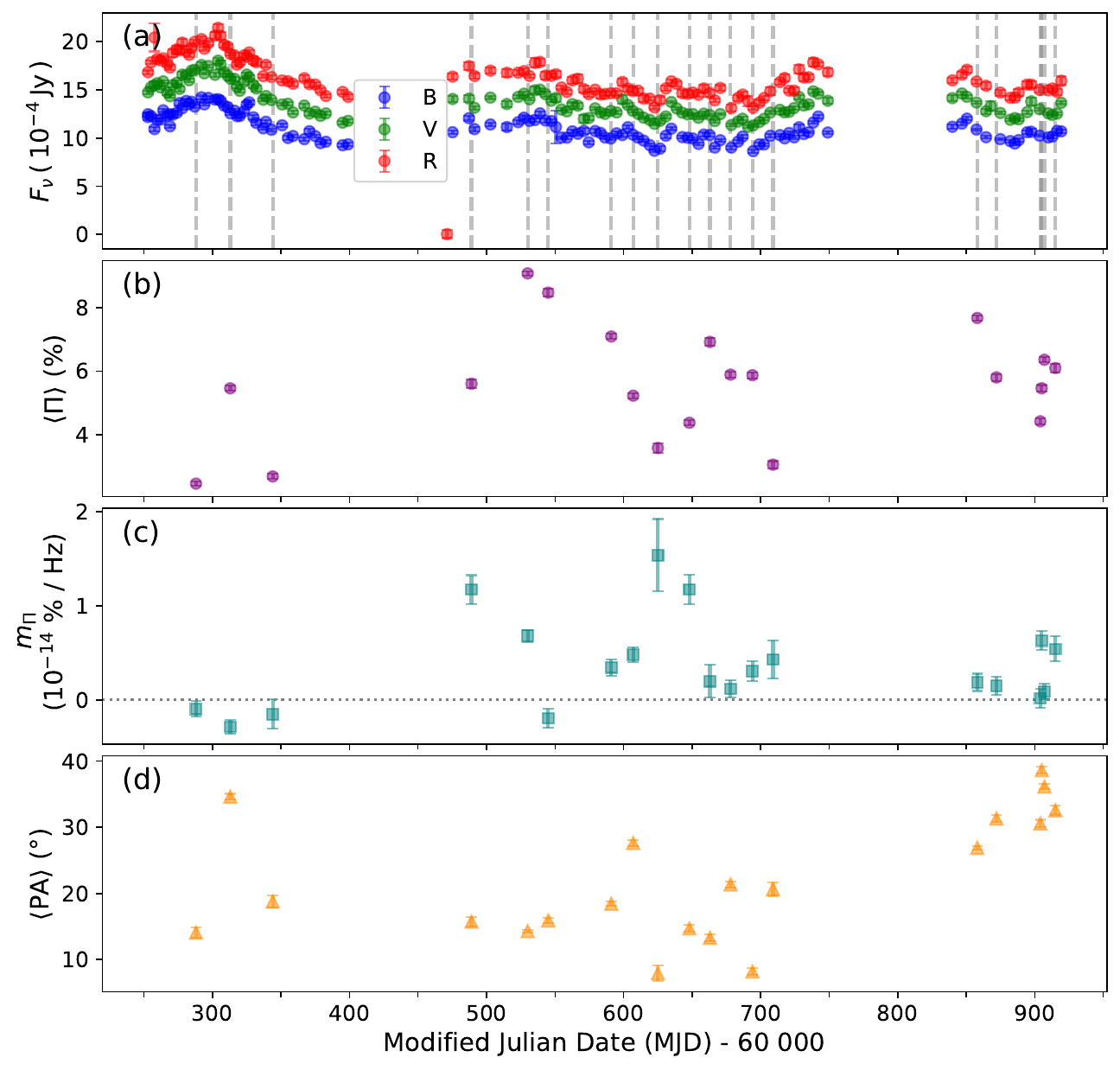}
        \caption{PKS 0301-243}
    \end{subfigure}
    \hfill
    \begin{subfigure}[t]{0.42\textwidth}
        \includegraphics[width=\linewidth]{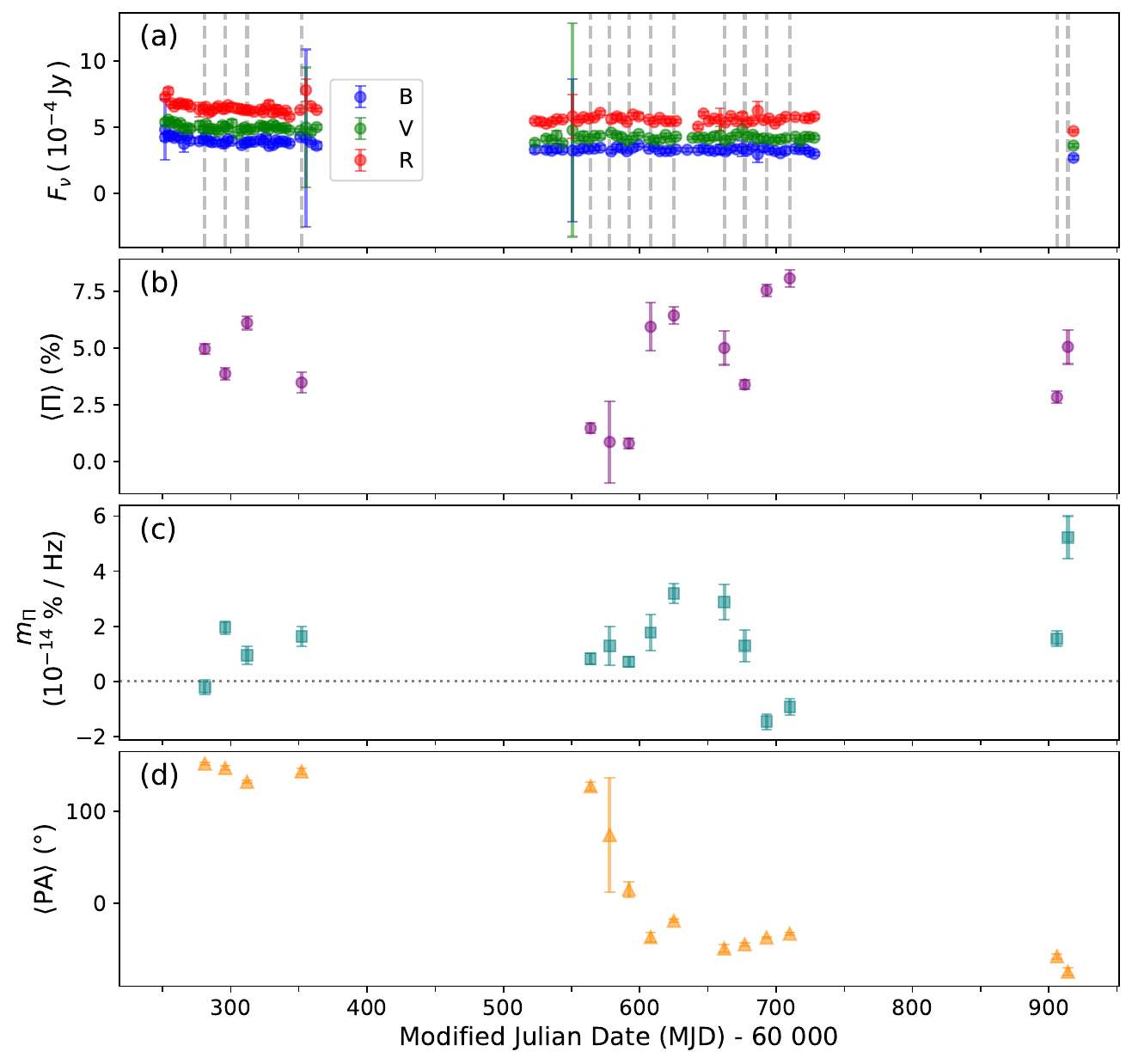}
        \caption{1ES 0414+009}
    \end{subfigure}

    \vspace{0.3cm}

    \begin{subfigure}[t]{0.42\textwidth}
        \includegraphics[width=\linewidth]{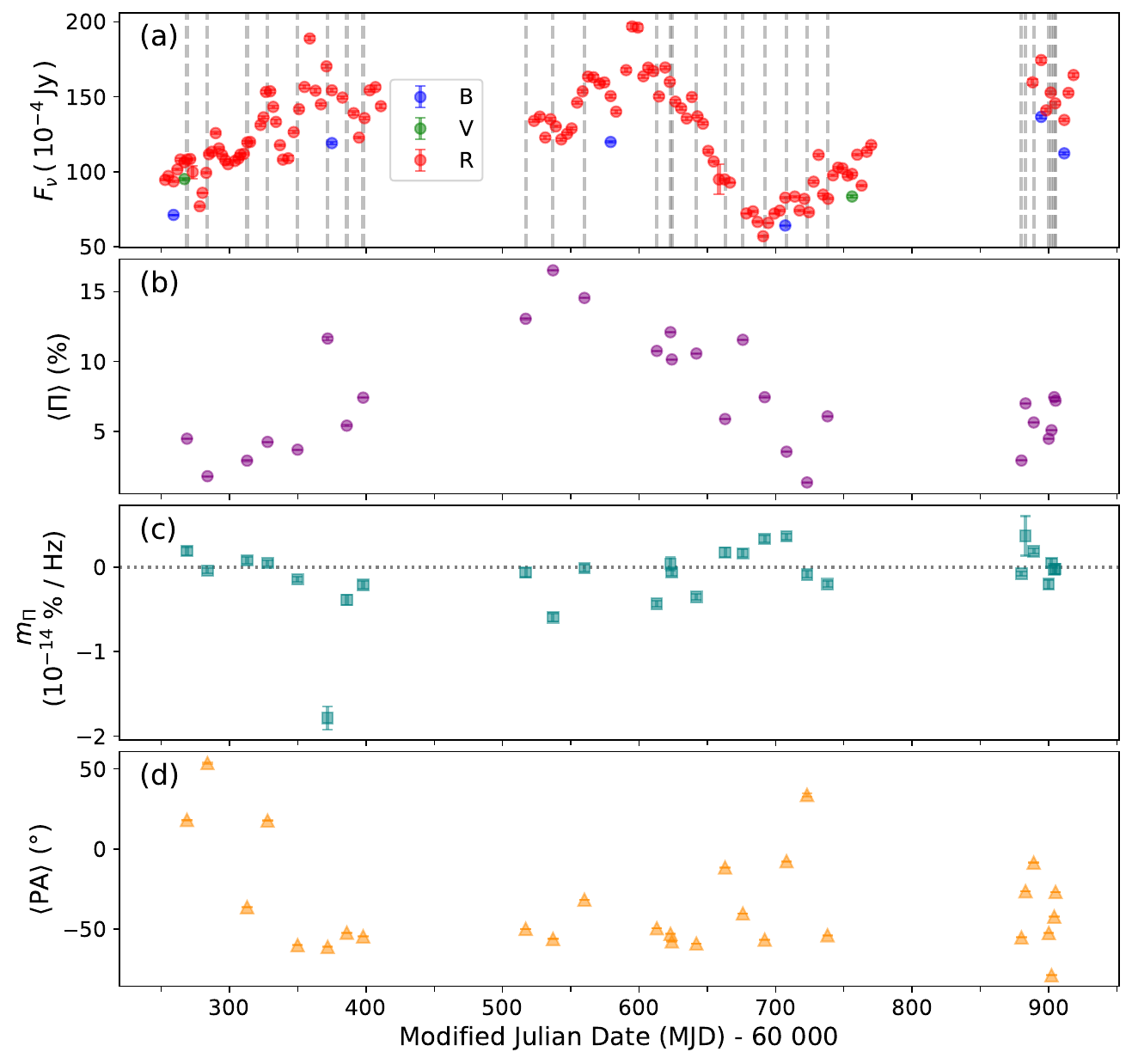}
        \caption{PKS 0447-349}
    \end{subfigure}
    \hfill
    \begin{subfigure}[t]{0.42\textwidth}
        \includegraphics[width=\linewidth]{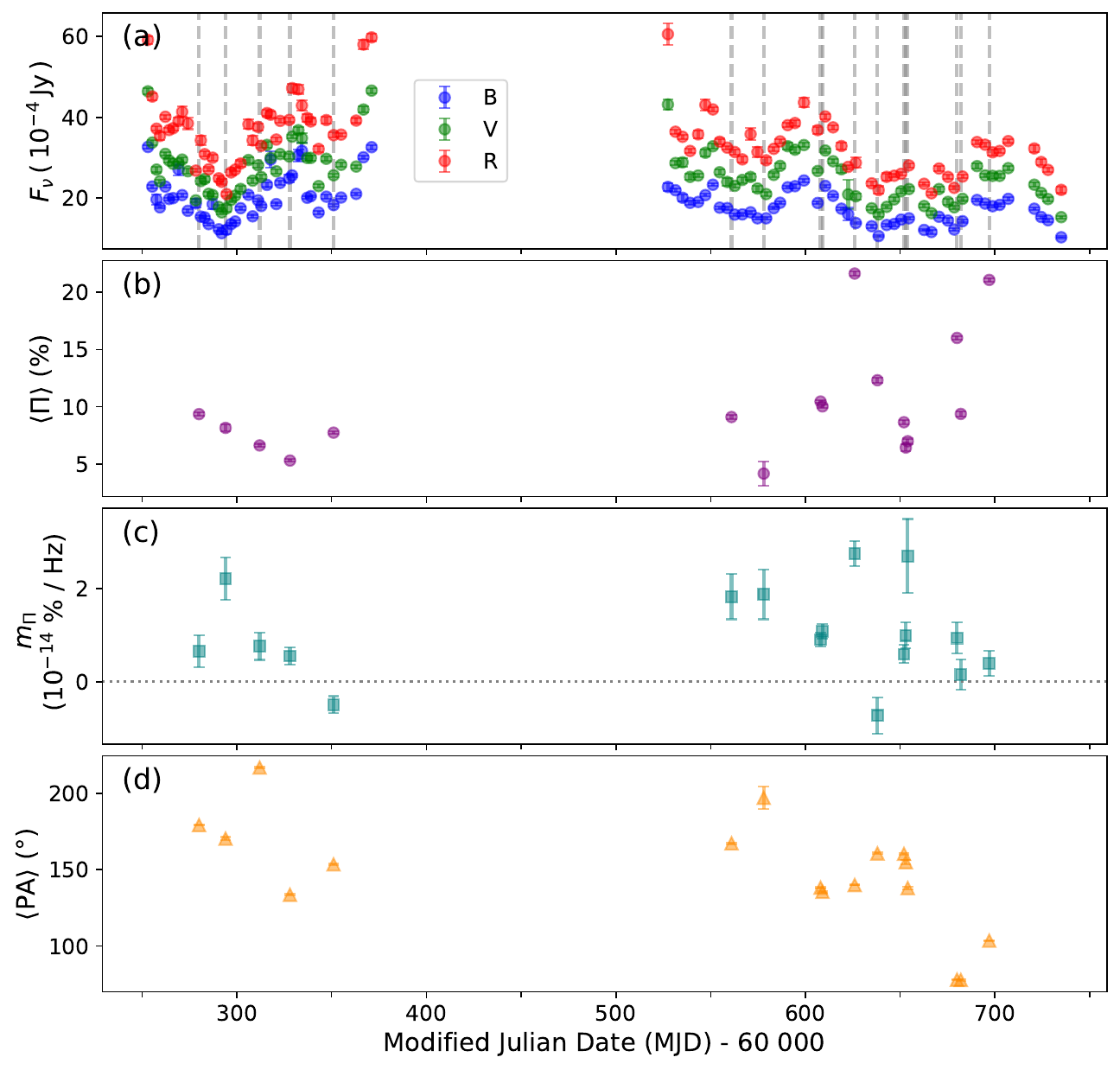}
        \caption{TXS 0506+056}
    \end{subfigure}
\end{figure*}

\begin{figure*}\ContinuedFloat
    \centering
    \vspace{0.3cm}

    \begin{subfigure}[t]{0.42\textwidth}
        \includegraphics[width=\linewidth]{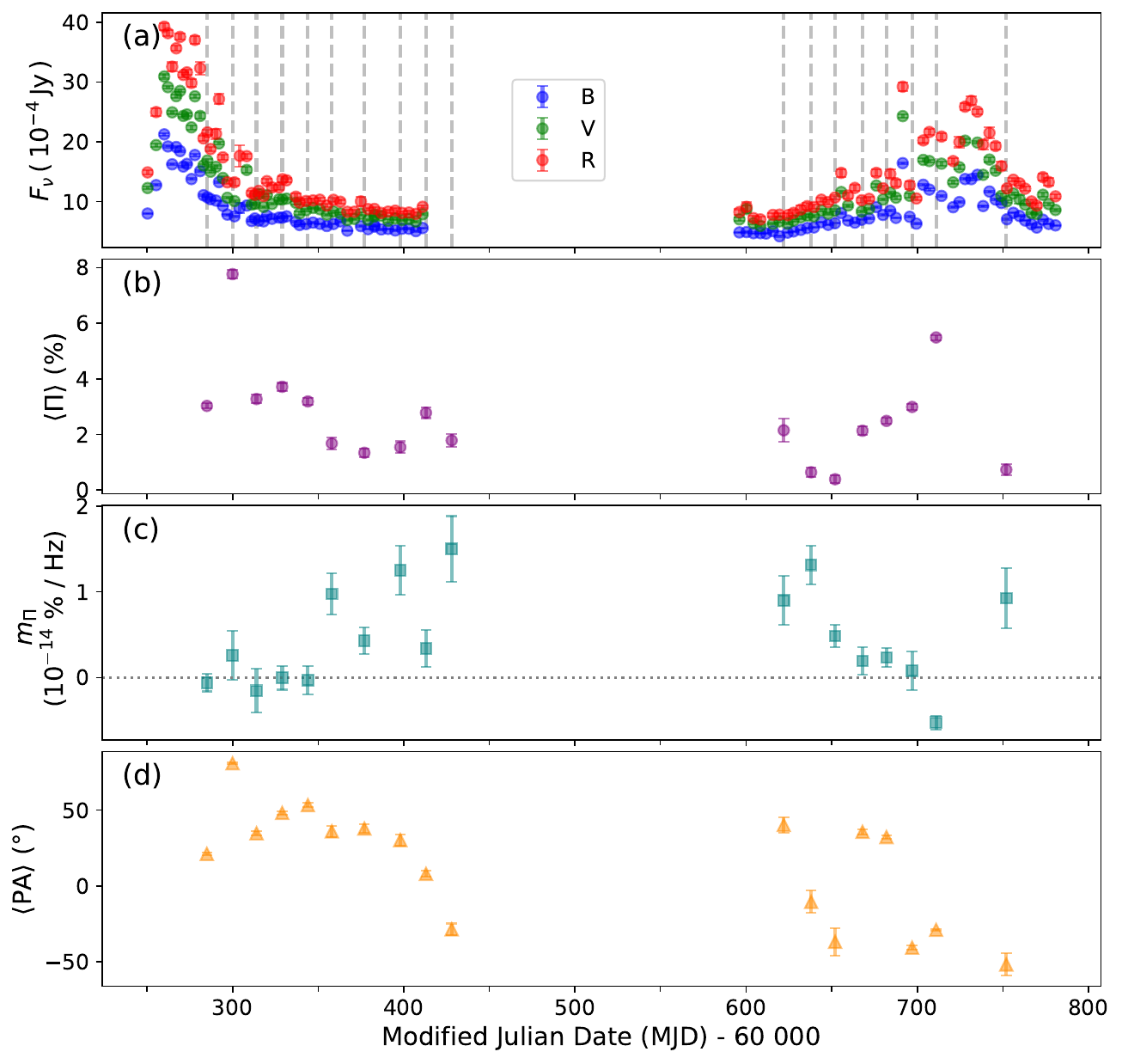}
        \caption{PKS 0736+017}
    \end{subfigure}
    \hfill
    \begin{subfigure}[t]{0.42\textwidth}
        \includegraphics[width=\linewidth]{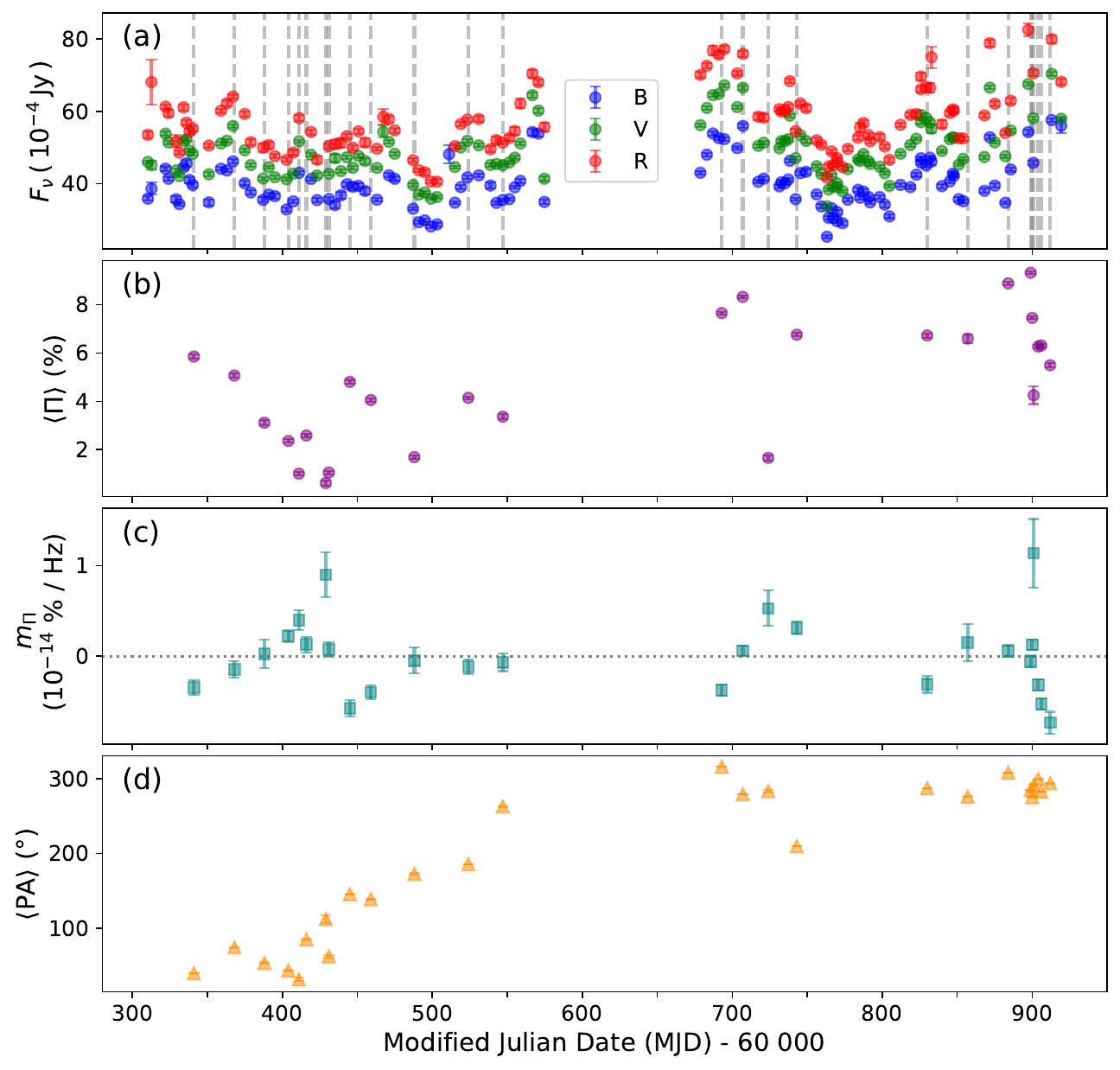}
        \caption{PKS 1440-389}
    \end{subfigure}

    \vspace{0.3cm}

    \begin{subfigure}[t]{0.42\textwidth}
        \includegraphics[width=\linewidth]{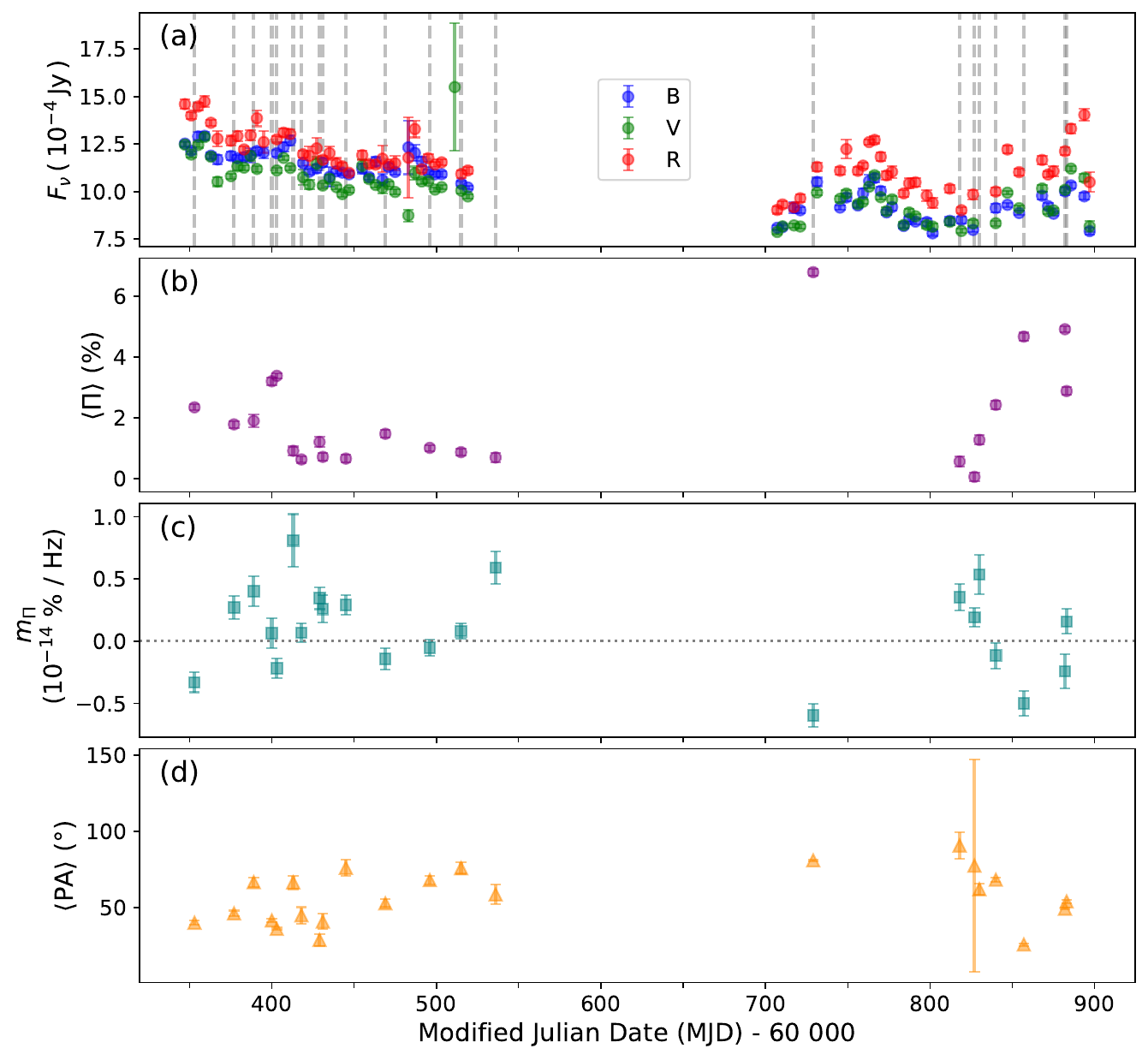}
        \caption{PKS 1510-089}
    \end{subfigure}
    \hfill
    \begin{subfigure}[t]{0.42\textwidth}
        \includegraphics[width=\linewidth]{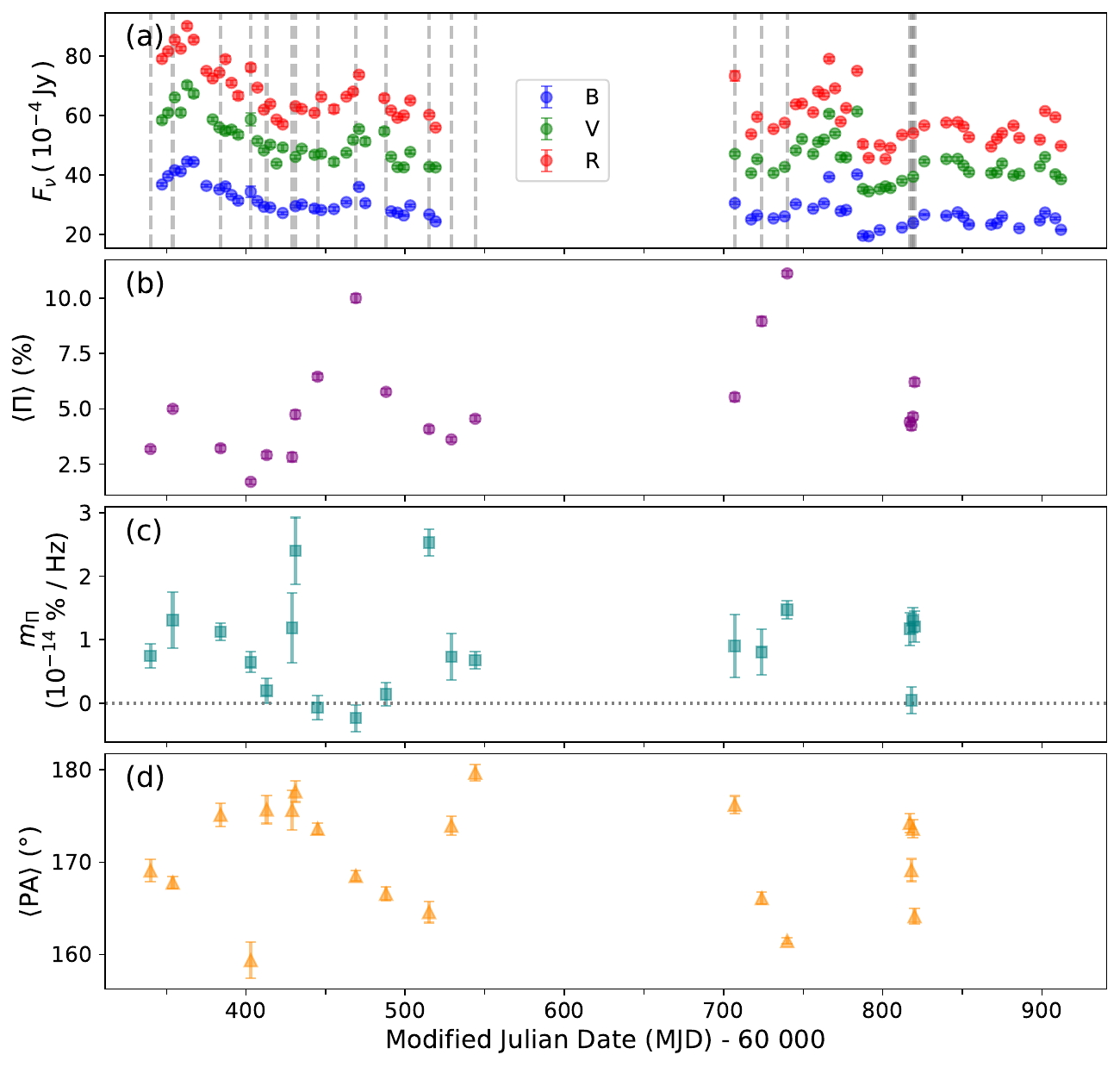}
        \caption{AP Lib}
    \end{subfigure}
    
    \vspace{0.3cm}

    \begin{subfigure}[t]{0.42\textwidth}
        \includegraphics[width=\linewidth]{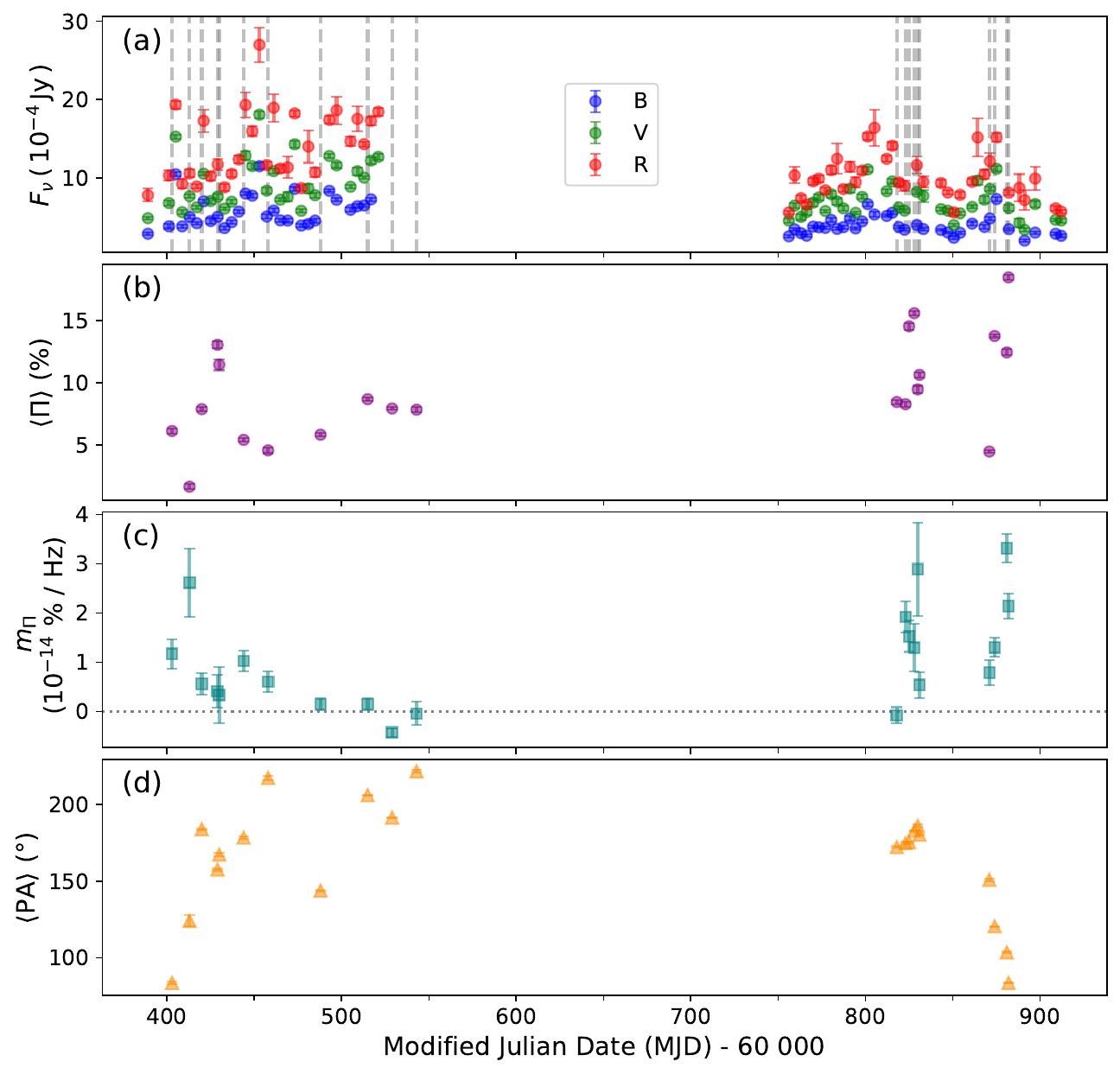}
        \caption{PKS 1749+096}
    \end{subfigure}
    \hfill
    \begin{subfigure}[t]{0.42\textwidth}
        \includegraphics[width=\linewidth]{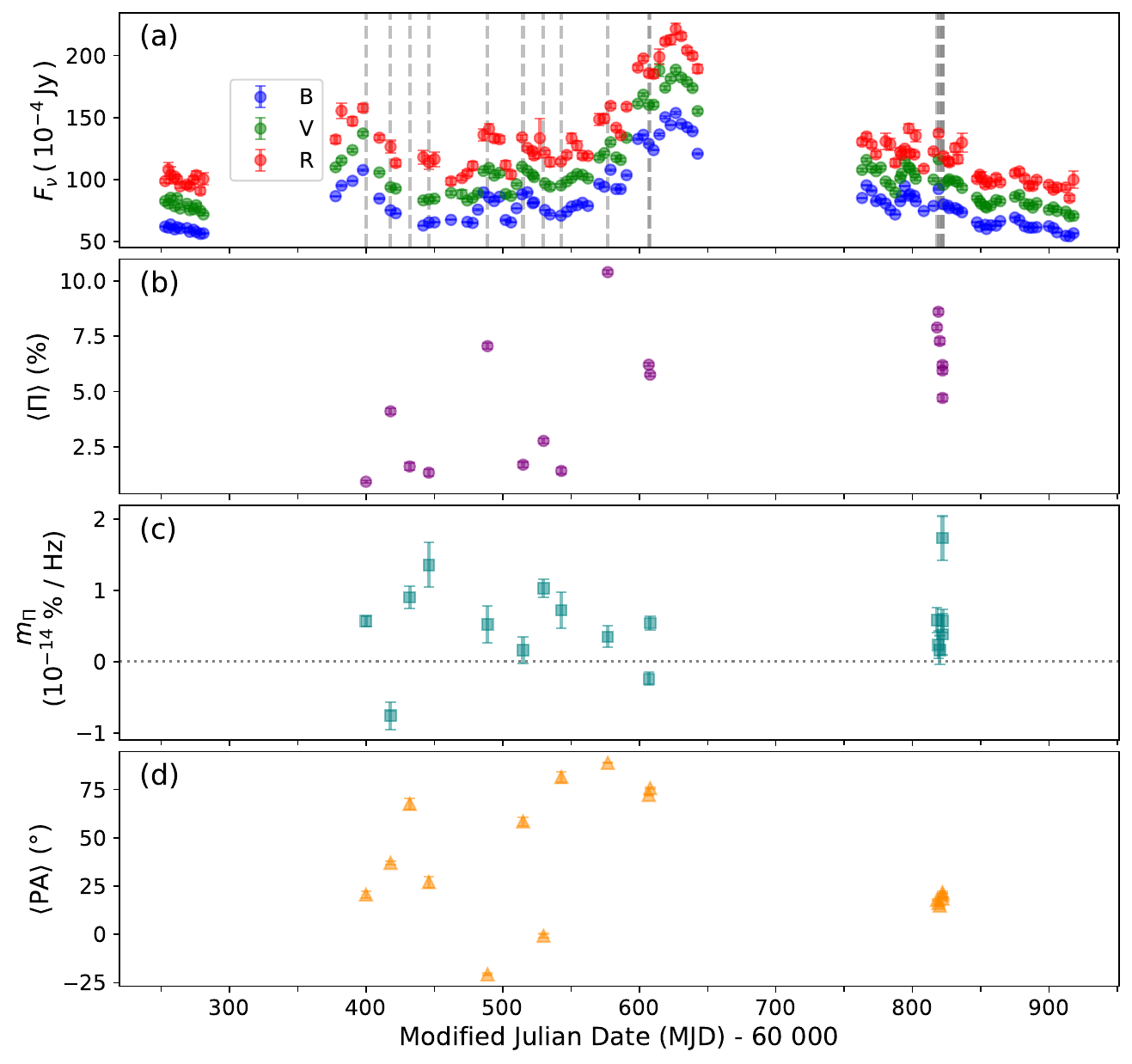}
        \caption{PKS 2005-489}
    \end{subfigure}
\end{figure*}

\begin{figure*}\ContinuedFloat
    \centering

    \vspace{0.3cm}

    \begin{subfigure}[t]{0.42\textwidth}
        \includegraphics[width=\linewidth]{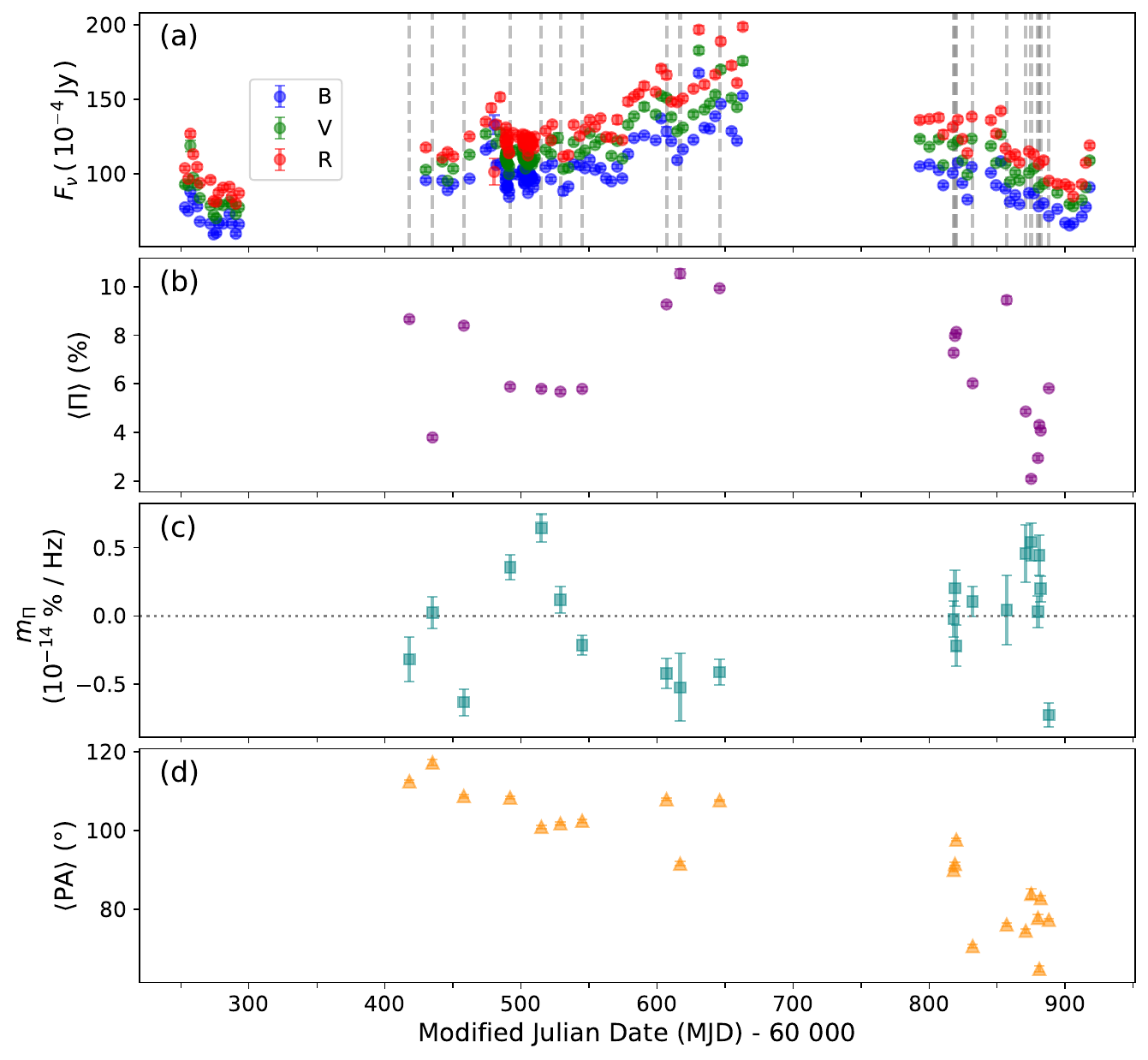}
        \caption{PKS 2155-304}
    \end{subfigure}
    \hfill
    \begin{subfigure}[t]{0.42\textwidth}
        \includegraphics[width=\linewidth]{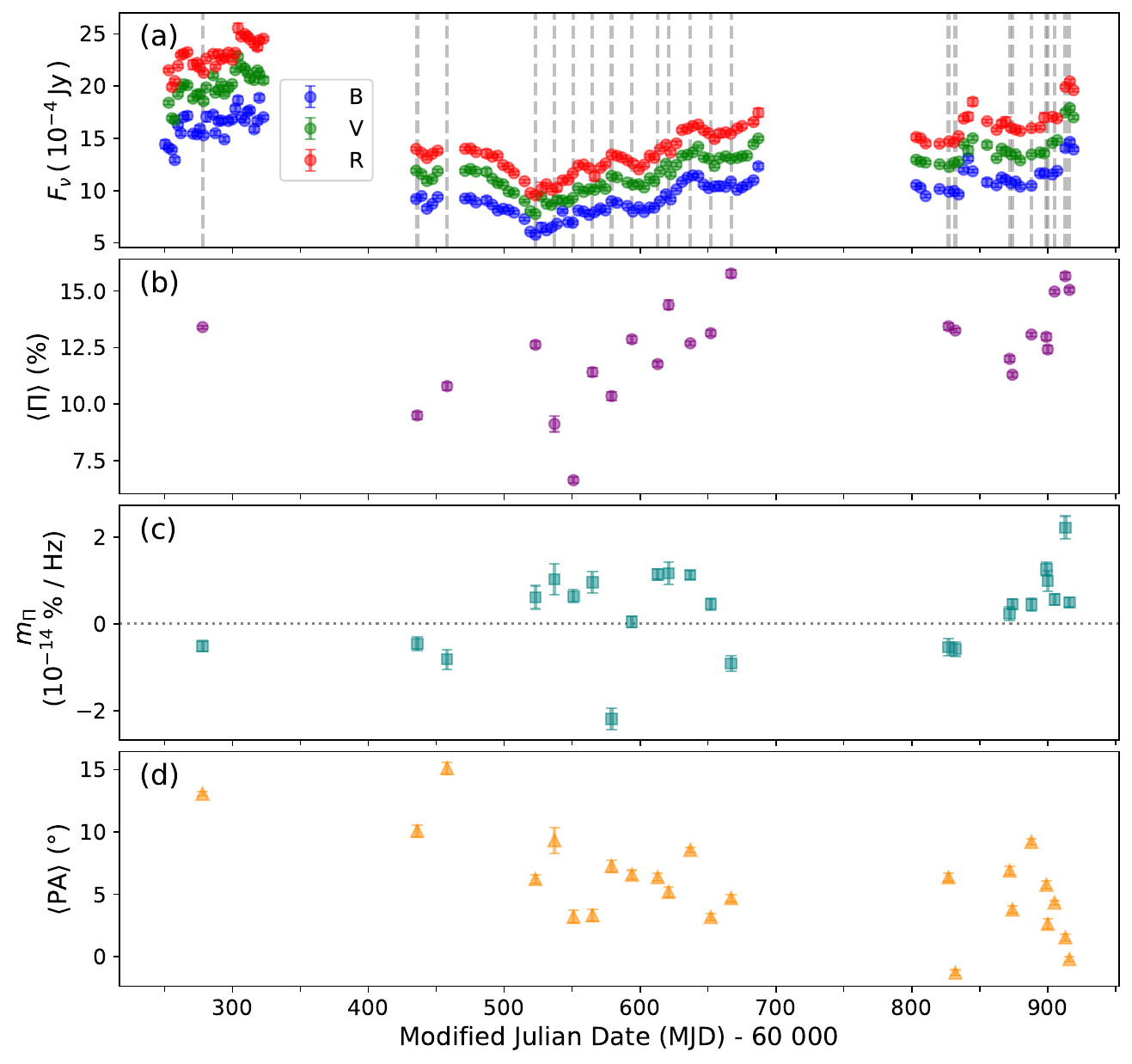}
        \caption{1ES 2322-409}
    \end{subfigure}
    \caption{The optical data of all the sources in the SPOTS monitoring campaign. The layout of these plots is as follows: \textit{a)} the optical photometric light curves of each source in B, V, and R filters (LCOGT data), indicated by the blue circles, green squares, and red triangles, respectively. The dashed gray lines indicate the dates on which SALT observations were taken. \textit{b)} The averaged degree of polarization, reported for each SALT observation. \textit{c)} The frequency dependence of the degree of polarization, where the grey dotted line indicates where the slope is flat. Below this line, the polarization decreases towards increasing frequency, and above it, the polarization increases toward increasing frequency. \textit{d)} The averaged equatorial polarization angle, reported for each SALT observation.}
    \label{fig:all_lcs}
\end{figure*}

\section{Estimating the expected wavelength dependence of the degree of polarization}

\label{app:polarization_approximation}

In order to estimate the expected degree of polarization as a function wavelength using only the optical spectroscopic data, we use the slope of the flux at different wavelengths and calculate the level of polarization that would be produced by a power-law distribution of electrons, that produce this spectral slope. We show examples of this estimate compared to the full calculation for two electron spectra, namely a power-law, 
\begin{equation}
    N(\gamma) = n_0 \gamma^{-p}  \quad \gamma_{\rm min} \leq \gamma \leq \gamma_{\rm max}
\end{equation}
and a broken-power-law with an exponential cut-off, 
\begin{equation}
N(\gamma) = n_0
\begin{cases}
\left( \dfrac{\gamma}{\gamma_b} \right)^{-p_1}
\exp\!\left(-\dfrac{\gamma_b}{\gamma_c}\right),
& \text{if } \gamma \le \gamma_b, \\[10pt]
\left( \dfrac{\gamma}{\gamma_b} \right)^{-p_2}
\exp\!\left(-\dfrac{\gamma}{\gamma_c}\right),
& \text{if } \gamma > \gamma_b.
\end{cases}
\end{equation}
where $\gamma$ is the energy of the electron ($\gamma = E/m_e c^2$), $\gamma_b$ is the break energy, $\gamma_c$ is the cut-off energy and the electron spectrum has a minimum and maximum value of $\gamma_{\rm min}$ and $\gamma_{\rm max}$, respectively. The synchrotron flux is calculated following the expression given in, for example, \citet{2008ApJ...686..181F}. The synchrotron polarization is calculated as 
\begin{equation}
\Pi_{\rm synch}(\nu) = 
\frac{ \int N(\gamma)\, G(x)\, d\gamma}
{ \int N(\gamma)\, F(x)\, d\gamma}.
\label{equ:synchrotron_polarization}
\end{equation}
where, 
\begin{equation}
    F(x) = x \int_x^\infty K_{5/3}(x')\,dx', \quad {\rm and} \quad
G(x) = x K_{2/3}(x). \\[6pt]
\end{equation}
Here, $K$ is the modified Bessel function and $x = \nu/\nu_c$. The critical frequency is given by 
\begin{equation}
\nu_c = \frac{3\, e\, B\, \gamma^2}{4\pi\, m_e\, c},
\label{eqn:critical_freq}
\end{equation}
where $e$ is the charge of an electron, $B$ is the magnetic field, $m_e$ is the mass of the electron, and $c$ is the speed of light.

The approximation is calculated by determining the spectral index, $\alpha$ as a function of frequency, $\nu$, as
\begin{equation}
    \alpha(\nu^\prime) = - \frac{d \log F'_\nu}{d \log \nu'}
\end{equation}
where the primed values are corrected for redshift to the galaxy frame. The polarization as a function of frequency is then estimated as
\begin{equation}
    \Pi(\nu^\prime) = \frac{p(\nu^\prime)+1}{p(\nu^\prime) + 7/3}
\end{equation}
where $p(\nu^\prime) = 2\alpha(\nu^\prime) +1$.

\begin{figure}
    \centering
        \includegraphics[width=\columnwidth]{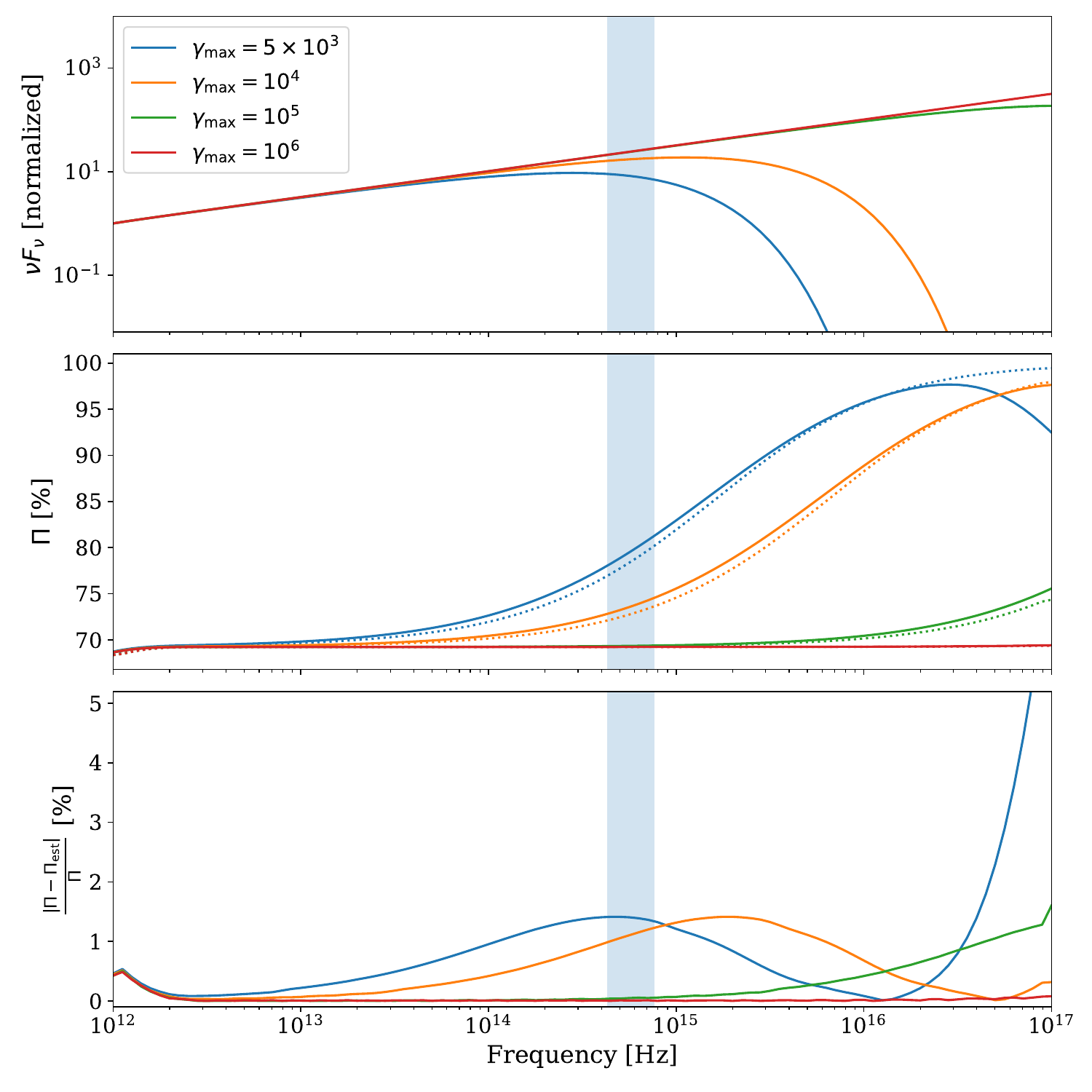}
        \caption{Comparison between the full calculated degree polarization and the approximation applied in this paper for a power-law electron spectrum. The value of $\gamma_{\rm max}$ is changed as indicated in the legend. {\it Top:} Normalized synchrotron SED for different values of $\gamma_{\rm max}$. {\it Middle:} degree of polarization for the full calculation (solid lines) and the estimate (dotted lines). {\it Bottom:} Fractional difference between the full calculation and the approximation. The shaded blue region indicates the frequencies covered by the SALT observations.}
    \label{fig:comparison_PL}
\end{figure}

Fig.~\ref{fig:comparison_PL} shows the example for a power-law electron spectrum where the value of $\gamma_{\rm max}$ is increased from $\gamma_{\rm max}=5\times10^3$ to $10^6$. The base model uses an electron index of $p=2.0$, a minimum energy of $\gamma_{\rm min} = 10^2$, a Doppler factor of $\delta=15$ for the emission region, and a redshift of $z=1$. The match between the approximation and the fully calculated polarization is within 2 per cent, and the accuracy increases if the considered region is not near to the high energy cut-off. 

\begin{figure}
    \centering
        \includegraphics[width=\columnwidth]{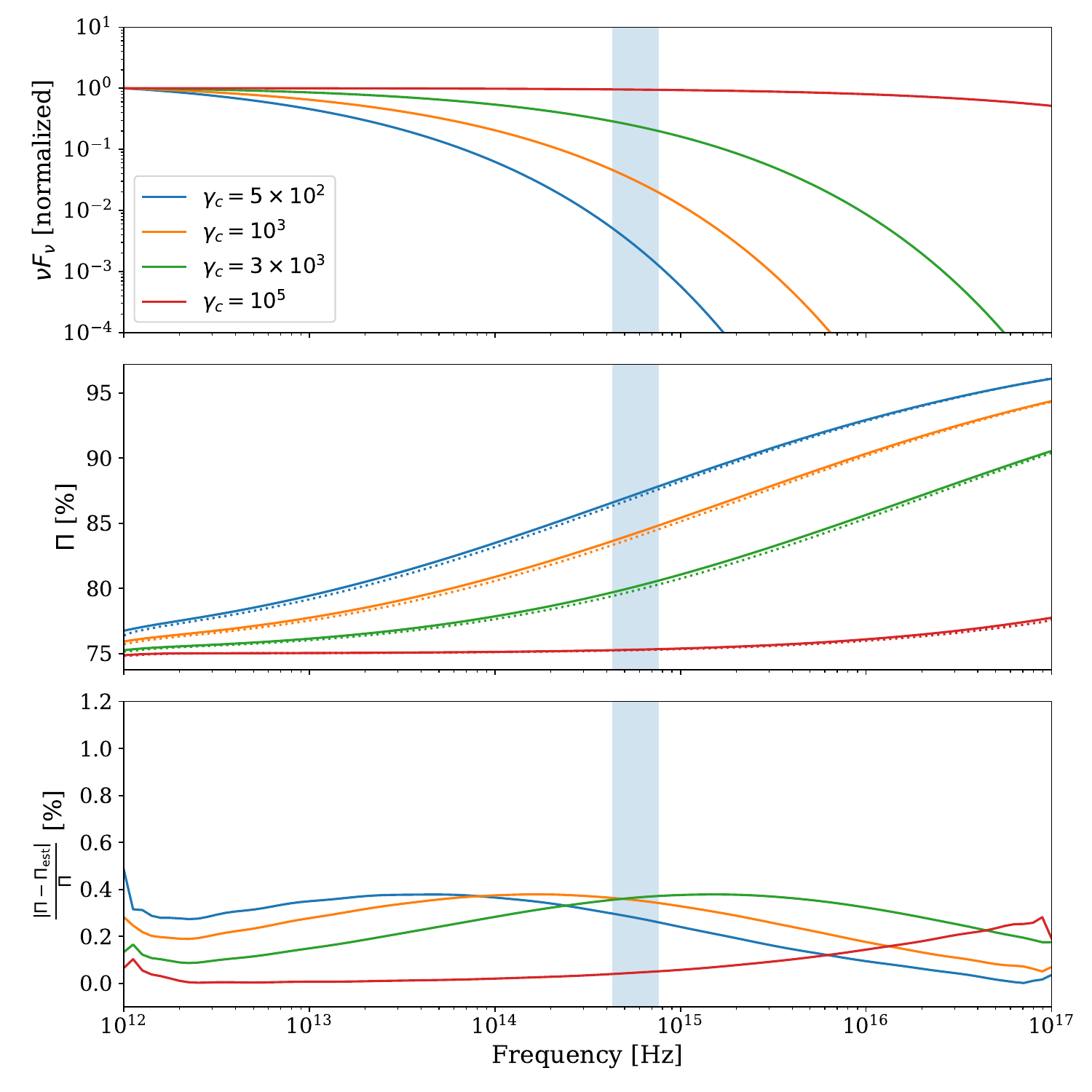}
        \caption{Comparison between the full calculated degree polarization and the approximation applied in this paper for a broken power-law with exponential cut-off electron spectrum. The value of $\gamma_{\rm c}$ is changed as indicated in the legend. {\it Top:} Normalized synchrotron SED for different values of $\gamma_{\rm max}$. {\it Middle:} degree of polarization for the full calculation (solid lines) and the estimate (dotted lines). {\it Bottom:} Fractional difference between the full calculation and the approximation. The shaded blue region indicates the frequencies covered by the SALT observations.}
    \label{fig:comparison_BPL}
\end{figure}

Fig.~\ref{fig:comparison_BPL} shows the example for a broken power-law with exponential cut-off spectrum electron spectrum where the energy of the cut-off, $\gamma_{\rm c}$ is increased from $\gamma_{\rm c}=5\times10^2$ to $10^5$. The base model uses electron indexes of $p_1=2$ and $p_2=3$, with the break held at $\gamma_b=10^2$, a minimum energy of $\gamma_{\rm min} = 10$, a maximum energy of $\gamma_{\rm max} = 10^7$, a Doppler factor of $\delta=15$ for the emission region, and a redshift of $z=1$. The match between the approximation and the full calculation is better than 1 per cent cent for all cases since the cut-off is due to the shape of the electron spectrum and not due to the maximum electron energy. 

%%%%%%%%%%%%%%%%%%%%%%%%%%%%%%%%%%%%%%%%%%%%%%%%%%

% Don't change these lines
\bsp	% typesetting comment
\label{lastpage}
\end{document}